\documentclass[a4paper,11pt]{article}
\usepackage{jheppub} 
\usepackage{amsmath,amssymb,graphicx,float,slashed,xcolor,multicol}
\usepackage{graphicx,tabularx}
\usepackage{url}
\usepackage{footmisc}
\usepackage{amsfonts}
\usepackage{cancel}
\usepackage{color}
\usepackage{xcolor}
\usepackage{multirow} 
\usepackage{amssymb}
\usepackage{pifont}
\usepackage{epstopdf}
\usepackage{slashed}
\usepackage{comment}
\usepackage{booktabs} 
\usepackage{natbib}
\usepackage{array}
\usepackage{mathrsfs}
\usepackage{subcaption}
\usepackage[toc,page]{appendix}
\usepackage{mathtools}
\usepackage{romannum}
\usepackage{ulem}
\usepackage{bbold}
\usepackage{enumitem}
\usepackage{multirow}
\usepackage{cleveref}
\usepackage{amsmath,amssymb,graphicx,float,slashed,xcolor,multicol}
\usepackage{caption,subcaption}
\usepackage{float}
\usepackage{multirow}
\usepackage{slashed}
\usepackage{upgreek}
\usepackage[toc,page]{appendix}
\usepackage{romannum}
\usepackage{soul,xcolor}

\newcommand*{\rom}[1]{\expandafter\@slowromancap\romannumeral #1@}

\def\ra{\rightarrow}

\def\beq{\begin{equation}}
\def\eeq{\end{equation}}
\def\bea{\begin{eqnarray}}
\def\eea{\end{eqnarray}}
\newcommand\rep\mathbf
\title{A Guide to Diagnosing Colored Resonances\\  at Hadron Colliders}
\author[a]{Tao Han,}
\author[a,b]{Ian M. Lewis,}
\author[c]{Hongkai Liu,}
\author[d]{Zhen Liu,}
\author[e]{Xing Wang}

\affiliation[a]{PITT PACC, Department of Physics and Astronomy, University of Pittsburgh, Pittsburgh, PA 15217, USA}
\affiliation[b]{Department of Physics and Astronomy, University of Kansas, Lawrence, KS 66045, USA}
\affiliation[c]{Physics Department, Technion – Israel Institute of Technology, Haifa 3200003, Israel}
\affiliation[d]{School of Physics and Astronomy, University of Minnesota, Minneapolis, MN 55455, USA}
\affiliation[e]{Department of Physics, University of California at San Diego, La Jolla, CA 92093, USA}

\emailAdd{than@pitt.edu}
\emailAdd{ian.lewis@ku.edu}
\emailAdd{liu.hongkai@campus.technion.ac.il}
\emailAdd{zliuphys@umn.edu}
\emailAdd{xiw006@physics.ucsd.edu}

\preprint{
\begin{flushright}
PITT-PACC-2216\\
UMN-TH-4216/23
\end{flushright}
}

\abstract{We present a comprehensive study on how to distinguish the properties of heavy dijet resonances at hadron colliders.  A variety of spins, chiral couplings, charges, and QCD color representations are considered.  Distinguishing the different color representations is particularly difficult at hadron colliders. To determine the QCD color structure, we consider a third jet radiated in a resonant dijet event. We show that the relative rates of three-jet versus two-jet processes are sensitive to the color representation of the resonance. We also show analytically that the antennae radiation pattern of soft radiation depends on the color structure of dijet events and develops an observable that is sensitive to the antennae patterns.
Finally, we exploit a Convolutional Neural Network with Machine Learning techniques to differentiate the radiation patterns from different colored resonances and find encouraging results to discriminate them.
We demonstrate our results numerically at a 14 TeV LHC, and the methodology presented here should be applicable to other future hadron colliders. 
}

\begin{document}
\allowdisplaybreaks
\titlepage

\maketitle

%\newpage

\flushbottom

%%%%%%%%%%%%%%%%%%%%%%%%%%%%%%%%%%

\section{Introduction}
\label{sec:intro}

Searching for new heavy particle resonances is a driving motivation at the energy frontier. Many extensions beyond the Standard Model (SM) predict the existence of new states in a variety of charge and color representations. The classification of dijet resonances and their underlying model correspondence, as well as the search strategy at the LHC have been laid out \cite{Han:2010rf}; the experimental searches are being actively carried out~\cite{CMS:2019gwf,CMS:2019mcu}. Those new states, if kinematically accessible, would lead to large production rates and quite distinctive kinematic signatures in their decay. 
Once a new resonance is discovered at a hadron collider, it will be ultimately important to  scrutinize the underlying dynamics and determine its properties, such as the mass, spin, parity, and gauge charges.
The most challenging of all is to determine a resonant particle's color quantum number. The color structure is extremely difficult to diagnose in a realistic experimental environment since quarks and gluons hadronize into color singlet bound states due to QCD confinement. The hadronization processes necessarily involve QCD soft physics, which renders the underlying dynamics elusive. It would be desirable to develop some techniques for diagnosing the underlying color structure for the signal events and to discriminate against QCD backgrounds effectively. 

Dijet resonances with different color structures have different color flows, leading to distinctive  radiation patterns. This radiation pattern has been used to propose observables to distinguish color octet and singlet resonances  \cite{Ellis:1996eu}.  For a color octet resonance, the initial and final state quarks are color connected whereas for a color singlet, they are not.  Hence, in the scattering plane formed by the beam and two hard final state jets, an octet resonance is expected to have more radiation than a singlet resonance. This observation was used in previous proposals to detect the color of particles.  In Ref.~\cite{Ellis:1996eu}, it was proposed to look at the antennae behavior of gluon radiation to determine if a resonance decaying into a quark-antiquark pair is a color singlet or octet.   Reference \cite{Gallicchio:2010sw} analyzed the radiation patterns inside jets to separate singlet from octet color flows.  Similar color flow ideas have been applied to distinguishing color octet and singlet dijet events~\cite{Ellis:1996eu}, top pair tagging~\cite{Hook:2011cq}, and searching for double Higgs production \cite{Kim:2019wns}.   Machine learning techniques and two-point correlators have also been used to distinguish pair-produced color singlets and octets decaying into quark-antiquark pairs~\cite{Chakraborty:2019imr}.

This paper provides a comprehensive guide to diagnose the properties of a singly produced colored heavy resonance at high-energy hadron colliders.  We move beyond the typical color singlet versus  octet classification and consider the various resonances classified in Ref.~\cite{Han:2010rf}.  That is, different color representations such as triplets and sextets; various spins such as scalars, fermions, vectors, and tensors; and resonances produced by and decaying into all possible partons: quark-antiquark, quark-gluon, and gluon-gluon.  We start by presenting the standard methodology used to determine 
the spin and couplings of resonance to quarks and gluons.  

To understand the color structure of events, we study dijet resonance events with an additional radiated gluon.  We present an analytical understanding of the antennae radiation pattern of soft gluons for the various resonances, extending the results of Ref.~\cite{Ellis:1996eu}.  As we show, a particularly powerful observable to distinguish different colored resonance is to compare the 3-jet and 2-jet rates.  This observable has been useful for testing SM predictions for pure QCD, vector boson plus jet, and Higgs plus jet~\cite{Ellis:1985vn,Englert:2011cg,Englert:2011pq,Gerwick:2011tm}. A similar approach has been proposed to use the ratio of the dijet cross section to the total width of the resonance, if the width could be measured, as the color discriminant variable~\cite{Atre:2013mja, SekharChivukula:2014kof, SekharChivukula:2014uoe, Chivukula:2015zma, Chivukula:2017nvl}.

From the analytical understanding of the antennae patterns, we develop a collider observable sensitive to the resonance's different color structures.  As we show, this observable can in principle distinguish the different color representations, and its behavior is largely independent of the spin of the resonance.  Hence, it provides a robust test of the color structure of the events. 
In the process of analyzing the large data sample at the LHC or future colliders, ``deep-learning'' (or machine-learning, ML) techniques have been well-developed and proved to be quite fruitful for exploring 
the rich physics and uncovering the subtle features otherwise inaccessible. Recent successful examples include Lorentz boosted boson tagging~\cite{deOliveira:2015xxd,Chen:2019uar,Ju:2020tbo}, top~\cite{Almeida:2015jua}, bottom~\cite{Aad:2015ydr}, and strange~\cite{Nakai:2020kuu} quarks tagging, and quark/gluon jet discrimination~\cite{Cogan:2014oua, Komiske:2016rsd}.  
We exploit the machine-learning techniques in the hope of improving the analyses and distinguishing different colored  resonances. We use a convolutional neural network (CNN) as an example to demonstrate how ML techniques can help.

The rest of the paper is organized as follows. In Section~\ref{sec:class}, we review our classifications of the different possible resonances, and lay out the standard techniques and ideas for determining the spin and chiral coupling as well as the color radiation pattern. In Section~\ref{sec:cut}, we perform a cut-based analysis to observe the LHC antennae radiation pattern. Section~\ref{sec:ml} analyzes  the ability to distinguish different colored resonance by using Machine Learning techniques at the LHC. We conclude in Section~\ref{sec:con}. Although our numerical results are shown for  a 14 TeV LHC, the methodology presented here should be applicable to other future hadron colliders. 

%%%%%%%%%%%%%%%%%%%%%%%%%%%%%%%%%%%%%%%%%%%%%%
\section{Classification and characteristics of color resonances}
\label{sec:class}

%%%%%%%%%%%%%%%%%%%%%%%%%%%%%%%
\subsection{Resonances and interactions}
\label{Models.sec}
%%%%%%%%%%%%%%%%%%%%%%%%%%%%%%%%%%%%%%%%%%%%%%
\begin{table}[tb]
\begin{center}
\begin{tabular}{|c|c|c|c|c|c|}  \hline
 Particle Names & $J$  & $SU(3)_{C}$  & $|Q_{e}|$ & $B$ & Related models \\
 (leading coupling) &  &  &  &  &  \\ \hline
$E_{3,6}^{(\mu)}\ (uu)$       &0(1)& ${\rep3},\ \overline{\rep6}$ &
$4\over3$ & $ {2\over3} $ & scalar/vector diquarks \\ \hline
$D_{3,6}^{(\mu)}\ (ud)$       &0(1)& ${\rep3}, \ \overline{\rep6}$ &
$1\over3$ & $ {2\over3} $ & scalar/vector diquarks; ${\tilde d}$ \\
\hline $U_{3,6}^{(\mu)}\ (dd)$       &0(1)& ${\rep3},\
\overline{\rep6}$ & $2\over3$ & $ {2\over3} $ & scalar/vector
diquarks; $\tilde u$ \\  \hline\hline
 $U^{*(\mu)}_{3,6}\ (ug)$ &$1\over2$($3\over2$) & $\rep3,\ \bar{\rep6}$ & ${2\over 3}$ & $ {1\over3} $ & excited $u$;
 quixes; stringy \\ \hline
  $D^{*(\mu)}_{3,6}\ (dg)$ &$1\over2$($3\over2$)& $\rep3,\ \bar{\rep6}$ & ${1\over 3}$ & $ {1\over3} $ & excited $d$;
 quixes; stringy \\  \hline\hline
 $S_{8}\ (gg)$   &0& $\rep8_{S}$ &  $0$ & $0$ & $\pi_{TC},\ \eta_{TC} $ \\ \hline
  $T_{8}\ (gg)$   &2& $\rep8_{S}$ &  $0$ & $0$ & stringy \\ \hline \hline
  $V_{8}\ (u\bar u,\ d\bar d)$      &1& $\rep8$          &  $0$   & $ 0 $  & axigluon; $g^{}_{KK},\ \rho_{TC}$; coloron \\ \hline
  $V_8^{\pm}\ (u\bar d)$  &1& $\rep8$          &  $1$   & $ 0 $  & $\rho^{\pm}_{TC}$; coloron \\ \hline
   $V_1\ (u\bar u)$  &1& $\rep1$          &  $0$   & $ 0 $  & $Z^{\prime}$ \\ \hline
\end{tabular}
\end{center}
\caption{Summary for resonant particle names, their quantum numbers, and possible underlying models \cite{Han:2010rf}.}
\label{qnum.tab2}
\end{table}

The dijet resonances are classified according to their electric and $SU(3)_C$ color charges. Here we briefly review possible resonances according to these two conserved quantum numbers.  Table~\ref{qnum.tab2} summarizes the different colored resonances discussed in this section. We list our notation for the different states along with the leading couplings to SM partons and spin, color representation, and electric charge of each state.  A more detailed discussion, including examples of specific realizations of the various resonances in existing literature, is given in Ref.~\cite{Han:2010rf}.

It is beneficial to consider the color-resonances according to their production mechanisms from the initial state partons. 
Quark-quark annihilation can produce color anti-triplet or sextet scalars~\cite{Pati:1974yy,Mohapatra:1980qe,Hewett:1988xc,Chacko:1998td,Atag:1998xq,Barbier:2004ez,Babu:2006wz,Han:2009ya} and vectors~\cite{Arik:2001bc,Cakir:2005iw,Berger:2010fy,Zhang:2010kr}, so-called ``diquarks".  
The possible scalar diquark are denoted as  $E_{N_D},~U_{N_D}$, and $D_{N_D}$ with electric charges $-4/3,\ 2/3,\ -1/3$ respectively.  The subscript $N_D=3,~6$ for the $\rep3$ and $\rep{\mathbf6}$ color representations, respectively.
Vector diquarks of spin-1 are represented with an additional Lorentz index $\mu$.
The interaction Lagrangian between quarks and diquarks is then 
\begin{eqnarray}
\mathcal{L}_{qqD} &=& K^j_{ab} \left[ \lambda^{E,\tau}_{\alpha\beta}
E^j_{N_D}\ \overline{u^C_{\alpha a}}P_\tau u_{\beta b}
+\lambda^{U,\tau}_{\alpha\beta} U^j_{N_D}\ \overline{d^{C}_{\alpha a}}P_\tau d_{\beta b}
+\lambda^{D,\tau}_{\alpha\beta} D^j_{N_D}\ \overline{d^C_{\alpha b}}P_\tau u_{\alpha a} \right.\nonumber \\
&
+&\lambda^{E',\tau}_{\alpha\beta} E^{j\mu}_{N_D}\  \overline{u^C_{\alpha a}}\gamma_\mu P_\tau u_{\beta b}
+ \lambda^{U'\tau}_{\alpha\beta}  U^{j\mu}_{N_D}\ \overline{d^C_{\alpha a}}\gamma_\mu P_\tau d_{\beta b}   
\left.
+\lambda^{D',\tau}_{\alpha\beta}\ D^{j\mu}_{N_D} \overline{u^C_{\alpha a}}\gamma_\mu P_\tau d_{\beta b}
\right]
+\rm{h.c.},
\label{eq:qq}
\end{eqnarray}
where $P_{\tau}={1\over2}(1\pm\gamma_5)$ with $\tau=R,L$ for the right- and left-chirality projection operators and
$K^j_{ab}$ are $SU(3)_{C}$ Clebsch-Gordan (CG) coefficients with  the quark color indices $a,b=1,2,3$, and the diquark color index $j=1,...,N_D$. Explicit forms of the CG coefficients and sextet representation matrices can be found in Ref.~\cite{Han:2009ya}.

Quarks and gluons annihilate into color triplet~\cite{Cullen:2000ef,Burikham:2004su,Anchordoqui:2008hi,Anchordoqui:2008di,Dong:2010jt,Kim:2018mks,Alhazmi:2018whk,Criado:2019mvu,Belyaev:2021zgq,Hassanain:2009at,Moussallam:1989nm,Stirling:2011ya,Dicus:2012uh,Christensen:2013aua,Hagiwara:2010pi,Burges:1983zg} or anti-sextet~\cite{Frampton:1987dn,Martin:1992aq,Celikel:1998dj} fermions with $1/2$ or $3/2$ spin.  It is possible to produce a $\rep15$-plet, but the existence of such a fermion would spoil asymptotic freedom~\cite{Chivukula:1990di}.  
The spin-$1/2$ ($3/2$) fermion states are denoted by $D^*_{N_D},U^*_{N_D}$ ($D^{*\mu}_{N_D},U^{*\mu}_{N_D}$) with electric charged $-1/3$ and $2/3$, respectively.
The lowest order gauge invariant interaction between a gluon, quark, and heavy spin-$1/2$ fermion is dimension-five:  
\begin{eqnarray}
\displaystyle\mathcal{L}_{qgF} &=  \displaystyle\frac{g_{s}}{\Lambda}F^{A,\alpha\beta}
%\hskip -0.2in
 &\left[\overline{U}_{N_D}^* {K_{N_D}^A} (\lambda^U_LP_L+\lambda^U_RP_R)\sigma_{\alpha\beta} u + \overline{D}^*_{N_D} {{K}_{N_D}^A} (\lambda^D_LP_L+\lambda^D_RP_R)\sigma_{\alpha\beta} d ~~~~\right. \nonumber\\
&&  + \left.\overline{U}^{*\mu}_{N_D} {K_{N_D}^A} \left(g_{\beta\mu}+z\,\gamma_\mu\gamma_\beta\right)\gamma_\alpha (\lambda^U_LP_L+\lambda^U_RP_R)u  \right.\nonumber\\
&&+ \left.\overline{D}^{*\mu}_{N_D} {K_{N_D}^A} \left(g_{\beta\mu}+z\,\gamma_\mu\gamma_\beta\right)\gamma_\alpha (\lambda^D_LP_L+\lambda^D_RP_R)d   \right ] + \rm{h.c.}
\label{eq:qstar}
\end{eqnarray}
where $A=1,...,8$ is the adjoint color index, $F^{A,\mu\nu}$ is the gluon field strength tensor, $\Lambda$ is the scale of new physics, $z$ is a constant which does not contribute for on-shell spin-3/2 particles, and $K_{N_D}^{A}$ are $3\times N_D$ CG coefficient matrices.  Explicit forms of these CG coefficients can be found in Appendix~\ref{app:CG}.  A spin-$3/2$ fermion is described by the Rarita-Schwinger spinor~\cite{Rarita:1941mf,Moldauer:1956zz,osti_4283250,Haberzettl:1998rw} and we give a review of the Lagrangian for spin-3/2 fields in Appendix~\ref{app:spin32}.

Gluon-gluon annihilation can result in many different representations. 
A complete list of the possible resonances from gluon-gluon annihilation can be found in Table 1 of Ref.~\cite{Han:2010rf}. 
Two possible color-octet resonances that can result from gluon-gluon annihilation are of particular interest: a scalar $S_8$~\cite{Eichten:1986eq,Hill:2002ap,Gresham:2007ri,Choi:2008ub,Plehn:2008ae,Idilbi:2009cc}, and a tensor $T_8$~\cite{Burikham:2004su,Anchordoqui:2008hi,Anchordoqui:2008di,Dong:2010jt}.  These interactions can be described in a gauge invariant way by dimension five operators: 
\begin{eqnarray}
\mathcal{L}_{gg8}=g_sd^{ABC}\bigg{(}\frac{\kappa_S}{\Lambda_S} S_{8}^{A} F^B_{\mu\nu}F^{C,\mu\nu}+\frac{\kappa_{T}}{\Lambda_T}({T^{A,\mu\sigma}_8}F^{B}_{\mu\nu}{F^{C}_{\sigma}}^{~\nu}+
f {T_{8\ \rho}^{A,\rho}} \
F^{B,\mu\nu}F^C_{\mu\nu})\bigg{)},
\label{tensscal.EQ}
\end{eqnarray}
where $\Lambda_{S,T}$ are the new physics scales, and the relative coupling factor $f$ is expected to be order one.  The symbol $d^{ABC}$ is fully symmetric and defined by the anti-commutation relations
\begin{equation}
\{T^A,T^B\}=\frac{1}{N_C}\delta^{AB}+d^{ABC}T^C,
\end{equation}
where $T^A$ are the $SU(3)_C$ fundamental representation matrices.  The subscript $S$ on $\rep8_S$ in Tab.~\ref{qnum.tab2} indicates that this color octet representation is the symmetric combination of two other octets, as shown in Eq.~(\ref{tensscal.EQ}).

Finally, quark-antiquark annihilation can produce color octet~\cite{Frampton:1987dn,Bagger:1987fz,Hill:1991at,Hill:1993hs,Chivukula:1996yr,Cullen:2000ef,Burikham:2004su,Agashe:2006hk,Lillie:2007ve,Lillie:2007yh,Anchordoqui:2008hi,Anchordoqui:2008di,Dong:2010jt} or singlet scalars and vectors with zero or unit charge.  The neutral vector-octet is denoted by $V_8$ and the  charged vector octet states $V^\pm_8$.  The interaction Lagrangian is then
\begin{eqnarray}
\mathcal{L}_{q\bar{q}V}&=&g_s\left[ {V_8}^{A,\mu}\ \bar{u} T^A \gamma_\mu (g^U_L P_L+g^U_{R}P_R)u+
{V_8}^{A,\mu}\ \bar{d} T^A \gamma_\mu (g^D_L P_L+g^D_{R}P_R)d \right .\nonumber \\
&+&\left .\left(V_8^{+,A,\mu}\ \bar{u} T^A \gamma_\mu (C_L V^{CKM}_LP_L+C_R V^{CKM}_R P_{R})d+\rm{h.c.}\right)\right ] ,
\label{qqV.EQ}
\end{eqnarray}
where $V^{CKM}_{L,R}$ are the left- and right-handed Cabibbo-Kobayashi-Maskawa (CKM) matrices, respectively.  To avoid constraints from flavor physics, it is assumed that the charged current interactions are proportional to the SM CKM matrices and that there is no tree-level flavor changing neutral currents, {\it i.e.}, $g^{U,D}_{L,R}$ and $C_{L,R}$ are flavor-diagonal.  To obtain the interactions with the color singlet vector bosons replace the representation matrices ${T^{A,a}}_b$ with the Kronecker delta ${\delta^a}_b$.  It is also more natural to write the coupling in terms of the weak coupling constant instead of the strong coupling constant.  
The neutral and charged color singlet vectors will be denoted as $V_1$ and $V_1^\pm$, respectively. %}  
The couplings between the octet and singlet scalar and light quarks are constrained to be small by minimal flavor violation~\cite{Manohar:2006ga}.  Hence, we ignore their contributions as $s$-channel resonances.

All resonances listed in Table~\ref{qnum.tab2} couple to SM partons and will contribute to dijet signals at the LHC.  If a dijet resonance is discovered it will be imperative to disentangle the properties of the resonant particle, such as mass, spin, and color representation. In the following subsections, we investigate what information can be gleaned from a dijet resonance.  We illustrate the methods to measure the spin, interactions, and color structure of a resonance, and comment on their limitations. 

\subsection{Initial-state couplings}
\label{rap.sec}
%%%%%%%%%%%%%%%%%%%%%%%%%%%%%%%%%%%%%%%%%%%%%%%

\begin{figure}[tb]
\centering
\includegraphics[clip,width=0.5\textwidth]{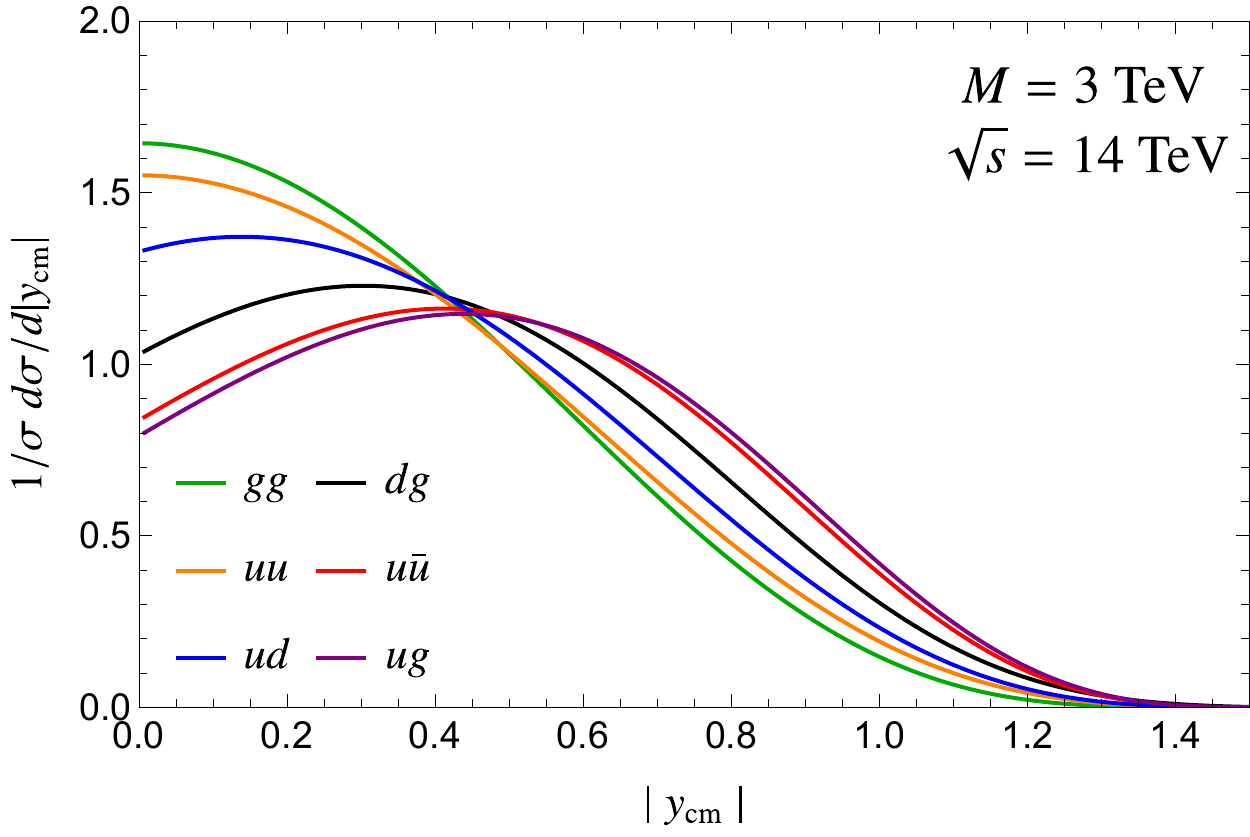}
\caption{Normalized rapidity distributions of different initial states at 3 TeV partonic center-of-momentum energy at the $14$ TeV LHC. 
}
\label{fig:rap}
\end{figure}

The species of initial state partons that produce the resonance may be probed by the rapidity distribution of the di-jet system. Defining the partonic center-of-momentum (c.m.)~system rapidity as 
\begin{equation}
y_{cm}={1\over 2}\ln{x_1\over x_2},
\end{equation}
where $x_{1,2}$ are the parton momentum fractions of the initial state. The distribution of the c.m. system rapidity reflects the imbalance between the momenta of the two incoming partons, governed by their parton distribution functions (PDFs). 
Gluons and sea quarks typically carry lower momentum fractions than valence quarks.  Hence, the $qg, q \bar q$ initial states with a valence quark tend to have a broader rapidity distribution, and may indeed have a peak at a non-zero value of $y$. In contrast, $gg$ and $qq$ initial states have a more concentrated distribution in rapidity, with the peak typically near zero, as $x_1\approx x_2$. We show some features of the rapidity distributions in Fig.~\ref{fig:rap} for $gg$, $dg$, $ug$, $uu$, $ud$, and $u\bar u$ initial states. Here and henceforth, for the sake of illustration, we choose a hard scattering partonic c.m. energy $\sqrt{\hat{s}}=M=3$ TeV at the $14$ TeV LHC, where $M$ is the resonance mass. 
As can be seen, the $gg$ rapidity is the most highly peaked  due to the symmetric $gg$ initial state. While, $ug$, with an asymmetric initial state, has the broadest rapidity distribution and is peaked at the highest rapidity. Those qualitative features may provide circumstantial information for the initial partonic states.

%%%%%%%%%%%%%%%%%%%%%%%%%%%%%%%%%%%%%%%%%%%%%%%
\subsection{Spin}
\label{angular.sec}
%%%%%%%%%%%%%%%%%%%%%%%%%%%%%%%%%%%%%%%%%%%%%%
 
The spin of an $s$-channel resonance determines the angular correlations between the initial and final states.  Hence, by analyzing the angular distributions of two-to-two processes near the invariant mass peak, the spin of the resonances can be determined. 
The two-to-two process is not sensitive to the color representation of a resonance. Hence, the relevant classification of resonances for angular distributions is according to spin: scalar, spin-$1/2$ fermion, vector, spin-$3/2$ fermion, and tensor.  
Details for the matrix element calculation and rates for each resonance see Appendix~\ref{HelAmp.app}.  We are only interested in the distributions, so we factor out the total partonic cross-section of a spin-$J$ resonance $\hat{\sigma}_{J}(\hat{s})$, which is governed by its coupling strength. 

%%%%%%%%%%%%%%%%%%%%%%%%%%%%%%%%%%%%%%%%%%%%%%%%
%\subsubsection{Analytical Results}
%%%%%%%%%%%%%%%%%%%%%%%%%%%%%%%%%%%%%%%%%%%%%%%%
There are four examples of scalar resonances listed in Section~\ref{Models.sec}: three scalar diquarks and an octet scalar. 
There are no spin correlations between the initial and final states in dijet events because of the scalar nature. Hence, there is no angular dependence in the partonic cross-section as it must be isotropic, and the differential cross-section $d\hat\sigma_0/d\cos\theta$ is just a flat distribution in the partonic c.m.~frame.

\begin{figure}[tb]
\centering
      \includegraphics[width=0.4\textwidth]{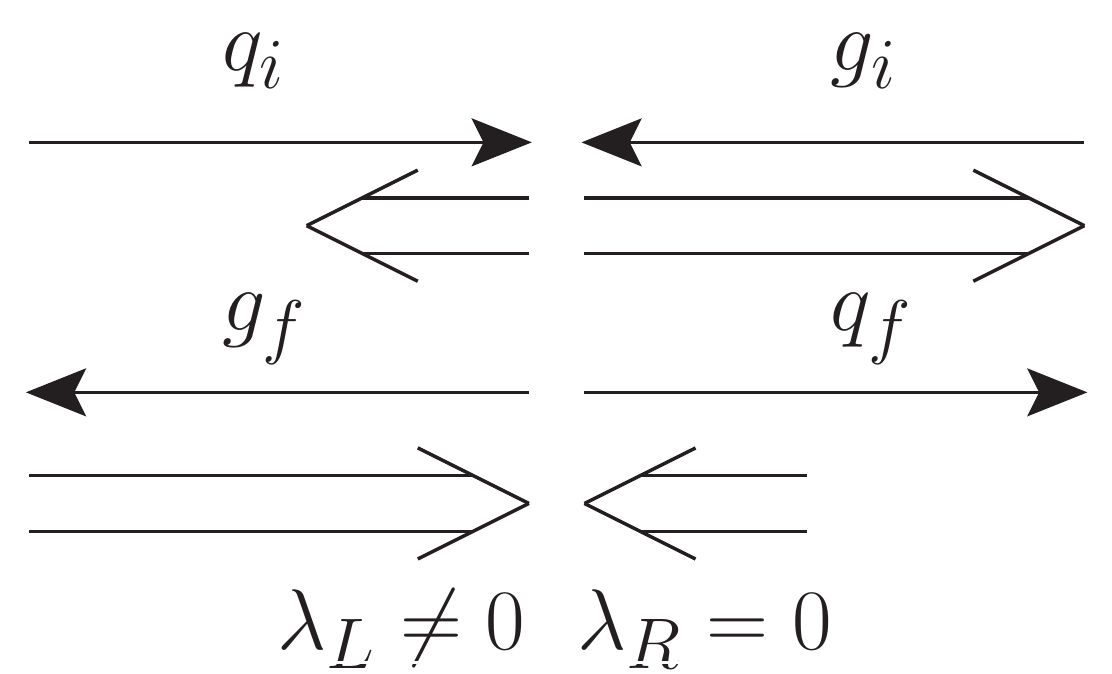}
      \includegraphics[width=0.4\textwidth]{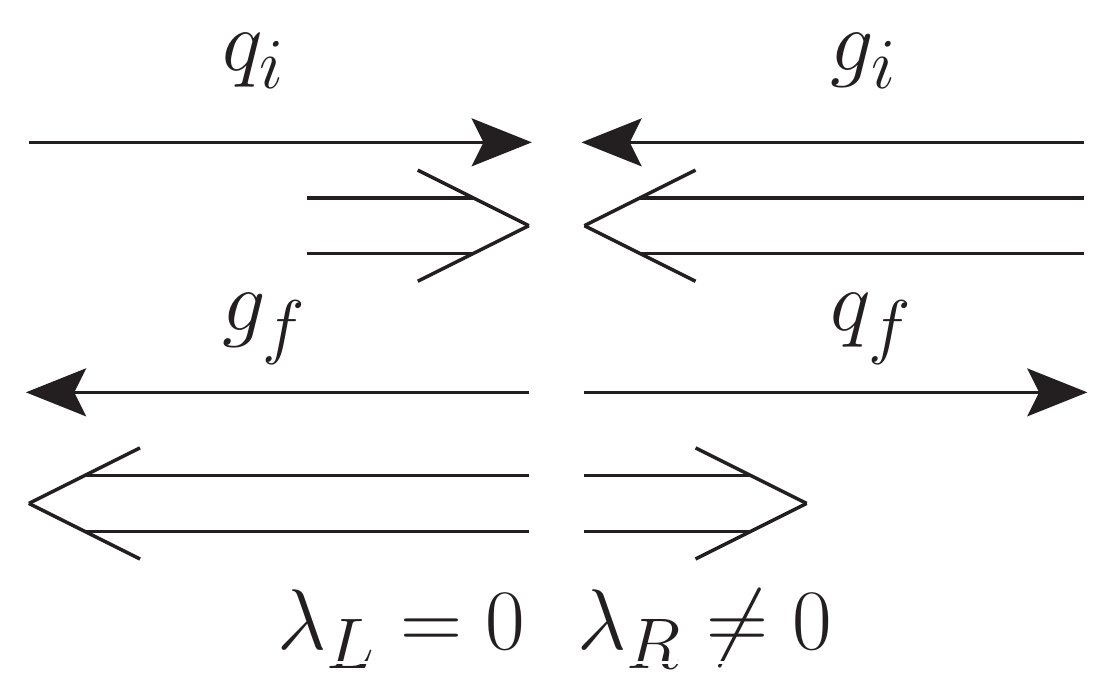}
\caption{Spin correlations for color sextet and anti-triplet spin-$1/2$ fermion resonance production assuming pure (left) left-handed and (right) right-handed couplings.  Single-arrowed lines represent momentum in the c.m. frame and double-arrowed lines represent spin.  The subscript $i$ ($f$) indicates initial (final) state particles.  Longer double arrows indicate spin-1 particles, shorter for spin-1/2.}
\label{qgqgspin.fig}
\end{figure}

We explore two examples of spin-1/2 and 3/2 resonances: the color sextet and antitriplet fermions, from a quark-gluon annihilation. Allowing for both left- and right-handed couplings in Eq.~(\ref{eq:qstar}) would 
result in large corrections to the fermionic magnetic moments \cite{Brodsky:1980zm}. 
Hence, we will assume that the couplings are either purely left- or right-handed. The spin correlations for $qg\rightarrow q^*\rightarrow qg$ with spin-1/2 resonance are illustrated in Fig.~\ref{qgqgspin.fig} for (left) left-handed and (right) right-handed couplings. For both types of couplings, the spin of the initial and final state quarks are in the same direction and, hence, the final state quark will preferentially move in the direction of the initial state quark.  Therefore, in the partonic c.m.~frame, the angular distribution for a spin-1/2 resonance is
\begin{equation}
\frac{d\hat\sigma_{1/2}}{d\cos\theta}=\frac{1}{2}\hat\sigma_{1/2}(\hat s=M^2)\left(1+\frac{|\lambda_{i,L}|^2-|\lambda_{i,R}|^2}{|\lambda_{i,L}|^2+|\lambda_{i,R}|^2}\frac{|\lambda_{f,L}|^2-|\lambda_{f,R}|^2}{|\lambda_{f,L}|^2+|\lambda_{f,R}|^2}\cos\theta\right),
\label{fermang.eq}
\end{equation}
where $\theta$ is the angle between the initial state and final state quarks, and the subscripts $i,j$ indicated initial and final state couplings, respectively. We note that for any chiral coupling $\lambda_L \ne \lambda_R$, there will be a forward-backward asymmetry linearly proportional to $\cos\theta$.

Directly analogous to the discussion for the spin-1/2 color sextet and antitriplet fermions, we also consider the spin-3/2 resonances. 
 The partonic angular distribution for an on-shell resonance is given as 
\begin{equation}
\frac{d\hat\sigma_{3/2}}{d\cos\theta}=\frac{1}{2}\hat\sigma_{3/2}(\hat s=M^2)\left[1+3\cos^2\theta+\frac{|\lambda_{i,L}|^2-|\lambda_{i,R}|^2}{|\lambda_{i,L}|^2+|\lambda_{i,R}|^2}\frac{|\lambda_{f,L}|^2-|\lambda_{f,R}|^2}{|\lambda_{f,L}|^2+|\lambda_{f,R}|^2}\cos\theta\left(3+\cos^2\theta\right)\right].~~~
\end{equation}
Again, for a chiral coupling $\lambda_L \ne \lambda_R$, 
there will be a forward-backward asymmetry scaled to $3\,\cos\theta + \cos^3\theta$.
A spin-$3/2$ resonance leads to higher powers in $\cos\theta$ due to the multiple partial-wave contributions.
To measure the above angular distributions, 
the directions of the initial state and final state quarks must be determined.  On average, valence quarks have a higher momentum fraction of the proton than sea quarks or gluons.  Previous studies~\cite{Langacker:1984dc} have utilized this momentum imbalance to identify the reconstructed partonic system direction with the initial state valence quark direction, as already discussed in the previous section. However, even if we are able to statistically determine the momentum direction of a valance quark, it is still a real challenge to identify the correlated momentum direction of the final state quark, i.e. distinguish the quark and gluon jets. There has been much previous work on measuring the differences between quark and gluon jets at LHC~\cite{Gallicchio:2011xq,Komiske:2016rsd,Komiske:2018cqr,Kasieczka:2018lwf,Larkoski:2019nwj}. These techniques are subtle and need experimental verification at the LHC. If we 
treat gluon and quark jets as indistinguishable, the jet angular distribution is symmetrized and the spin-$1/2$ angular distribution has no difference from the scalar angular distribution.

\begin{figure}[tb]
\centering
      \includegraphics[width=0.4\textwidth]{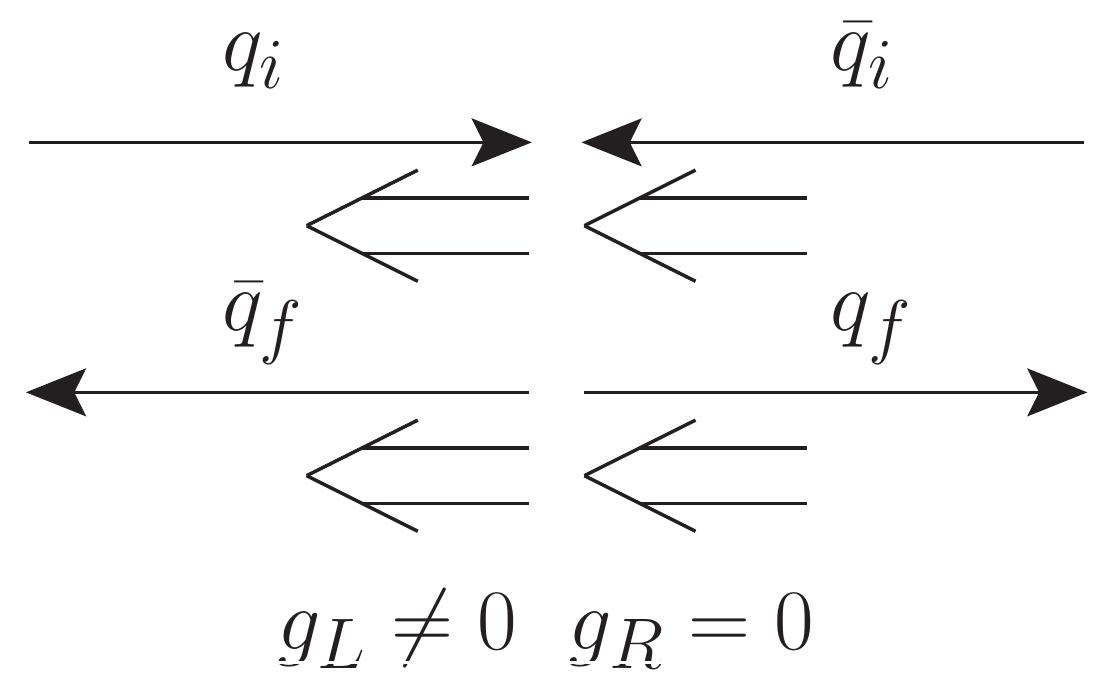}
      \includegraphics[width=0.4\textwidth]{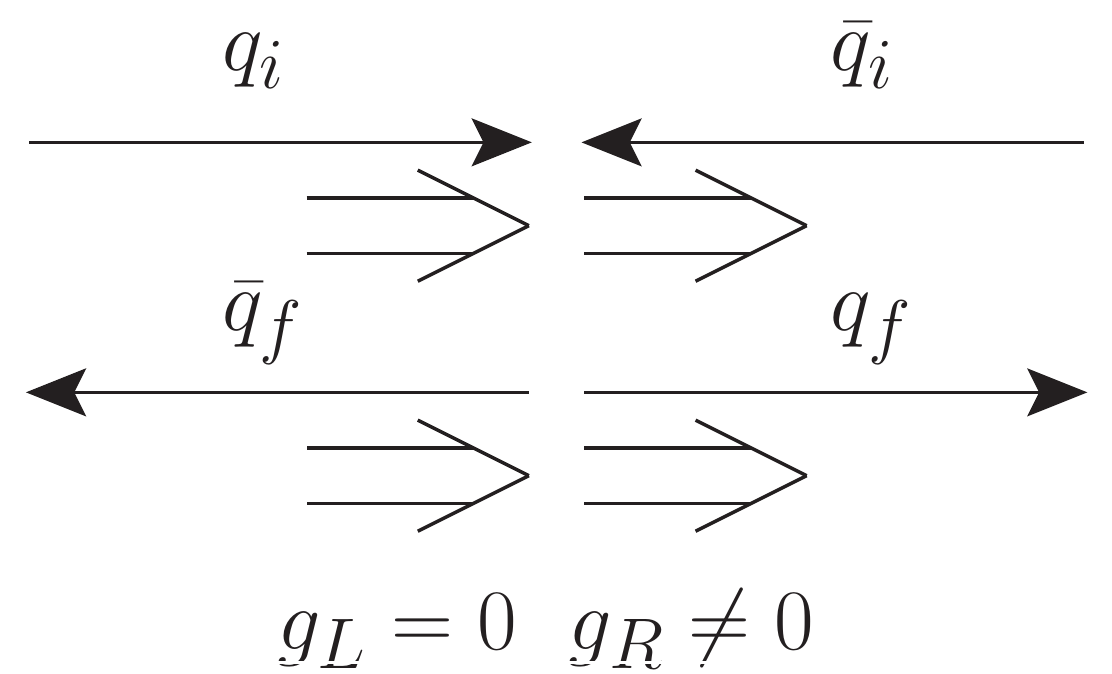}
\caption{Spin correlations for color singlet and octet vector resonance production assuming pure (left) left-handed and (right) right-handed couplings.  Single-arrowed lines represent momentum in the c.m. frame and double-arrowed lines represent spin.  The subscript $i$ ($f$) indicates initial (final) state particles.}
\label{qqqqspin.fig}
\end{figure}

In Section~\ref{Models.sec} four vector resonances with different color representations were introduced: color singlet, triplet, anti-sextet, and octet. For the color triplet and anti-sextet diquark vector production the dijet process is $qq\rightarrow qq$ while for the color singlet and octet vectors the process is $q\bar{q}\rightarrow q\bar{q}$.  In general, the angular distribution of those spin-1 states is
\begin{equation}
\frac{d\hat\sigma_1}{d\cos\theta}=\frac{3}{8}\hat\sigma_1(\hat s)\left(
1+\cos^2\theta+2\frac{g_{i,L}^2-g_{i,R}^2}{g_{i,L}^2+g_{i,R}^2}\frac{g_{f,L}^2-g_{f,R}^2}{g_{f,L}^2+g_{f,R}^2}  \cos\theta \right),
\label{eq:PV}
\end{equation}
where $\theta$ is the angle between the initial state and final state quarks, $g_{i,L},g_{i,R}$ are initial state chiral couplings, and $g_{f,L},g_{f,R}$ are final state chiral couplings.  
In general, chiral couplings $\lambda_L \ne \lambda_R$ lead to a forward-backward asymmetry that is linearly proportional to $\cos\theta$.
In the diquark vector cases, either final state quark can be spin correlated with either initial state quark.  Hence, the left- and right-chiral couplings are equal and the angular distribution is completely symmetrized at the partonic level. %However, t
The characteristic distribution for a spin-1 vector state remains and it leads to the well-known non-chiral symmetric form
\begin{equation}
\left.\frac{d\hat\sigma_1}{d\cos\theta}\right|_{g_L=g_R}=\frac{3}{8}\hat\sigma_1(\hat s)(1+\cos^2\theta).
\label{eq:smear}
\end{equation}
For the color-octet and singlet vectors, their spin correlations and chiral couplings are not necessarily equal.  Figure \ref{qqqqspin.fig} illustrates the spin correlations for (left) pure left-handed and (right) right-handed couplings.  If the chiral couplings are purely right-handed or purely left-handed the spin of the initial state quark (antiquark) is correlated with the final state quark (antiquark), and the final state quark preferentially moves in the direction of initial state quarks.  In the case of both left- and right-handed chiral couplings, the helicity of the initial state and final state quarks (antiquarks) are not necessarily the same.  Hence, in principle, the shape of the angular distribution contains information about the relative strengths of the chiral couplings, similar to that in Eq.~(\ref{eq:PV}). Unfortunately,
it is very difficult to distinguish a quark jet from an anti-quark jet~\cite{Field:1977fa,Krohn:2012fg,Waalewijn:2012sv,Fraser:2018ieu}. In the indistinguishable limit, the observed angular distribution will be of the symmetric form as in Eq.~(\ref{eq:smear}).

\begin{figure}[tb]
\centering
      \includegraphics[width=0.4\textwidth]{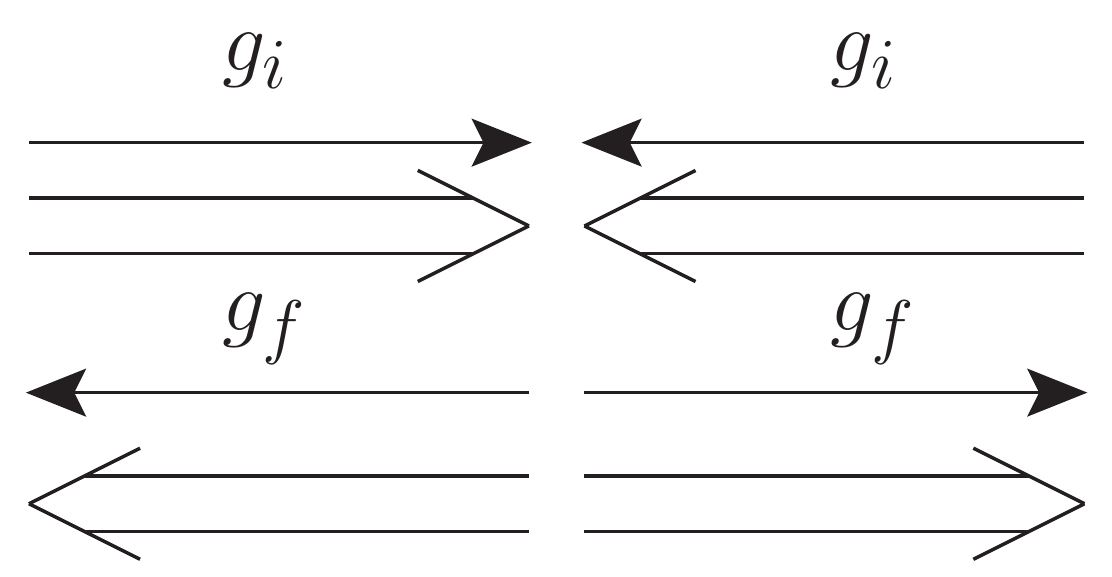}
\label{gRgR.fig}
      \includegraphics[width=0.4\textwidth]{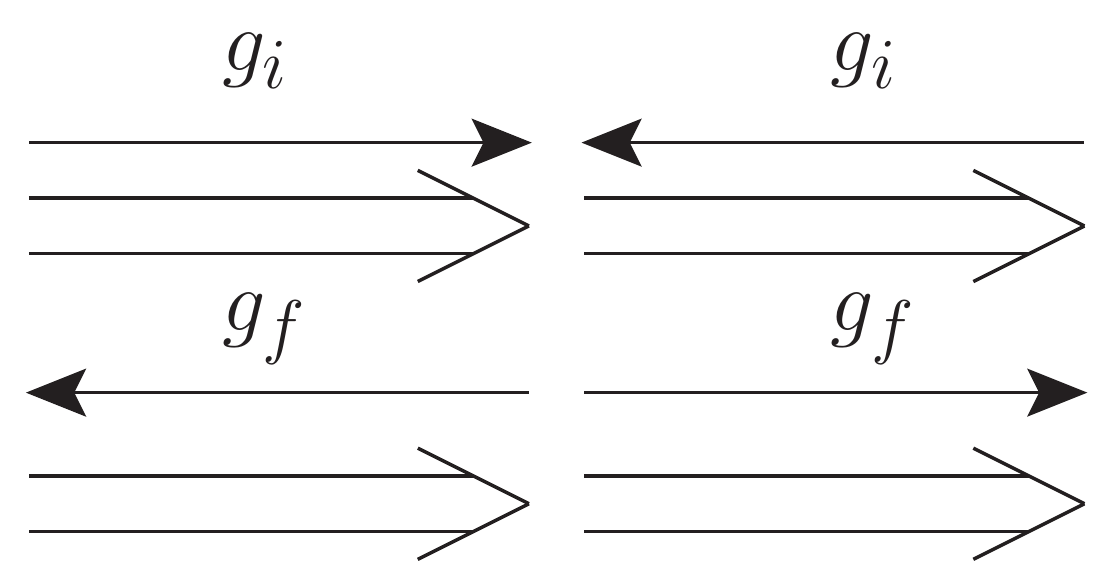}
\label{gRgL.fig}
\caption{Representative spin correlations for color tensor octet production.  single-arrowed lines represent momentum in the c.m. frame and double-arrowed lines represent spin. The subscript $i$ ($f$) indicates initial (final) state particles.}
\label{ggggspin.fig}
\end{figure}

Finally, we consider a tensor octet.  Figure \ref{ggggspin.fig} depicts some representative spin correlations for initial state gluons with (left) the same helicity and (right) opposite helicities.  
The final and initial states are individually symmetric under the exchange of gluons. The angular distribution is then of the symmetric form 
\begin{equation}
\frac{d\hat\sigma_2}{d\cos\theta}\sim (1+\cos\theta)^4+(1-\cos\theta)^4+\frac{4}{9}\left (2+\frac{\hat s}{M^2_T}\right )^2\left(1-\frac{\hat s}{M^2_T}\right )^2(1+4f)^2,
\end{equation}
where $M_T$ is the mass of the octet tensor, $f$ is the relative coupling factor defined in Eq.~(\ref{tensscal.EQ}). 
For an on-shell tensor $\hat s \to M^2_T$, the angular distribution reduces to 
\begin{equation}
\frac{d\hat\sigma_2}{d\cos\theta}=\frac{5}{32}\sigma_2(\hat{s})\left( 1+6\cos^2\theta+\cos^4\theta\right). 
\end{equation}

\begin{figure}[tb]
\centering
\includegraphics[clip,width=0.8\textwidth]{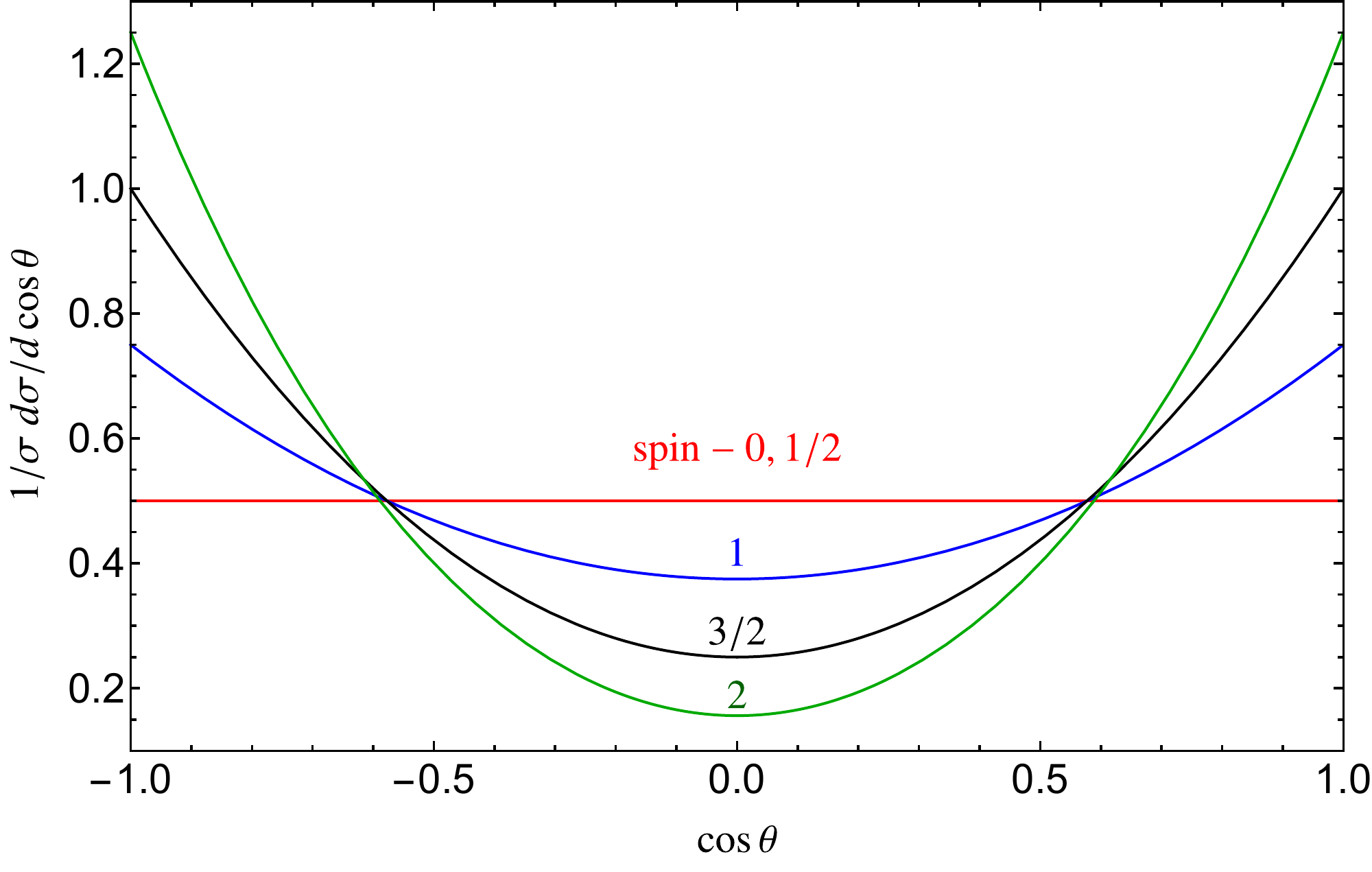}
\caption{Symmetrized di-jet angular distributions for different resonant signals in the partonic center-of-momentum frame. The invariant mass of the partonic system is $\sqrt{\hat s}=M$, where $M$ is the resonance mass.
}
\label{AngDist.fig}
\end{figure}

In Fig.~\ref{AngDist.fig} we present the angular distributions for scalars, spin-$1/2$ (3/2) fermions, vectors, and tensors.  
Not distinguishing the final state jets from $q$, $\bar{q}$, and $g$, all the distributions of spin-$1/2$ (3/2) fermions  and scalar/vector would be symmetric. We still see the shape difference
for a vector and spin-3/2 resonance. 
The angular distribution of the tensor resonance is most forward due to the higher power dependence of $\cos\theta$. 
We reiterate that for chiral couplings $\lambda_L \ne \lambda_R$, there will be forward-backward asymmetries for spin-1/2, spin-3/2, and spin-1 resonances, which, if observable, would provide crucial information for their underlying couplings.

%%%%%%%%%%%%%%%%%%%%%%%%%%%%%%%%%%%%%%%%%%%%%%
\subsection{Color representation}
\label{Ant.sec}
%%%%%%%%%%%%%%%%%%%%%%%%%%%%%%%%%%%%%%%%%%%%%%
%%%%%%%%%%%%%%%%%%%%%%%%%%%%%%%%%%%%%%%%%

\begin{figure}[t]
		\centering
		\begin{subfigure}{.73\textwidth}
			\centering
		\includegraphics[width=\textwidth]{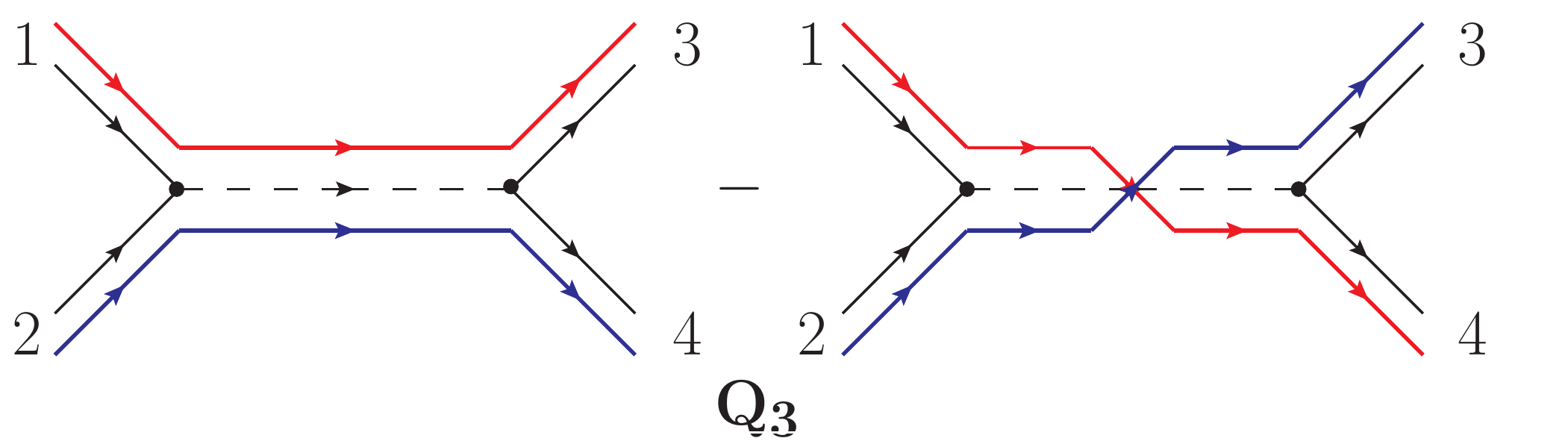}
			\caption{}
			\label{Col333b.fig}
		\end{subfigure}
		\begin{subfigure}{.73\textwidth}
			\centering
		\includegraphics[width=\textwidth]{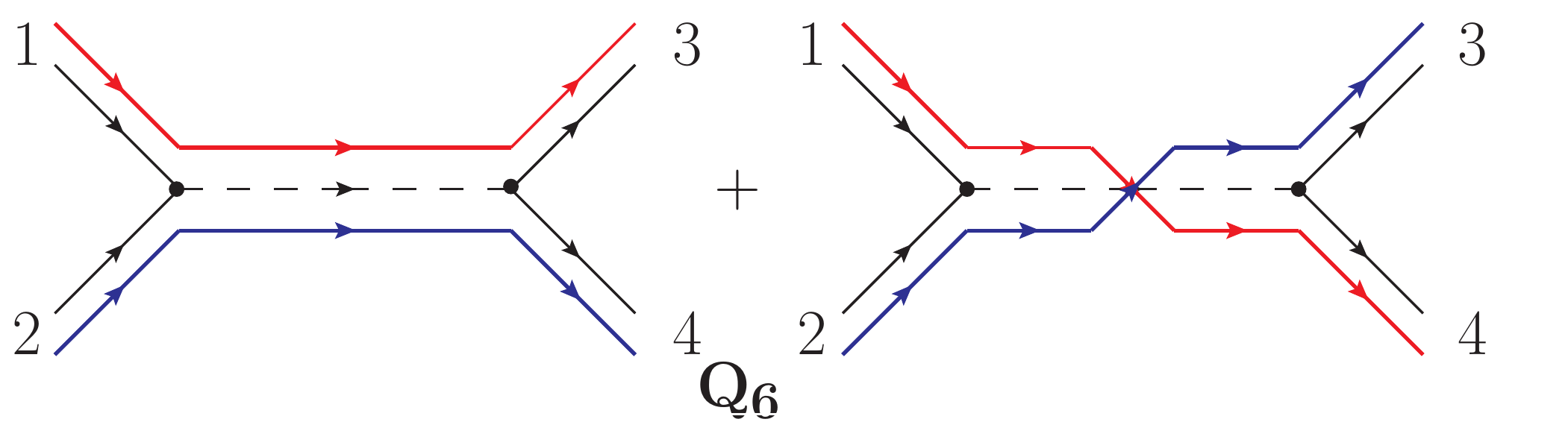}
			\caption{}
			\label{Col336.fig}
		\end{subfigure}
		\begin{subfigure}{.38\textwidth}
			\centering
		\includegraphics[width=\textwidth]{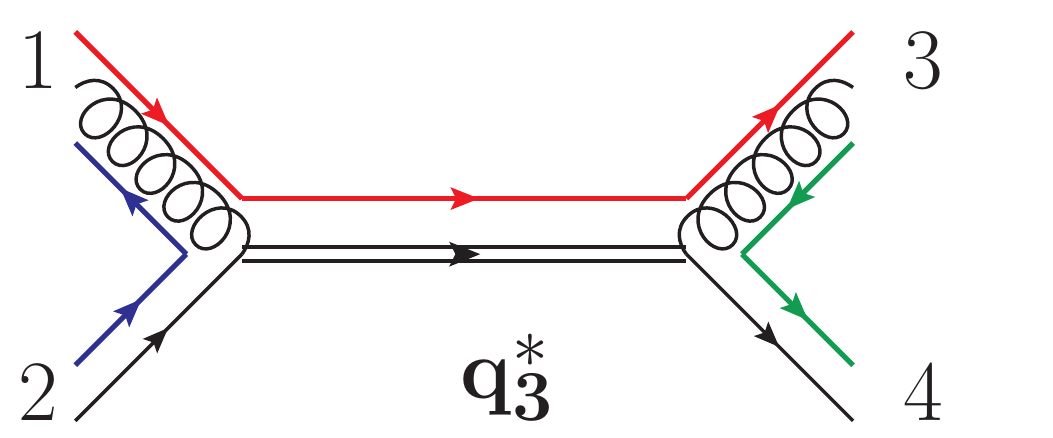}
			\caption{}
			\label{Col383.fig}
		\end{subfigure}
		\begin{subfigure}{.38\textwidth}
			\centering
		\includegraphics[width=\textwidth]{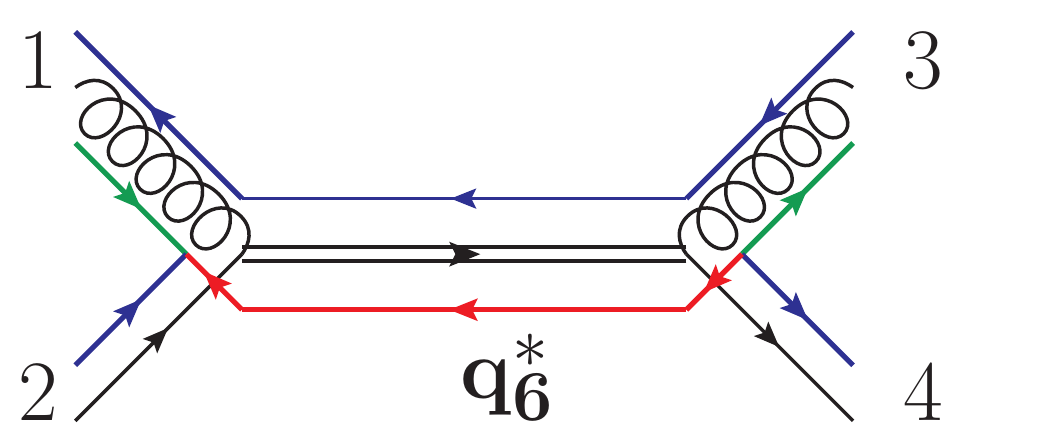}
			\caption{}
			\label{Col386b.fig}
		\end{subfigure}
		\begin{subfigure}{.38\textwidth}
			\centering
		\includegraphics[width=\textwidth]{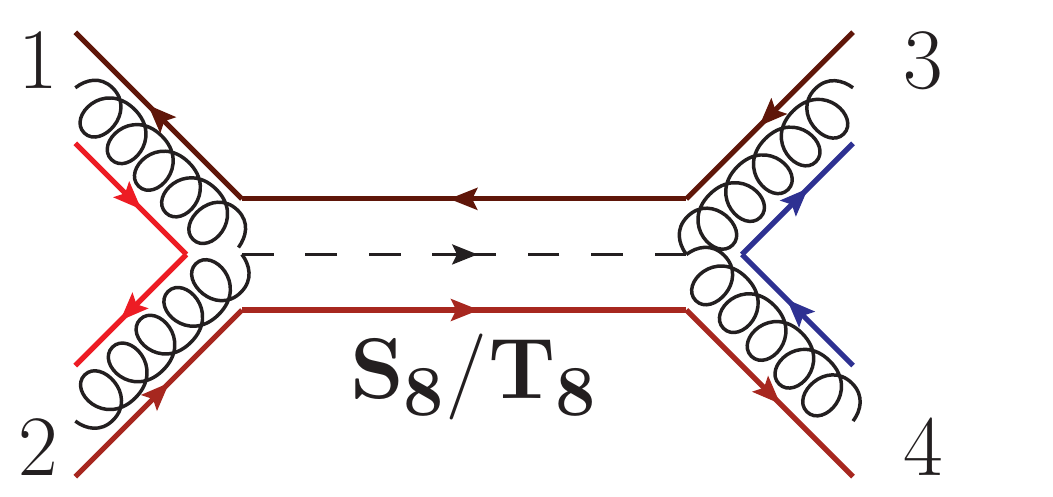}
			\caption{}
			\label{Col888.fig}
		\end{subfigure}
		\begin{subfigure}{.38\textwidth}
			\centering
		\includegraphics[width=\textwidth]{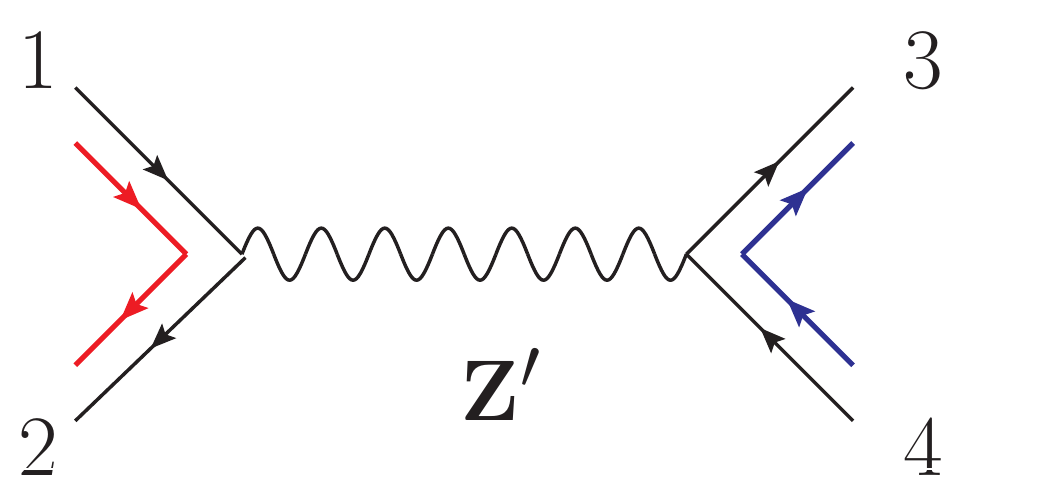}
			\caption{}
			\label{ColSing.fig}
		\end{subfigure}
		\begin{subfigure}{.38\textwidth}
			\centering
		\includegraphics[width=\textwidth]{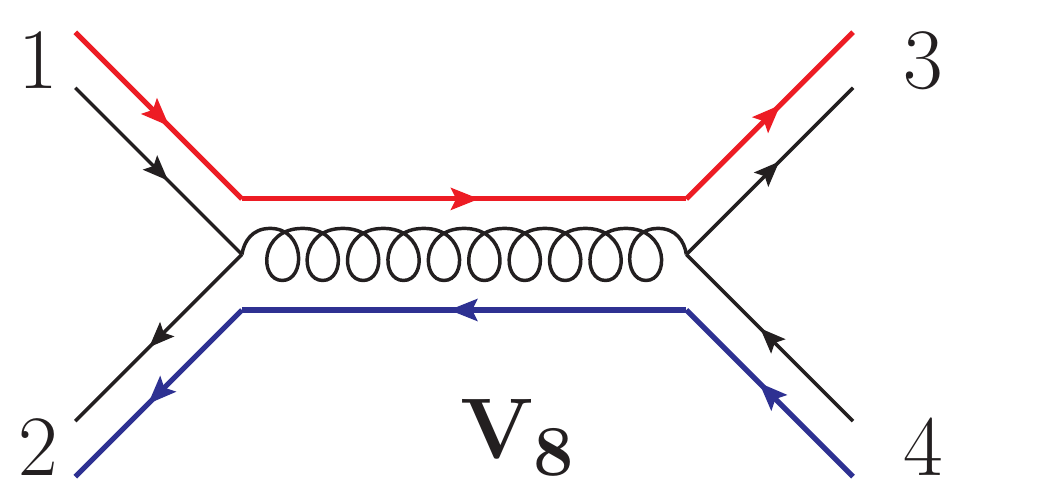}
			\caption{}
			\label{ColOct.fig}
		\end{subfigure}
		\caption{The leading color flow diagrams for (a) $\rep3\otimes\rep3\rightarrow \rep3$, (b) $\rep3\otimes\rep3\rightarrow\overline{\rep6}$, (c) $\rep3\otimes\rep8\rightarrow \rep3$, (d) $\rep3\otimes\rep8\rightarrow\overline{\rep6}$, (e)$\rep8\otimes\rep8\rightarrow \rep8$, (f) $\rep3\otimes\overline{\rep3}\rightarrow \rep1$, and (g) $\rep3\otimes\overline{\rep3}\rightarrow \rep8$.  The solid, colored arrowed lines  along the particle lines represent the flow of fundamental color charge.  $Z'$ denotes a color singlet vector.}
		\label{fig:ColAll}
	\end{figure}

It will ultimately be essential to probe the color quantum number for a resonance after discovery. Radiation patterns of gluons can be instrumental in identifying the color representation of the resonant particle. Representative examples of the leading color flow for the resonances outlined in Sec.~\ref{Models.sec} are shown in Fig.~\ref{fig:ColAll} for (a) antitriplet vectors and scalar, (b) sextet vectors and scalars, (c) triplet fermions, (d) antisextet fermions, (e) octet scalars and tensors, (f) singlet vectors, and (g) octet vectors.  The solid arrowed lines along the particle lines represent the flow of fundamental color charge.  As can be seen, for different spins and color representations, different initial and final state partons are color connected and the color flow of a given resonance depends on the resonance's color representation and the representations of the SM partons it couples to. 

%%%%%%%%
\label{colflow.sec}

%%%%%%%%
In the large $N_C$ limit, only gluons radiated off color-connected lines will interfere.  Hence, the radiation pattern of gluons can be instrumental in detecting the color representation of the resonant particle.  For example, consider a singlet versus octet resonance. In the octet case, the initial state partons are color connected to the final state partons. In contrast, the initial state and final state partons are separately color-connected in the singlet case.  Hence, in the plane defined by the two hard final state jets containing the colliding beams, an octet resonance is expected to have more radiation between the beams and the jets than a singlet resonance, {\it i.e.}, where the phase space of gluons radiated off the initial and final states overlap.  This observation has been used in previous proposals to identify the color of particles.  Ref.~\cite{Gallicchio:2010sw} analyzed the radiation patterns inside jets to separate singlet from octet color flows.  Similar color flow ideas have been applied to distinguishing color octet and singlet dijet events~\cite{Ellis:1996eu} and top pair tagging~\cite{Hook:2011cq} and many other new physics searches~\cite{Kim:2018cxf,Kim:2019wns,Filipek:2021qbe,Kumar:2023zjj}. This is also the basic idea of the rapidity gap~\cite{Sung:2009iq}. 

In this section, we study the radiation patterns and cross-sections of a single gluon radiated off the hard di-jets.  We provide analytical results of the gluon radiation pattern, so-called ``hadronic antennae" patterns, for the different colored resonances,
as previously studied analytically in Ref.~\cite{Ellis:1996eu}.

%%%%%%%%%%%%%%%%%%%%%%%%%%%%%%%%%%%%%%%%%%%%%%
\subsubsection{Antennae patterns} 
\label{soft.sec}
%%%%%%%%%%%%%%%%%%%%%%%%%%%%%%%%%%%%%%%%%%%%%
First, we present analytical results for dijet events with an additional radiated gluon.  As the discussion of color flows made apparent, the interference of the gluon radiated off of different external legs is sensitive to the color representation of the $s$-channel resonance.  Hence, we work in the soft limit where the interference of different diagrams is dominant.  Under the soft approximation, the matrix element $\mathcal{M}_{2\rightarrow3}$ of a $2\rightarrow2$ process plus a soft gluon radiated off of an external colored particle is related to the matrix element without radiated gluon, $\mathcal{M}_{2\rightarrow2}$, by 
\begin{eqnarray}
\mathcal{M}_{2\rightarrow3}\sim g_s\frac{\varepsilon(l)\cdot q}{l\cdot q}\mathcal{M}_{2\rightarrow 2},
\label{Mrad.EQ}
\end{eqnarray}
where $\varepsilon$ is the gluon polarization vector, $l$ is the gluon momentum, $q$ is the momentum of the parton radiating the gluon, and $g_s$ is the strong coupling constant. This  factorized form is valid for any soft gluon radiation.
For massless partons, the square of the matrix element in Eq.~(\ref{Mrad.EQ}) with the gluon spin summed is zero.
Also, for a process mediated by a heavy resonance, width effects regularize any soft divergence.   As a result, in the soft limit, interference terms between gluons radiated off of external legs are dominant for massless external partons.  This interference pattern is precisely the effect needed to detect the color flow for the different resonances.

\begin{figure}[tb]
\centering
\includegraphics[clip,width=0.5\textwidth]{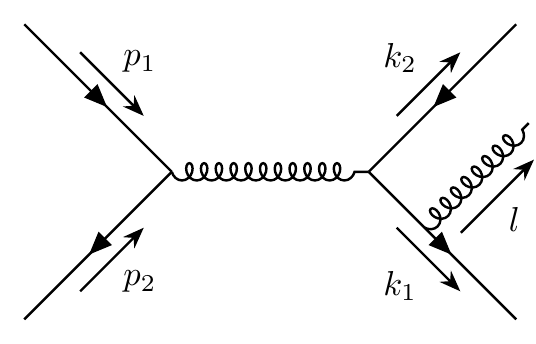}
\caption{Representative Feynman diagram of a $2\ra 3$ process with soft gluon radiation in the final state quark.}
\label{fig:radiation}
\end{figure}

Motivated by this observation, we calculate the matrix elements of our various resonances in the soft gluon limit and only consider radiation off of the initial and final state partons.  These calculations have been performed before for vector color-singlet resonances and the process $q\bar{q}\rightarrow g\rightarrow q'\bar{q'}$ in the large $N_C$ limit~\cite{Ellis:1996eu}.  For the following matrix elements, we adopt the notation $\mathcal{M}_{2\rightarrow R\rightarrow n}$ for the $2$ to $n$ process through the resonance $R$.  
Also, we generically label initial state momenta as $p_{1,2}$ and final state momenta as $k_{1,2}$. We depict the representative Feynman diagrams with their momentum assignments in Fig.~\ref{fig:radiation}.

First, we present the antennae patterns of a vector resonance for the color-singlet, $V_1$, and color-octet, $V_8$,
\bea
\label{radi36.eq}
\frac{|\mathcal{M}_{2\rightarrow {V_1}\rightarrow 3}|^2}{|\mathcal{M}_{2\rightarrow V_1\rightarrow 2}|^2}&\propto& g^2_s([p_1p_2]+[k_1k_2]),\\
\frac{|\mathcal{M}_{2\rightarrow V_8\rightarrow 3}|^2}{|\mathcal{M}_{2\rightarrow V_8\rightarrow 2}|^2}&\propto& g^2_s\bigg{\{}\left(1-\frac{2}{N^2_C}\right)\left([p_1 k_1]+[p_2 k_2]\right)+\frac{2}{N^2_C}\left([p_1k_2]+[p_2k_1]\right)\nonumber\\
&&-\frac{1}{N^2_C}\left([p_1p_2]+[k_1k_2]\right)\bigg{\}},
\eea
where $p_1$ ($k_1$) is the momentum of the initial (final) state quark and $p_2$ ($k_2$) the initial (final) state antiquark.  The square brackets indicate ``antennas", which specify the interference between radiations from different primary partons, defined by
\bea
\label{pattern.eq}
[k_ik_j]=\frac{k_i\cdot k_j}{l\cdot k_i \, l\cdot k_j }, 
\eea
where $l$ is the radiated gluon momentum.
As predicted by the color flow diagrams 
in Figs.~\ref{fig:ColAll}(f) and (g), the most significant interferences are between the initial state and final state partons separately for the singlet and the initial state quark (antiquark) and final state quark (antiquark) for the octet.  The color connections and antennae patterns for a singlet (octet) scalar coupling to quarks are the same as those for a singlet (octet) vector.

One can explicitly evaluate the hadronic antennae in a fixed frame.  
Working in the partonic c.m.-frame, for $p_1+p_2 \to k_1+k_2+l$ in the soft gluon approximation $l_T \ll E_T$, we have 
\begin{eqnarray}
p_1 &=&E_T(\cosh\eta,0,0,\cosh\eta),
\qquad\quad  \quad 
p_2 =E_T(\cosh\eta,0,0,-\cosh\eta)\\
k_1 &=&E_T(\cosh\eta,\sin\phi,\cos\phi,\sinh\eta), \ \ \ 
k_2 =E_T(\cosh\eta,-\sin\phi,-\cos\phi,-\sinh\eta)\\
l &=&l_T (\cosh\eta_g,\sin\phi_g,\cos\phi_g,\sinh\eta_g).%,
\end{eqnarray}
Using these momenta, we find the hadronic antennae
\begin{eqnarray}
[p_1\,p_2]&=&\frac{2}{l^2_T}, \qquad
[k_1\,k_2]=\frac{2}{l^2_T}\ \frac{\cosh^2\eta}{\cosh^2\eta_g\cosh^2\eta-(\cos(\phi-\phi_g)+\sinh\eta\sinh\eta_g)^2}\nonumber\\~
[p\,k]&=&\frac{4}{l^2_T}\frac{\cosh^2\eta+\cosh^2\eta_g-1-\cos(\phi-\phi_g)\sinh\eta\sinh\eta_g}{\cosh^2\eta_g\cosh^2\eta-(\cos(\phi-\phi_g)+\sinh\eta\sinh\eta_g)^2},
\label{antennae.EQ}
\end{eqnarray}
where, again, $[p_1\,p_2]$, $[k_1\,k_2]$, and $[p\,k]\equiv\sum_{i,j=1,2}[p_i\,k_j]$ are associated with interference of gluons radiated from initial state partons, final state partons, and between initial and final state partons.  

Next, the antennae patterns for scalar diquarks are
\bea
\frac{|\mathcal{M}_{2\rightarrow Q_{N_D}\rightarrow 3}|^2}{|\mathcal{M}_{2\rightarrow Q_{N_D}\rightarrow 2}|^2}&\propto& g^2_s\bigg{\{}[p_1k_1]+[p_2k_2]+[p_1k_2]+[p_2k_1]\mp2\frac{C_FN_C}{C_DN_D}\left([p_1p_2]+[k_1k_2]\right)\bigg{\}},
\label{eq:33colorsoft1}
\eea
where the upper sign is for the sextet case, and the lower is for the triplet, and $C_D$ is the quadratic Casimir operator for the sextet or triplet representation. For the sextet, $C_D=10/3$ and for the triplet $C_D=C_F=4/3$.  Using the values of $C_D$, one can find the sextet case
\bea
\label{radi6.eq}
\displaystyle\frac{|\mathcal{M}_{2\rightarrow Q_{6}\rightarrow 3}|^2}{|\mathcal{M}_{2\rightarrow Q_6\rightarrow 2}|^2}\propto g^2_s&\bigg{\{}[p_1k_1]+[p_2k_2]+[p_1k_2]+[p_2k_1]-\displaystyle\frac{2}{5}\left([p_1p_2]+[k_1k_2]\right)\bigg{\}},
\label{eq:33colorsoft2}
\eea
and for the triplet
\bea
\label{radi3.eq}
\displaystyle\frac{|\mathcal{M}_{2\rightarrow Q_{3}\rightarrow 3}|^2}{|\mathcal{M}_{2\rightarrow Q_3\rightarrow 2}|^2}\propto g^2_s\bigg{\{}[p_1k_1]+[p_2k_2]+[p_1k_2]+[p_2k_1]+2\left([p_1p_2]+[k_1k_2]\right)\bigg{\}}.
\label{eq:33colorsoft3}
\eea
Note that the interference between initial (final) state quarks is suppressed by $2/5$ relative to the initial and final state interference for the sextet. In contrast, for the triplet, there is no such suppression.  Since a sextet is the symmetric combination of two triplets and an antitriplet the antisymmetric combination, there is destructive and constructive interference between the two possible color flow for triplet and sextet diquarks, respectively.  This effect is shown in Figs.~\ref{Col333b.fig} and~\ref{Col336.fig}.  Hence, for the triplet, the interference between the initial and final state quarks is suppressed relative to the sextet case.  As can be seen in Eq.~(\ref{radi3.eq}), the destructive interference in the triplet case causes all possible interferences to make roughly equal contributions, even though, in the large $N_C$ limit, the interference between the two initial (final) state quarks is subdominant.  Whereas, for the sextet case, the interference between color-connected partons remains dominant, as shown in Eq.~(\ref{radi6.eq}).

For the triplet and sextet fermions:
\bea
\frac{|\mathcal{M}_{2\rightarrow Q^*_3\rightarrow 3}|^2}{|\mathcal{M}_{2\rightarrow Q^*_3\rightarrow 2}|^2}&\propto& g^2_s\bigg{\{}[p_1k_1]+\frac{8}{9}\left([p_2p_1]+[k_2k_1]\right)-\frac{1}{9}\left([p_2k_1]+[k_2p_1]\right)+\frac{1}{81}[p_2k_2]\bigg{\}},\\
\frac{|\mathcal{M}_{2\rightarrow Q^*_6\rightarrow 3}|^2}{|\mathcal{M}_{2\rightarrow Q^*_6\rightarrow 2}|^2}&\propto& g^2_s\bigg{\{}[p_1k_1]+\frac{1}{3}\left([p_2k_1]+[k_2p_1]\right)+\frac{4}{15}\left([p_2p_1]+[k_2k_1]\right)+\frac{1}{9}[p_2k_2]\bigg{\}},
\eea
where $p_1$ ($k_1$) is the momentum of the initial (final) state gluon and $p_2$ ($k_2$) is the momentum of the initial (final) state quark.  
As can be seen in Fig.~\ref{Col383.fig}, for the triplet fermion the initial and final state gluons, the initial state gluon and quark, and the final state gluon and quark are color connected.  Hence, the interferences between these pairs are the dominant contribution to the antennae behavior.  The sextet fermion is much more difficult to interpret, although interference between initial and final state gluons are the dominant contribution.

Finally, we have the scalar octet coupling to two gluons:
\bea
\frac{|\mathcal{M}_{2\rightarrow S_8\rightarrow 3}|^2}{|\mathcal{M}_{2\rightarrow S_8\rightarrow 2}|^2}\propto g^2_s\bigg{\{}[p_1p_2]+[k_1k_2]+\frac{1}{2}\left([p_1k_1]+[p_2k_1]+[p_1k_2]+[p_2k_2]\right)\bigg{\}},
\eea
Since all gluons are color connected and the matrix elements are symmetric under exchanges $k_1\leftrightarrow k_2$ and $p_1\leftrightarrow p_2$, all possible interferences are significant.  The antennae pattern for the tensor octet is the same since the color flow is identical.

%%%%%%%%%%%%%%%%%%%%%%%%%%%%%%%%%%%%%%%%%%%%%
\subsubsection{Cross-section ratios}
\label{ratio.sec}
%%%%%%%%%%%%%%%%%%%%%%%%%%%%%%%%%%%%%%%%%%%%%

\begin{table}[tb]
\centering

\begin{tabular}{|c|c|c|c|c|}
  \hline
    Initial& Color & Spin & Type & $R_{3/2}$\\ \hline
   &  & $0$ & $D_3$    & $ 0.41$  \\ \cline{3-5}
   &  &  & $E_3^\mu$   &  $0.41$ \\ \cline{4-5}
   & \raisebox{1.6ex}[0pt]{$3$} & $1$ & $D_3^\mu$ & $0.40$  \\ \cline{4-5}
   &  &  & $U_3^\mu$ & $0.39$  \\ \cline{2-5}
   &  &  & $E_6$ & $0.29$   \\ \cline{4-5}
 \raisebox{1.6ex}[0pt]{$3\otimes3$}  &  & $0$ & $D_6$             & $0.29$  \\ \cline{4-5}
   &  &  & $U_6$  & $0.28$  \\ \cline{3-5}
   & \raisebox{1.6ex}[0pt]{$6$} &  & $E_6^\mu$ &  $0.29$  \\ \cline{4-5}
   &  & $1$ & $D_6^\mu$   & $0.28$ \\ \cline{4-5}
   &  &  & $U_6^\mu$  & $0.27$ \\ \hline \hline
   &  &  & $U^*_3$ &  $0.61$ \\  \cline{4-5}
 \raisebox{1.6ex}[0pt]{$3\otimes8$}  & \raisebox{1.6ex}[0pt]{$3$} & \raisebox{1.6ex}[0pt]{$1\over 2 $} & $D^*_3$ & $0.59$ \\ \hline \hline
   &  & $0$ & $S_8$     & $0.69$  \\ \cline{3-5}
 \raisebox{1.6ex}[0pt]{$8\otimes8$}  & \raisebox{1.6ex}[0pt]{$8_S$} & $2$ & $T_8$ & $0.70$ \\ \hline \hline
   &  &  & $V_8$ & $0.27$  \\ \cline{4-5}
 \raisebox{1.6ex}[0pt]{$3\otimes\bar3$}  & \raisebox{1.6ex}[0pt]{$8_A$} & \raisebox{1.6ex}[0pt]{$1$} & $V_8^{\pm}$ &  $0.26$ \\ \hline
  \hline
\end{tabular}
\caption{Ratios of $2\rightarrow3$ resonant production cross-section over $2\rightarrow2$ processes at parton level with $p_T^j > 200$~GeV, $|\eta_j|< 3.0$, and $\Delta R_{jj} > 0.4$ at the 14 TeV LHC. The mass of all color resonances is set to be 3 TeV and the width is set to be 30 GeV. 
\label{ratio.tab}
}
\end{table}

An important quantity in  understanding QCD dynamics is the scaling of cross-sections with additional jets.  The scaling is broadly defined as
\begin{eqnarray}
R_{(n+1)/n}=\frac{\sigma_{2 \ra n+1}}{\sigma_{2 \ra n}},
\end{eqnarray}
where $\sigma_{2\rightarrow n}$ is the hadronic cross-section of an underlying process with $n$ observed hadronic jets. Naively, the ratio goes like $\sim \alpha_s$, but it depends on the color structure of a specific process and the jet selection procedure. This property has been studied at hadron colliders for the Drell-Yan processes $W+$jets and $Z+$jets~\cite{Ellis:1985vn,Englert:2011cg}, pure QCD jet production~\cite{Englert:2011cg}, direct photon+jets~\cite{Englert:2011pq}, and Higgs production~\cite{Gerwick:2011tm}.

The ratio of the three-jet cross-section to that of a di-jet resonance depends on many factors, such as the di-jet system invariant mass, the di-jet initial state composition, and, especially to our interest, the color representation of the $s$-channel resonance. We have presented the analytical expressions for their matrix elements in the previous Section~\ref{soft.sec}. 
Table~\ref{ratio.tab} shows the cross-section ratios $R_{3/2}$. 
\texttt{MadGraph5\_aMC@NLO}~\cite{Alwall:2011uj} was used to calculate these rates using model files generated via \texttt{FeynRules}~\cite{Christensen:2008py}.  As can be clearly seen, $R_{3/2}$ is strongly dependent on the color of the initial state partons as well as the color charge of the intermediate resonance. For instance, when comparing the $R_{3/2}$ between different initial states, such as $3\otimes3$, $3\otimes8$, $8\otimes8$, $3\otimes \bar 3$, a general trend is gluonic initial states provides stronger radiation and thus larger $R_{3/2}$. Within the same set of initial states, we can see that $R_{3/2}$ can distinguish the color representation of the resonance. For example, for $3\otimes 3$ the color sextet resonance generically has a smaller $R_{3/2}$ than the anti-triplet resonance.  This comes from the sizable change from constructive interference for the triplet to destructive interference for the sextet between radiation from the initial parton and final dijet pairs. The effect can be understood using the soft-emission approximation, {\it e.g.,} in Eq.~(\ref{eq:33colorsoft1}), Eq.~(\ref{eq:33colorsoft2}), and Eq.~(\ref{eq:33colorsoft3}). It is striking that particles with different spins but the same color flow have similar ratios, while particles with the same spin but different color flows can have very different values of $R_{3/2}$.

%%%%%%%%%%%%%%%%%%%%%%%%%%%%%%%%%%%%%%%%%%%%%
\section{Cut-based analysis for the antenna pattern}
\label{sec:cut}
%%%%%%%%%%%%%%%%%%%%%%%%%%%%%%%%%%%%%%%%%%%%%
Extracting the underlying color structure is a challenging task when it comes to collider signals. We now present our analysis with the ``interference spectrum''. Instead of looking into the total energy distribution of an event after hadronization, we focus on just one additional radiated jet to probe the color.  
As can be seen in the analytical results of Section~\ref{soft.sec}, the radiation pattern of a gluon depends on the underlying colored resonance. Specifically, the relative strengths of terms related to the interference of gluons radiated off the two initial state partons $[p_1p_2]$, the two final state partons [$k_1k_2$], and initial and final state partons $[p_i k_j]$ differ among the various color representations. Hence, if these relative strengths can be measured, then information on the color representation of the new resonance can be obtained.  With this intuition, we now develop a cut-based strategy to measure these effects.  The next section will use a machine learning-based strategy to disentangle the color representations.

\begin{figure}[tb]
\centering
\includegraphics[width=0.49\textwidth]{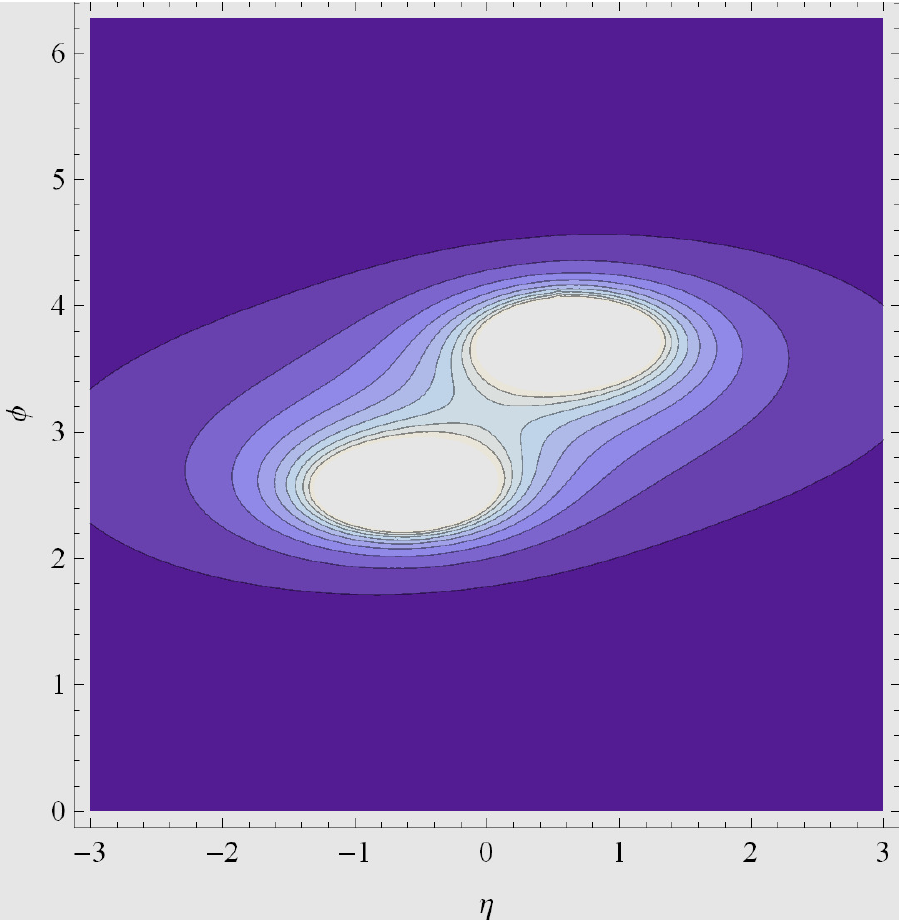}
\includegraphics[width=0.49\textwidth]{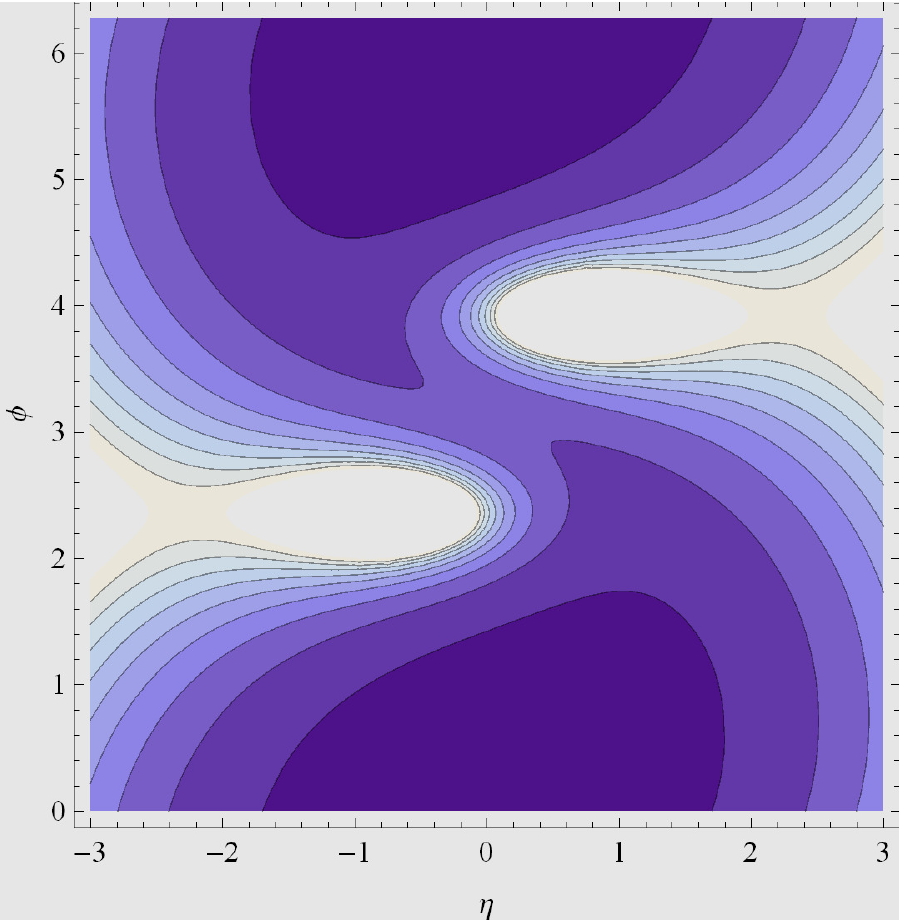}
\caption{Interference pattern between different jets/partons. (a) Interference between two final state jets; (b) interference between final state jets and initial state partons.}
\label{pattern.fig}
\end{figure}

Figure \ref{pattern.fig}  
shows contours as a dimensionless probability density (with arbitrary normalization) of radiations in the $\eta-\phi$ plane of the hadronic antennae between (a) final state partons, $l^2_T [k_1\,k_2]$, and (b) initial and final state partons, $l^2_T [p\, k]$,  with the ubiquitous factor of $l^2_T$ factored. To make the illustration more transparent rather than the uncharacteristic spread in the $\eta$-$\phi$ plane, we boosted the dijet and radiation system in the transverse plane, so that the two leading jets are not back-to-back. 
We do not show the pure initial state radiation interference pattern $l^2_T [p_1\,p_2]$, since it is trivial and just a constant.  The brighter regions indicate the larger numerical values of the antennae patterns, Eq.~(\ref{pattern.eq}), which represents a higher probability for the radiation to take place. Both figures share highlighted areas near the final state jets since the radiation tends to be collinear with its parent jet. It is this region that shall not be considered in our cut-based analysis due to the low signal (from interference) to the background (from colinear emission) ratio.

The difference in the radiation for different antennae patterns is very pronounced in Fig.~\ref{pattern.fig}.  
Namely, gluons tend to be radiated between the final state jets for $l_T^2[k_1\,k_2]$ for the color singlet and between the final state jets and the beam direction for $l_T^2[p\,k]$ for the color-octet.  These are the distinct regions that distinguish the different patterns.  Many works have focused on the region between the two interfering jets both theoretically \cite{Ellis:1996eu,Hook:2011cq,Gallicchio:2010sw} and experimentally \cite{Abazov:2011vh}.

From the discussion above, our observables should be sensitive to the interference regions, but insensitive to radiation collinear with the final state jets.  A generic antennae pattern can be rewritten as
\beq
\label{valley.eq}
[p_A p_B]=\frac {p_A \cdot p_B} {p_A \cdot l - p_B \cdot l} \left(\frac 1 {p_B \cdot l}- \frac 1 {p_A \cdot l}\right). 
\eeq
The two terms in the parentheses represent the regions where the radiated gluon is collinear with one of the parent partons. When the radiated gluon has the same angular separation with both parent partons, the denominator in the pre-factor  $\left(p_A \cdot l - p_B \cdot l\right)$ is zero.  Hence, this denominator is the characteristic feature of the interference.  
It is intuitive to choose the fractional factor
\beq
\frac {p_A \cdot p_B}{p_A \cdot l - p_B \cdot l}
\label{valley2.eq}
\eeq
to define the ``valley'' region where interference is maximized since this fraction is large when the radiated gluon is between the two parent partons.  This fraction scales as the energy scale of the dijets to the radiated gluon.  Hence, to place a cut on Eq.~(\ref{valley2.eq}), we use the transverse momentum of the radiation $l_T$ and the sum of transverse momentum magnitudes of all final state jets, $h_T$:
\beq
\label{region.eq}
\frac {p_A \cdot p_B}{|p_A \cdot l - p_B \cdot l|}\geq \frac{h_T}{l_T}\quad\Rightarrow\quad |p_A\cdot l-p_B\cdot l| \leq \frac {l_T} {h_T} \ p_A \cdot p_B.
\eeq
This condition is invariant under boosts along the beam direction and is robust at hadron colliders.

\begin{figure}[tb]
\centering
\includegraphics[width=0.49\textwidth]{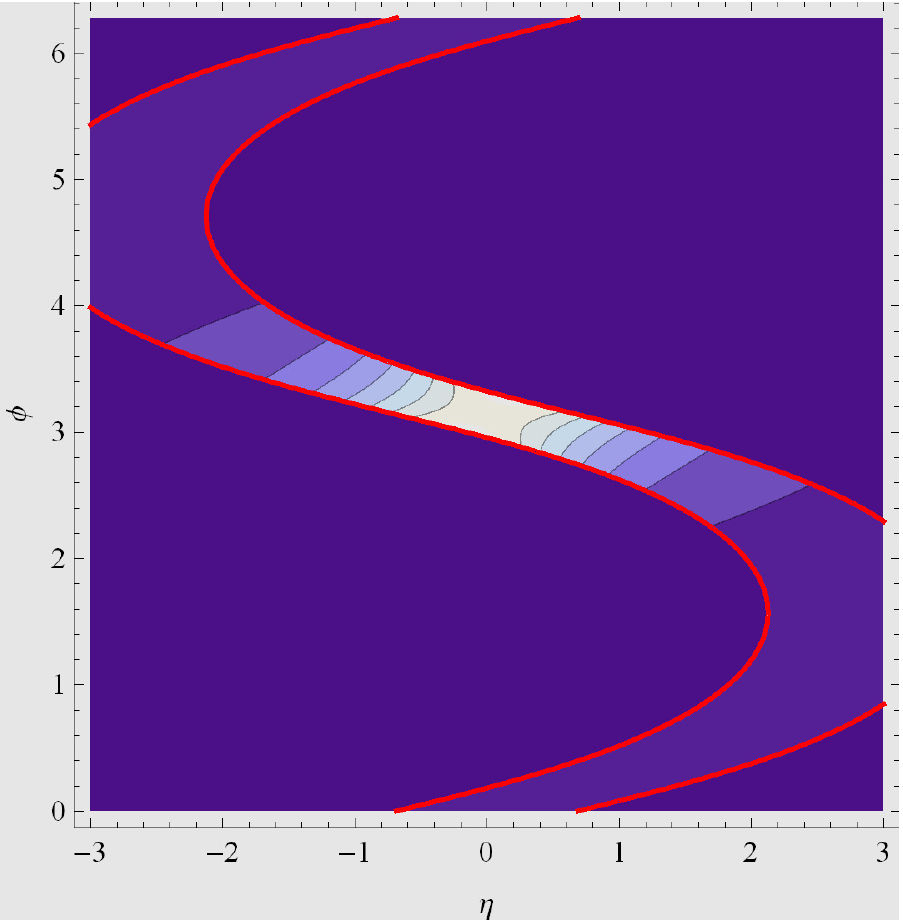}
\includegraphics[width=0.49\textwidth]{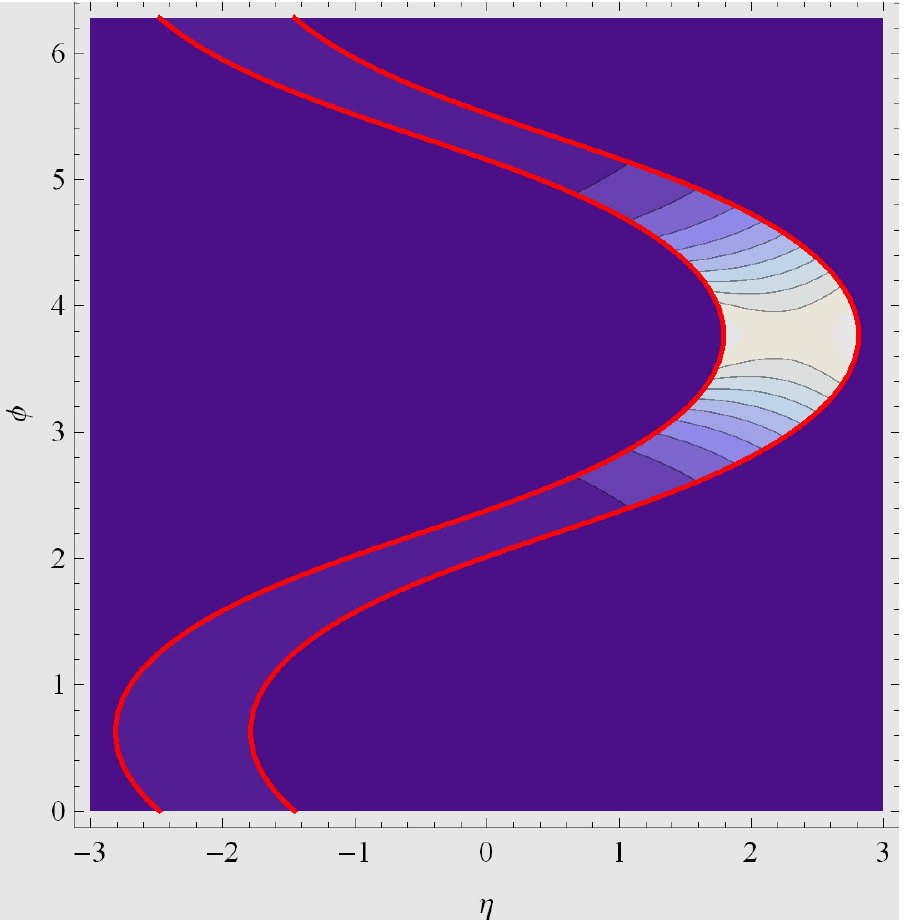}
\caption{The valley regions of Eq.~(\ref{region.eq}) that characterize interference effect for (a) interference between two final state jets, and (b) interference between one pair of final state jets and initial state partons.  
}
\label{region.fig}
\end{figure}

The region satisfying Eq.~(\ref{region.eq}) is illustrated in Fig.~\ref{region.fig}, with the same kinematic configuration as Fig.~\ref{pattern.fig}. In this figure, we set the event weight outside the valley region to be zero. It is clear that this region contains most of the interfering radiation and excludes the common collinear regions shared by all interference patterns that share the same jet/parton. 

 To test our observables beyond the idealized setup, a reliable event simulation is needed. \texttt{FeynRules}~\cite{Christensen:2008py} was used to implement the interactions of the colored resonances in \texttt{MadGraph5}~\cite{Alwall:2011uj} via UFO model files~\cite{Degrande:2011ua}.  \texttt{MadGraph5} is then used to simulate three jet events at a $14$~TeV LHC. Here we simulate all resonances with a mass of 3~TeV.
 We require all three jets to be hard. This requirement can be relaxed to allow for a soft 3rd jet or a fat jet analysis, and our ML-based analysis in the next section shows the complimentary analysis results. These hard jet requirements ensure events pass the LHC triggers and also avoid the subtlety of multiple soft emissions whose analysis is complicated by pile-up and underlying events.  To isolate the interference region, one jet is allowed to be softer than the other two. Hence, the following minimum cuts are applied~\cite{Aad:2011aj}
\bea
p^{j_1}_T>500~{\rm GeV},~~p^{j_2,j_3}_T>200~{\rm GeV},~~{\rm and},~~|\eta^j|<3
\eea
where $p_T$ is transverse momentum and $\eta$ is rapidity.\footnote{We work at the partonic level for the kinematics and ignore the detector effects.} The jets, $j_i$ are labeled numerically in the order of decreasing transverse momenta, with $j_1$ being the hardest. Due to the parent jets being of high energy, their radiation $j_3$ can also take high energy. Since we expect the base event to be two jets with equal $p_T$ that then radiate a jet, we also require that $p_T$ of the second hardest jet is near the $p_T$ of the hardest jet
\beq
p_T^{j_2}\geq 0.8\,p_T^{j_1}.
\eeq
Finally, to resolve three jets, a minimum separation is required for all jet combinations
\bea
\Delta R_{jj}>0.4
\eea
where $\Delta R_{jj}=\sqrt{(\Delta\phi_{jj})^2+(\Delta\eta_{jj})^2}$ is the angular separation of two jets in the azimuthal angle, $\phi$, and rapidity plane. 
This hard isolation cut affects our ability to extract the hadronic antennae pattern, and we shall see how machine learning can improve it in the next section.

We now determine how each of the six possible interference patterns is populated for different colored resonance as the partonic level. For every $2\rightarrow 3$ event, the jet with the smallest $p_T$ is considered as the radiation with momentum $l$. For a more realistic calculation at hadron colliders, unlike the previous section, the momenta $p_1,\,p_2,\,k_1,\,$ and $k_2$ are ordered according to their momentum. From the total reconstructed momentum and invariant mass of the three final state jets, the momentum of the initial state partons can be reconstructed.  The parton that carries a larger fraction $x$ of proton momentum is labeled $p_1$, and the smaller one is $p_2$. This step sets up our positive z-direction to be the same as the direction of the partonic center-of-mass reference frame with respect to the lab frame. Meanwhile, the jet with the largest transverse momentum is treated as $k_1$, and the second largest transverse momentum is $k_2$. Without the above steps, there would be no difference between the interference patterns  $[p_1 k_1]$, $[p_1 k_2]$, $[p_2 k_1]$ and $[p_2, k_2]$ unless techniques to distinguish jets originating from quarks, anti-quarks, or gluons are employed. This is not so important for resonances that come from identical initial states like $S_8(gg)$, $T_8(gg)$, and some diquarks from $uu$ and $dd$ initial states. When we encounter those resonances from different initial states like $V_8^{0,\pm}(q\bar q)$ and $Q^{*(\mu)}_{3,6}(qg)$ these steps of defining $k_1,~k_2,~p_1,~\text{and}~p_2$ will statistically differ the four would-be identical interference patterns and thus shed light on the color structures.

To calculate the ``valley'' condition of Eq.~(\ref{region.eq}), $p_A$ and $p_B$ are a combination of the two hardest final jets and initial state partons.  For each event, we then calculate all six possible combinations of the valley condition. The third jet (which is interpreted as the radiation with momentum $l$) can be counted as one radiation in each valley condition it satisfies. By this means, we have an ``interference spectrum'' of six different modes for every resonant signal. They should have various strengths for different color structures. 

\begin{figure}[htb]
\centering
\includegraphics[width=0.98\textwidth]{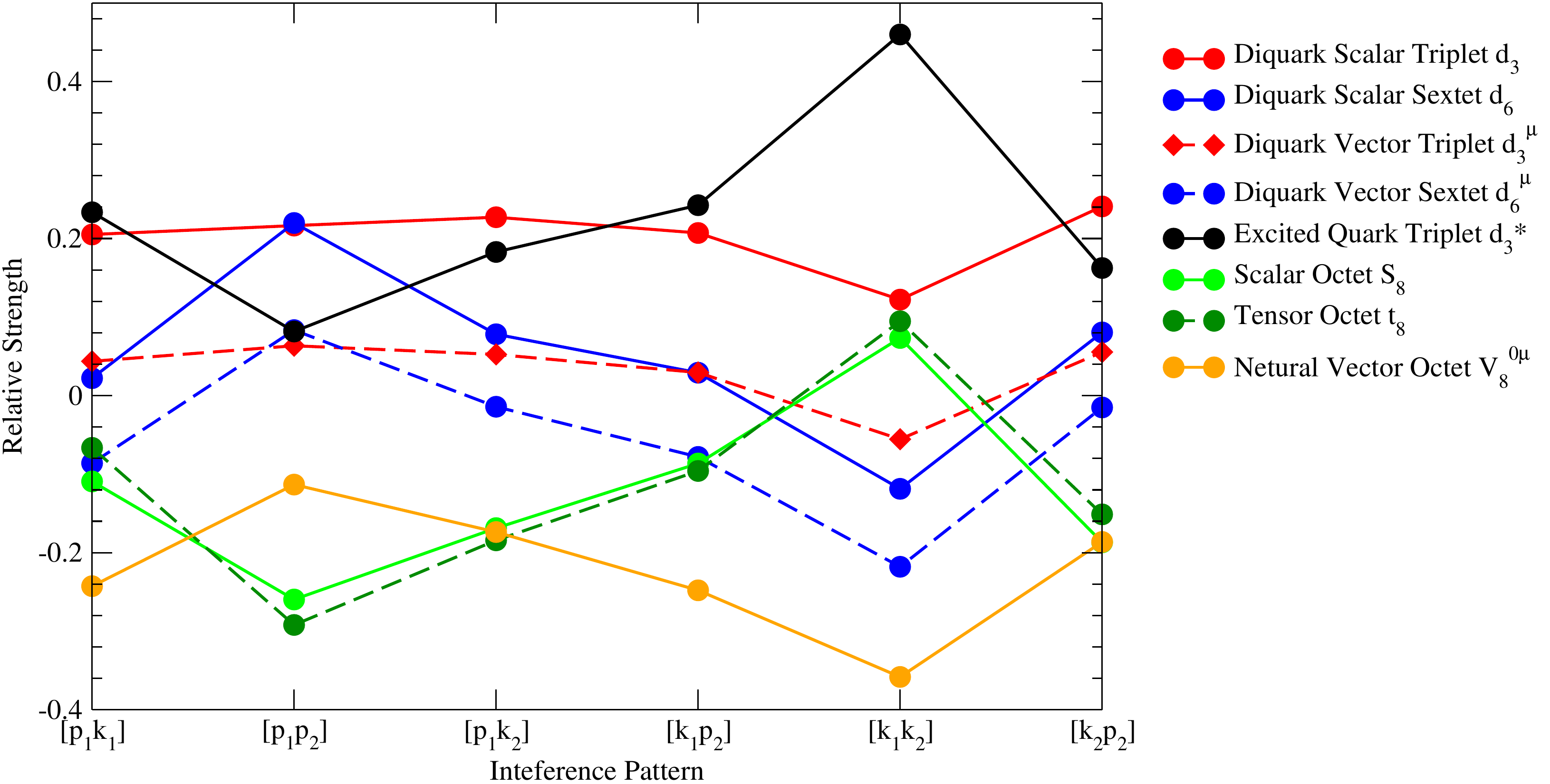}
\caption{Hadronic antenna ``Interference Spectrum" of different underlying colored resonances. The horizontal axis is the six different interference sources in a trijet (dijet plus a radiated jet) resonance event. The vertical axis is the normalized counting of radiation that comes from the interference. Different underlying $SU(3)$ color representations are shown in different colors. The overall shape difference between different underlying color representations shows that we are able to extract the color interference pattern using our cut-based method. A more detailed discussion can be found in the text.
\label{spectrum.fig}}
\end{figure} 

Figure \ref{spectrum.fig} shows the results of the interference spectrum through the procedure described above. The horizontal axis is the six interference patterns.  The vertical axis is then the relative difference for each pattern, defined by the number of events for a given resonance divided by the average number of events across all resonances and then subtracted by unity.
Identical color structures but different spins are labeled with the same color with solid or dashed lines. In the order of the resonance classification in Table 1, the red dots and diamonds are for color-triplet scalars and vectors, respectively; the blue for color sextet scalars and vectors; green for symmetric color octet scalars and tensors; black for color triplet fermions; and orange for octet vectors. 
Two critical features make this Fig.~\ref{spectrum.fig} very intriguing. First, the relative strength of the six patterns is very different for each color structure, providing the possibility of discriminating the color information. 
Second, the same color structures have almost overlapping or parallel event counts. In other words, we find an observable that is color structure-sensitive but insensitive to the initial parton, boost, or spin of the resonance. This is a surprisingly nice result, given that so many factors affect these distributions. 

This spectrum is obtained around $100,000$ generated events, and there are over $9,000$ counts for each pattern. The statistical uncertainty associated with our counting would be around $1\%$ of the value represented by the points, which is larger than the dots' size in Fig.~\ref{spectrum.fig}. We also tested the statistics by randomly generating another set of events and found our results robust. 
There are some resonances not presented in this figure. The diquarks from other initial states ($uu$, $dd$) all have the same behavior as diquarks from $ud$ initial states. The situation is similar for vector octets. 
Essential features of dijets, including the color information carried by the interference pattern, can be more effectively captured using modern machine-learning techniques. We explore aspects of machine learning in the next section.

%%%%%%%%%%%%%%%%%%%%%%%%%%%%%%%%%%%
\section{Diagnostic studies with deep learning}
\label{sec:ml}

While the interference patterns can already provide us with distinctive features for different color resonances, the ML techniques have the potential to make optimal use of all the information available. Various ML techniques have already been proven useful in collider physics (for recent reviews see Refs.~\cite{Schwartz:2021ftp,Feickert:2021ajf,Plehn:2022ftl,Shanahan:2022ifi}). 
We use a convolutional neural network (CNN)~\cite{cnn} to demonstrate the capability of ML in distinguishing different color resonances.

\subsection{General features of the signal processes}
\label{sec:pro}

As in previous sections, we make use of the topology of a resonance R decaying to two hard jets plus an additional radiated jet off the initial, final, or resonant states:
\beq
p\;p\rightarrow R (j) + {\rm remnants}  \rightarrow j j j + {\rm remnants},
\eeq
where $(j)$ in the intermediate step indicate the possibility of initial state radiation.  
In this section, we study the vector color singlet $V_1 (u\bar{u})$, vector color octet $V_8 (u\bar{u})$, diquark vector color sextet $E^\mu_6 (uu)$, and digluon scalar octet $S_8 (gg)$ as the representative resonances.   
The events simulated via $\texttt{MadGraph5\_aMC@NLO}$~\cite{Alwall:2014hca} with 3 jets in the final states at the 14 TeV LHC.  The model files were generated with $\texttt{FeynRules}$~\cite{Christensen:2008py}.  We set the masses of the resonances to be $M=3$~TeV and the widths to be narrow such that the apparent width in the experimental signature is dominated by detector resolution not the resonance's intrinsic width.  
For simplicity, we adopt the non-chiral couplings for the resonances in the rest of the presentations.
The generator-level cuts on the transverse momentum and pseudo-rapidity are set to be
\beq
p_T^{j_1} > 600~\text{GeV}, \quad p_T^{j_2} > 500~\text{GeV},\quad p_T^{j_3} > 100~\text{GeV},\quad{\rm and}\quad \mid\eta_j\mid < 3,
\eeq
where, as in the previous section, the jets are ordered according to their transverse momentum.  
\texttt{Pythia 8.1}~\cite{Sjostrand:2006za, Sjostrand:2007gs} is used for parton showering and hadronization. The jets are clustered by using the  $\text{anti-}k_T$ algorithm~\cite{Cacciari:2008gp} with $R=0.4$.

A jet is a collimated spray of particles, resulting from the parton showering and hadronization of high-energy quarks and gluons. Each jet can be defined as a calorimeter energy deposition in the 2D angular plane $\phi$-$\eta$.    
Pixelating jets in the $\phi$-$\eta$ plane can form jet images with the intensity of pixels being observables such as transverse momentum, energy, particle multiplicity, etc. These calorimeter images can then be used in a CNN.  To maximally utilize all radiation information, in addition to the three hard jets generated with $\texttt{MadGraph5\_aMC@NLO}$, we include all the jets satisfying $p_T^j > 100$ GeV and $|\eta_j| < 3$   in the jet images.  Some of these jets may be generated via parton showering and hadronization.

\begin{figure}[t!]
	\begin{subfigure}{.5\textwidth}
		\centering
		\includegraphics[width=\textwidth]{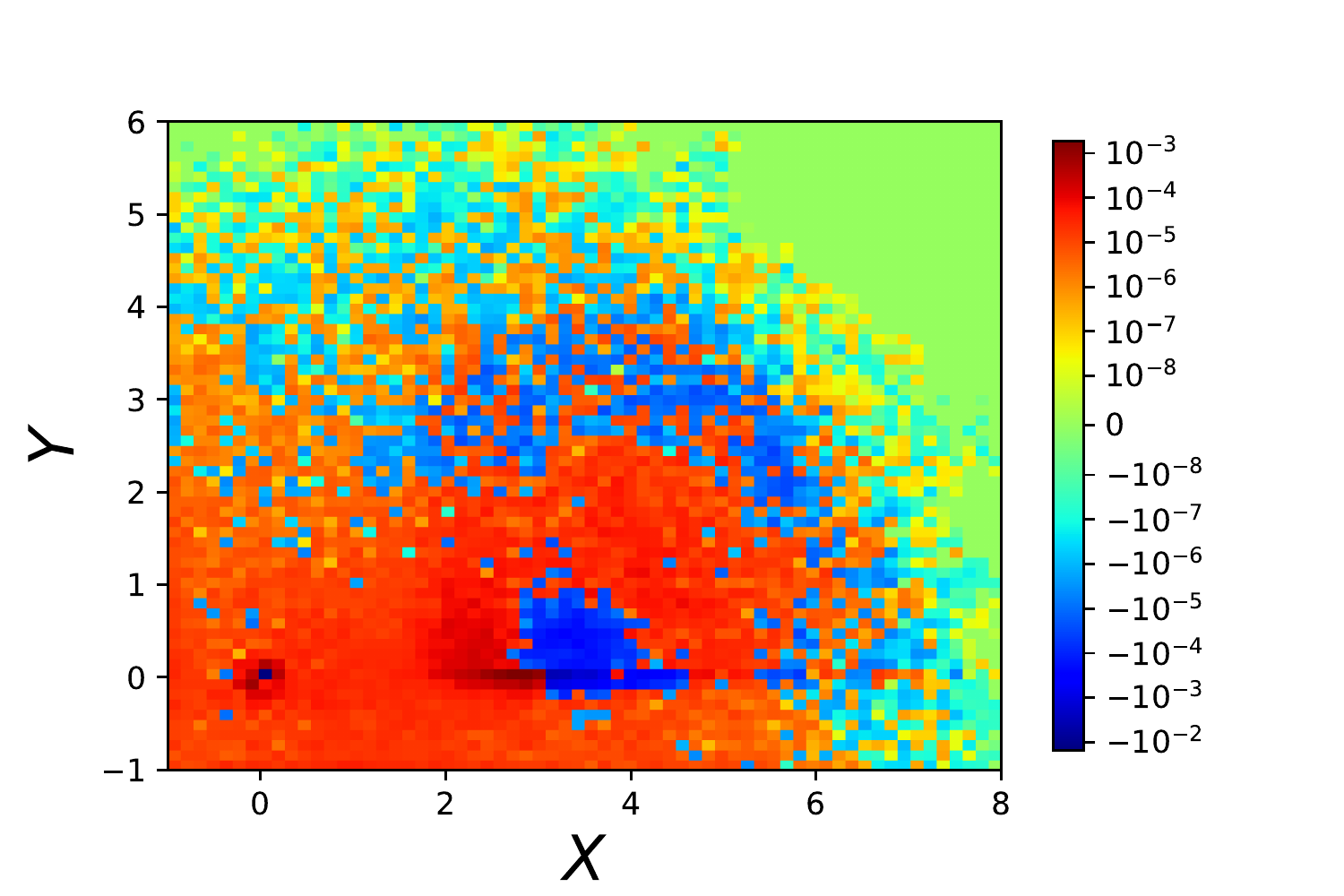}
		\caption{$V_1$ - $V_8$\label{fig:diff_V1V8}}
	\end{subfigure}
	\begin{subfigure}{.5\textwidth}
		\centering
		\includegraphics[width=\textwidth]{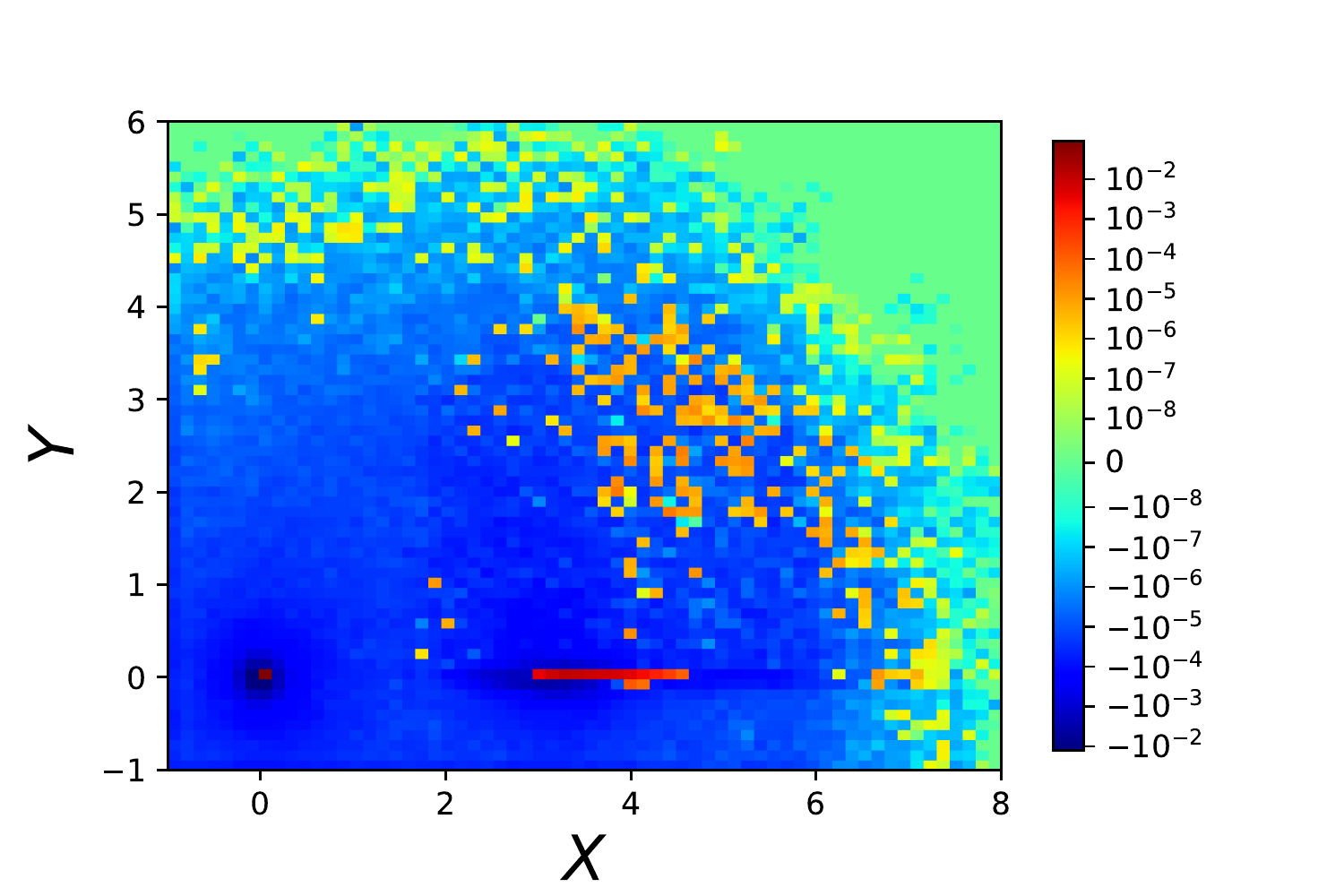}
		\caption{$V_1$ - $S_8$\label{fig:diff_V1S8}}
	\end{subfigure}
	\caption{The difference of stacked jet images in the generalized coordinate system $X$-$Y$ plane for $V_1 - V_8$ (left) and $V_1 - S_8$ (right).}
	\label{fig:diff} 
\end{figure}

%%%%%%%%%%%%%
\subsection{Data pre-processing}
The initial jet image is defined in the $\phi$-$\eta$ plane.  
In this study, we use four input channels as the intensity of the pixels:  
\begin{enumerate}
	\item transverse momenta of positively charged particles,
	\item transverse momenta of negatively charged particles,
	\item transverse momenta of neutral particles, and
	\item charged-particle multiplicity.
\end{enumerate}
The CNN is trained simultaneously on all four input channels.

To maximize the CNN learning performance, the images are pre-processed for faster training.  First, the images are rotated and reflected to change the jet image axes from $\phi$-$\eta$ to a generalized dimensionless coordinate system $X$-$Y$, while keeping $\Delta R$ invariant.  In this way, the geometry of the two jets identified as originating from the resonance are the same for each event and the jet images are sensitive to additional radiation.  In this case, we use the dominance of collinear radiation to identify the most distant of the three hardest jets as one of the jets originating from the resonant decay.  The hardest of the remaining two jets is identified as the second jet ``originating'' from the resonance decay.  The final jet is identified as radiation.   Additionally, we remove information that may superficially separate the resonances such as overall rates, absolute pixel intensity, resonant mass, etc. Considering the three hardest jets, the pre-processing steps applied to the jet images are:
\begin{enumerate}
	\item Shift the most distant jet\footnote{We identify the closest jet pair and define the remaining jet as the most distant one.}  of the first three leading jets to the origin of the coordinate system.
	\item Rotate the jet with higher transverse momentum of the remaining two jets to the positive $X$-axis. 
	\item Flip the third jet in the first quadrant.
	\item Digitize the jet image with $64 \times 64$ pixels in the range $X \in (-1, 9)$ and $Y \in (-1, 7)$. 
	\item Normalize the pixel intensities such that $\sum_{ij} I_{ij} = 1$ across the image, where $i$ and $j$ index over all pixels.  The intensity $I_{ij}$ of each pixel is the magnitude of transverse momentum or charged particle multiplicity depending on the input channel. 
	\item Subtract the mean $\mu_{ij}$ of the normalized images (the average intensity of pixel $(i,j)$ across all the data set) from each image, 
   transforming each pixel intensity as $I_{ij} \rightarrow I_{ij} - \mu_{ij}$. 
	\item Divide each pixel value by the standard deviation $\sigma_{ij}$ of that pixel value in the normalized dataset, $I_{ij} \rightarrow I_{ij}/\sigma_{ij}$. 
\end{enumerate}
\begin{figure}[t!]
	\centering
	\includegraphics[width=0.8\textwidth]{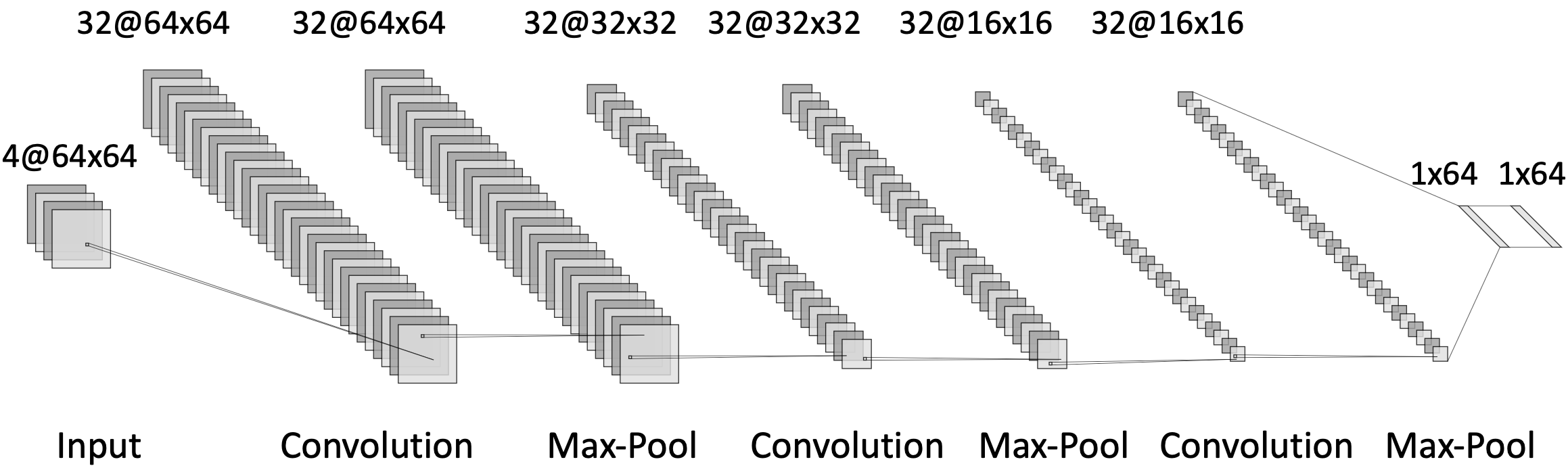}
	\caption{The structure of CNN model.}
	\label{fig:nn} 
\end{figure}
After the first three steps, we fixed the  two jets identified as originating from the resonance decay at the origin and along the $X$-axis, as well as the relative position of the soft radiation. 
The last three steps follow Ref.~\cite{Komiske:2016rsd}. In Fig.~\ref{fig:diff} we present the results of the pre-processing.  To obtain this figure, we stack 80,000 pre-processed images using the first three input channels, {\it i.e.}, the transverse momentum input channels.  The resonances considered are the color singlet vector $V_1$, the color octet vector $V_8$, and the color octet scalar $S_8$.   Figure \ref{fig:diff_V1V8} shows the results with the color octet vector pre-processed images subtracted from the color singlet vector pre-processed images.  The positive intensity pixels (red) have larger $V_1$ intensity, while negative intensity pixels (blue) have larger $V_8$ intensity.  This shows that, indeed, the radiation from $V_1$ occurs mostly near the $X$-axis, while radiation from the $V_8$ resonance occurs far away from the two jets identified as originating from the resonance.

The results are even more striking in Fig.~\ref{fig:diff_V1S8}, where we subtract the pre-processed color octet scalar $S_8$ images from the color singlet vector $V_1$. There is a strong positive intensity peak along the $X$-axis, while the negative intensity is more uniform.  This clearly reflects that the radiation from the color singlet preferentially occurs between the final state jets, while the color octet does not have a preferential radiation pattern in the pre-processed images, and gluon jets have more radiation than quark jets.

\subsection{CNN architecture and training}

 We use the pre-processed data as inputs. The deep convolutional network architecture used in this study consisted of three iterations of a convolutional layer with a ReLU activation and a max-pooling layer. Two dense hidden layers consisted of 64 units following the three convolutional layers.  An output layer of two units with softmax activation is fully connected to the final dense hidden layer. To avoid overfitting, the dropout rate was taken to be ${0.25,\ 0.25,\ 0.5}$ after the three convolutional layers, respectively, such that the CNN model only picks up the general features rather than the random fluctuations in the training samples.  Each convolutional layer consisted of 32 filters, with filter sizes of $3\time 3$. The max-pooling layers performed a $2\times 2$ down-sampling with a stride length of 1 to extract the most prominent features from the previous layer. We use zero padding in the convolution layer to keep the convolutional outputs from reducing in size. The structure of our CNN model is shown in Fig.~\ref{fig:nn}. We explored several CNN models with different architectures and filter sizes, and ultimately selected the best-performing model for our analysis.
 
 The CNN was trained using the Adam algorithm with categorical cross-entropy as the loss function.  The training used a batch size of 128 over 15 epochs. The data consisted of the 120,000 jet images, partitioned into 100,000 training images and 20,000 test images. An additional 10\% of the training images were used as validation data during the training.

\subsection{CNN results}

\begin{figure}[t!]
	\centering
	\includegraphics[width=.6\textwidth]{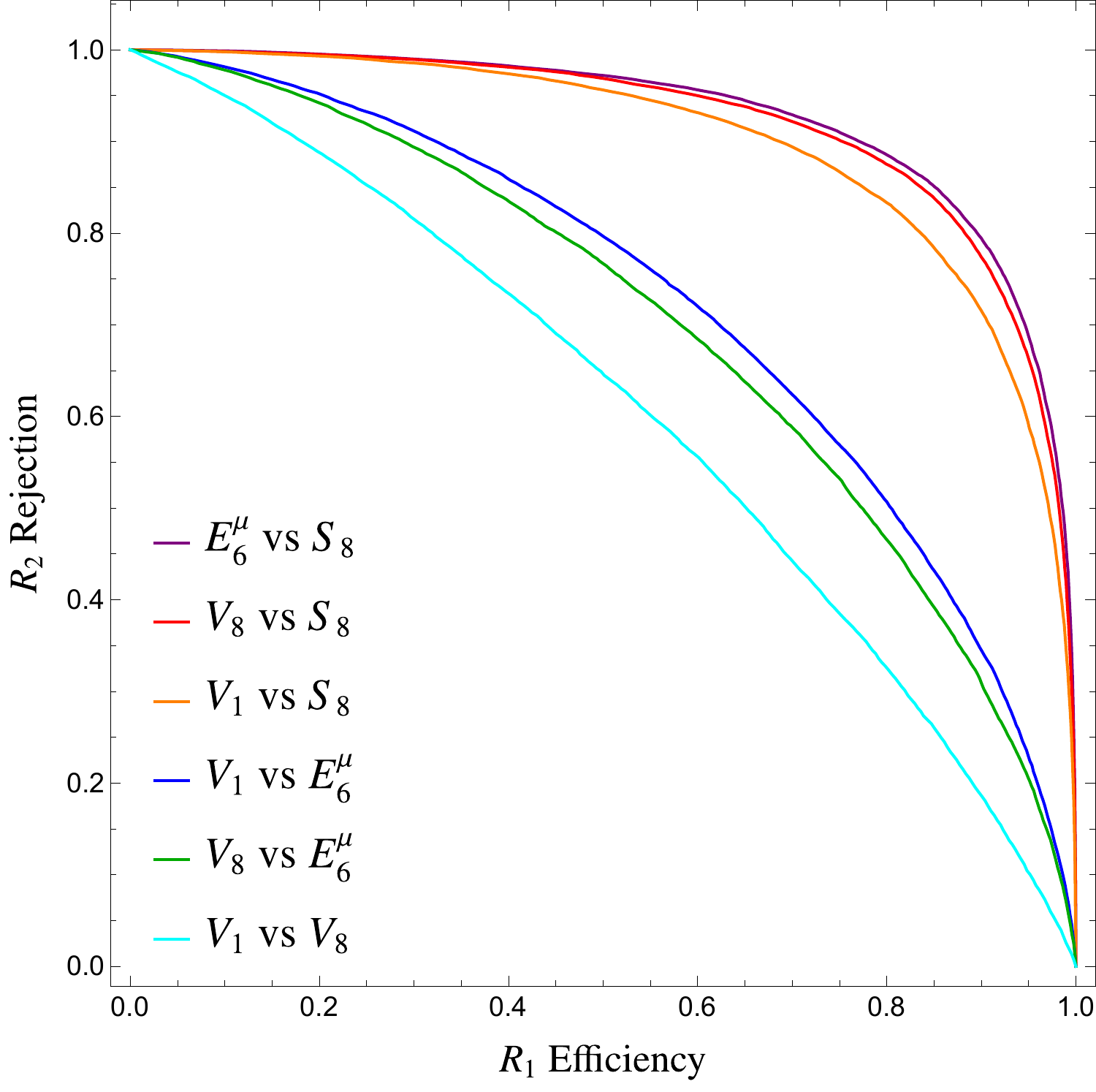}
	\label{fig:CNN_PFNC_S1}
	\caption{The ROC curves of the CNN model with 4 input channels for various resonances $R_1$ versus $R_2$.}
	\label{fig:roc}
\end{figure}

 To show the CNN performance using all four input channels in distinguishing two different signals, we plot the results in Fig.~\ref{fig:roc} for the receiver operating characteristic (ROC). The ROC curve shows the rejection rate of resonance $R_2$ as a function of the acceptance efficiency of resonance $R_1$, i.e., the power of the CNN to discriminate between two resonances $R_1$ and $R_2$. The area under the ROC curves (AUC) is shown in Table~\ref{Table:AUC}.  A larger AUC indicates that the CNN is more effective in distinguishing between the two resonances. This is because a curve with a larger AUC demonstrates a greater ability to reject $R_2$ resonances relative to the $R_1$ acceptance rates. As another measure of the CNN ability to distinguish resonances, in Table~\ref{Table:eff_CNN} we show the acceptance efficiencies\footnote{Here
 the acceptance efficiency is $100\%$ minus the rejection rate.} of various resonances $R_2$ when the acceptance efficiencies of another resonance $R_1$ is set to 50\%. For a fixed 50\% efficiency for $R_1$, the smaller the acceptance for $R_2$ the better the CNN can distinguish between the two resonances.  If the acceptance efficiencies of both resonances are 50\% they are indistinguishable.
 
 \begin{table}
	\centering
	\begin{tabular}{|c|c|c|}
		\hline
		\begin{tabular}{@{}c@{}} $R_1$ VS $R_2$  \\ AUC \end{tabular}  & \begin{tabular}{@{}c@{}} CNN \end{tabular}  \\[15pt]
		\hline
		$V_1$ ($u\bar{u}$) vs. $V_8$ ($u\bar{u}$) & 0.61 \\
		\hline
		$V_1$ ($u\bar{u}$) vs. $E^\mu_6$ ($uu$)& 0.73  \\
		\hline
		$V_1$ ($u\bar{u}$) vs. $S_8$ ($gg$)  & 0.90   \\
		\hline
		$V_8$ ($u\bar{u}$) vs. $E^\mu_6$ ($uu$)& 0.70  \\
		\hline
		$V_8$ ($u\bar u$) vs. $S_8$ ($gg$)& 0.92 \\
		\hline
		$E^\mu_6$ ($uu$) vs. $S_8$ ($gg$)& 0.93   \\
		\hline

	\end{tabular}
	\caption{AUC with $\rm{CNN}$ implementations.}
	\label{Table:AUC}
\end{table}

\begin{table}
	\centering
	\begin{tabular}{|c|c|c|c|c|}
		\hline
		\begin{tabular}{@{}c@{}} $R_{2}$ efficiency (\%) \\ at  50\% $R_1$ acceptance \end{tabular}  & $R_2: V_1$ ($u\bar u$)  & $R_2: V_8$ ($u\bar u$)  & $R_2: E^\mu_6$ ($uu$)  & $R_2: S_8$ ($gg$)  \\[15pt]
		\hline
		$R_1: V_1$ ($u\bar u$)  & 50\% & 35\% & 20\% & 4.4\% \\
		\hline
		$R_1: V_8$ ($u\bar{u}$) & 35\% & 50\% & 23\% & 3.1\%  \\
		\hline
		$R_1: E^\mu_6$ ($uu$)& 20\% & 23\% & 50\% & 2.8\%  \\
		\hline
		$R_1: S_8$ ($gg$)& 2.8\% & 1.8\% & 1.4\% & 50\%  \\
		\hline
	\end{tabular}
	\caption{$R_2$ acceptance efficiencies at 50\% $R_1$ acceptance with $\rm{CNN}$ implementations.}
	\label{Table:eff_CNN}
\end{table}

Among all the models we consider, the digluon scalar octet $S_8$ can be most easily distinguished from the others.  The AUC for any $R_1$ versus $S_8$ is 90\% or greater.  Also when the efficiency of $S_8$ is 50\% the acceptance efficiencies of other resonances are percent level and vice versa. This observation emphasizes the critical role played by the nature of the jet, whether it is composed of quarks or gluons. 
The easiest to distinguish resonances are $S_8$ and $E_6^\mu$, in which the spin, QED charge, color structure, and initial and final states are different. The efficiency of $E_6^\mu~(uu)$ can reach as low as $1.4$\% at the 50\% $S_8$ acceptance.  Although, $S_8$ and $V_8$ are nearly as easy to distinguish as $S_8$ and $E_6^\mu$.  In the case of $S_8$ versus $V_8$, the QED charges are the same but the spin, color structure, and initial and final states are different.  
This indicates that in the instance of distinguishing $S_8$ and other resonances, factors other than the charge of the resonance are the dominating factors.

The $V_1$ acceptance efficiency at 50\% $E_6^\mu$ acceptance is significantly lower than that at 50\% $V_8$ acceptance. The AUC for $V_1$ versus $V_8$ is significantly lower than $V_1$ versus $E_6^\mu$,  as well.  Similarly, the $V_8$ efficiency at 50\% $E_6^\mu$ acceptance is significantly lower than the $V_8$ efficiency at 50\% $V_1$ acceptance.  In all these cases, the spin is the same and the initial and final states consist of quarks and/or antiquarks.  The differences are the color structure and for $V_1/V_8$ versus $E_6^\mu$ the QED charge.  This is a hint that the QED charge information of the final state jets is important in distinguishing these resonances.
The most challenging case for CNN is to distinguish $V_1$ from $V_8$, where they are only differed by their color structure. Because the resonances are quite heavy, the two leading jets in the final state are almost back-to-back, so that the color interference is rather small. However, this issue can be potentially mitigated in the hadron collider with larger center-of-mass energy, where the produced heavy resonance can be significantly boosted. 

%%%%%%%%%%%%%%%%%%%%%%%%%%%%

\section{Summary and conclusions}
\label{sec:con}

If a new resonance is discovered at a hadron collider such as the LHC, it would be ultimately important to learn the underlying dynamics by determining its quantum numbers, such as the spin, couplings and gauge charges.
In this paper, we studied the characteristics of heavy resonances with a variety of spins, QED charges, and charges under the QCD color. 
We discussed the rapidity distribution of the dijet system to infer the information of the resonance coupling via the initial state partons. 
We presented the analytical expressions of the polar angle distributions in the resonance rest frame for the chiral couplings of resonances with a spin-$0,\ 1/2,\ 1,\ 3/2,$ and 2, and showed that the spin of resonance can be determined by measuring the angular distribution of its decay products.
Resonances with different color structures have different color flows, leading to distinctive radiation patterns. We showed that the ratio of the base two-to-two dijet resonance production and that with an additional radiation is quite powerful in distinguishing the color flow of different resonances.  Additionally, we presented analytical expressions for the hadronic antennae patterns of various dijet resonances with an additional soft gluon radiation, clearly showing the distinctions in these patterns.  We carried out a parton level cut-based analysis to exploit the  antenna radiation patterns in Fig.~\ref{spectrum.fig} for a variety of color resonance states. Those differences in the radiation patterns of different colored resonances can be used in the deep-learning techniques to distinguish them from each other. 

We then exploit the machine-learning techniques to improve the signal identification and to distinguish different colored resonances, by adopting a Conventional Neural Network (CNN). In the CNN model, the inputs are jet images and we exploit four input channels: charged particle multiplicity and the energy depositions of positively charged, negatively charged, and neutral particles.  
We study the heavy color resonance which is the most challenging scenario, as the decaying two leading jets are almost back-to-back and the color interference is small. We generate a 3-jet final state at the parton level with the minimum transverse momentum to be 600, 500, and 100 GeV, respectively. The softest parton can be an ISR or FSR. To fully make use of the radiation pattern from different color connections, we include all the jets after showering and hadronization via Pythia for CNN training. 
Our main results are shown 
in Fig.~\ref{fig:roc} and 
Table \ref{Table:eff_CNN}, summarized as follows: 
\begin{itemize}
\item 
We find excellent performance in distinguishing $S_8$ versus other resonances. 
In the CNN model, $V_1$ ($V_8$) efficiency at 50\% $S_8$ acceptance is 2.84 (1.77)\%, which implies a very low misidentification between the states. 
The performance in $S_8$ against $E^\mu_6$ is even better than against $V_8$.  When the $E_6^\mu$ efficiency is at 50\%, the $S_8$ acceptance is 1.40\%. These small improvements probably come from the difference in charge distribution and color structure. 
\item 
Distinguishing $E_6^\mu$ from $V_1$ and $V_8$ seems to be promising, as 
shown with 20\% and 23\% mis-identification acceptance, respectively. 
\item The most challenging channel is $V_1$ versus $V_8$, in which the only difference is the color quantum number. When the $V_1$ identification efficiency is at 50\%, the $V_8$ acceptance is only 34.9\% for the CNN model. 
It is interesting to note that 
distinguishing $V_1$ versus $E_6^\mu$ is better than $V_1$ versus $V_8$. This is because we separate the positive and negative charges in the CNN model, the difference in the charge distribution plays a significant role in this channel.
\end{itemize}
Our study shows that machine learning techniques can play an essential role in identifying different heavy color resonances at the LHC by exploiting the color information from the additional QCD radiation. 

Overall, we systematically studied the feasibility to determine the properties and quantum numbers of a heavy resonance if observed at a hadron collider. We found encouraging results including exploring the color structure of the events. 
Although we demonstrated our results numerically at the LHC, the methodology presented here should be applicable to other future hadron colliders.

\acknowledgments
The work of TH was supported in part by the U.S.~Department of Energy under grant No.~DE-SC0007914 and in part by the Pitt PACC. 
HL was supported by ISF, BSF and Azrieli Foundation.
IML was supported in part by U.S. Department of Energy grant number DE-SC0017988.  
ZL is supported in part by the U.S. Department of Energy (DOE) under grant No. DE-SC0022345 and DE-SC0011842. TH and ZL would like to thank the Aspen Center for Physics, where part of this work is complete, which is supported by the National Science Foundation (NSF) grant PHY-1607611. XW was supported by the National Science Foundation under Grant No.~PHY-1915147.  The data to generate the numerical results is available upon request.

%%%%%%%%%%%%%%%%%%%%%%%%%%%
\appendix
%%%%%%%%%%%%%%%%%%%%%%%%%%%

\section{Excited quark Clebsch-Gordan coefficients}
\label{app:CG}
Here we give the Clebsch-Gordan (CG) coefficients for the excited quark interactions in Eq.~(\ref{eq:qstar}).  For the triplet excited quark, the CG coefficients are proportional to the fundamental representation matrix:
\begin{eqnarray}
K_6^A=\sqrt{2}\,T^A.
\end{eqnarray}

For the color sextet excited quark the CG coefficients are
\begin{eqnarray}
&\displaystyle K^1_{6}=\frac{1}{\sqrt{6}}\begin{pmatrix}0&0&\sqrt{2}\\0 &0&0\\0&0&-\sqrt{2}\\0 & 1 & 0\\0&0&0\\-1&0&0\end{pmatrix},\quad
K^2_6=\frac{1}{\sqrt{6}}\begin{pmatrix}0&0&i\,\sqrt{2}\\0&0&0\\0&0&i\,\sqrt{2}\\0&-i&0\\0&0&0\\-i&0&0\end{pmatrix},\quad 
K_6^3=\frac{1}{\sqrt{6}}\begin{pmatrix}0&0&0\\0&0&-2\\0&0&0\\1&0&0\\0&0&0\\0&1&0\end{pmatrix},\\
&\displaystyle K_6^4=\frac{1}{\sqrt{6}}\begin{pmatrix}0&-\sqrt{2}&0\\1&0&0\\0&0&0\\0&0&-1\\0&\sqrt{2}&0\\0&0&0\end{pmatrix},\quad 
K_6^5=\frac{1}{\sqrt{6}}\begin{pmatrix}0&-i\,\sqrt{2}&0\\i&0&0\\0&0&0\\0&0&i\\0&-i\,\sqrt{2}&0\\0&0&0\end{pmatrix},\quad 
K_6^6=\frac{1}{\sqrt{6}}\begin{pmatrix}0&0&0\\0&-1&0\\\sqrt{2}&0&0\\0&0&0\\-\sqrt{2}&0&0\\0&0&1\end{pmatrix}\nonumber\\
&K_6^7=\displaystyle\frac{1}{\sqrt{6}}\begin{pmatrix}0&0&0\\0&-i&0\\i\sqrt{2}&0&0\\0&0&0\\i\sqrt{2}&0&0\\0&0&-i\end{pmatrix},\quad
K_6^8=\displaystyle\frac{1}{\sqrt{2}}\begin{pmatrix}0&0&0\\0&0&0\\0&0&0\\-1&0&0\\0&0&0\\0&1&0\end{pmatrix}.\nonumber
\end{eqnarray}
In deriving these CG coefficients, we have used sextet representation matrices of Ref.~\cite{Han:2009ya}

The bar notation is defined as $\bar{K}_{N_D,A}=(K^A_{N_D})^\dagger$. Then obey the orthonormality relationship
\begin{eqnarray}
{\rm Tr}\, \bar{K}_{N_D,A} K^{B}_{N_D}=\delta_A^B.
\end{eqnarray}

%%%%%%%%%%%%%%%%%%%%%%%%%%%%%%%
\section{Spin-3/2 Lagrangian}
\label{app:spin32}
Here we summarize results on the spin-3/2 Lagrangian as can be found in Ref.~\cite{Moldauer:1956zz,Haberzettl:1998rw,Benmerrouche:1989uc,osti_4283250,Rarita:1941mf,Hagiwara:2010pi,Christensen:2013aua,Stirling:2011ya,Alhazmi:2018whk,Dicus:2012uh,Nath:1971wp,Burges:1983zg}.  Start with a general form for the free field spin 3/2 Lagrangian
\begin{eqnarray}
\mathcal{L}=\bar{\psi}_\mu \Lambda^{\mu\nu}\psi_\nu,\label{eq:Spin32Lagrang}
\end{eqnarray}
In order to project out the two Dirac fermions (Lorentz representations of $(1/2,0)$ and $(0,1/2)$) that live in $\psi^\mu$, the on-shell spinors must obey the equalities~\cite{Moldauer:1956zz,osti_4283250}
\begin{eqnarray}
\gamma^\mu \psi_\mu=\partial^\mu \psi_\mu =0.\label{eq:Spin32Equalities}.
\end{eqnarray}
Additionally, the spinors should obey the Dirac equation
\begin{eqnarray}
\left(\slashed{p}-M\right)\psi_\mu=0.\label{eq:Dirac}
\end{eqnarray}
The equations of motion will be
\begin{eqnarray}
\Lambda^{\mu\nu}\psi_\nu = 0.
\end{eqnarray}
The conditions in Eqs.~(\ref{eq:Spin32Equalities}) and (\ref{eq:Dirac}) are invariant under the transformation
\begin{eqnarray}
\psi^\mu\rightarrow {\psi'}^\mu&=&\left(g^{\mu\nu}+\kappa \gamma^\mu\gamma^\nu\right)\psi_\nu.\label{eq:psiTrans}
\end{eqnarray}
Hence, we demand the Lagrangian to be invariant under this transformation as well.  It can then be found that the most general bilinear is~\cite{Moldauer:1956zz,osti_4283250,Nath:1971wp,Haberzettl:1998rw}
\begin{eqnarray}
\Lambda^{\mu\nu}=\left(\slashed{p}-M\right)g^{\mu\nu}+A\left(\gamma^\mu\,p^\nu+p^\mu\gamma^\nu\right)+\frac{B}{2}\gamma^\mu\slashed{p}\gamma^\nu+C\,M\,\gamma^{\mu}\gamma^{\nu},\label{eq:Lammunu}
\end{eqnarray}
where $B=3\,A^2+2\,A+1$, $C=3\,A^2+3\,A+1$, and $A$ has the transformation
\begin{eqnarray}
A\rightarrow A'=\frac{A-2\,\kappa}{1+4\,\kappa}.\label{eq:ATrans}
\end{eqnarray}
This transformation does not introduce any additional Lorentz structure to Eqs.~(\ref{eq:Spin32Lagrang}) and (\ref{eq:Lammunu}).  
The constant $A$ is unphysical and the Rarita-Schwinger Lagrangian~\cite{Rarita:1941mf} is a special case with $A=-1$.  

The most general propagator is
\begin{eqnarray}
S^{\mu\nu}=\frac{-i\,\Pi^{\mu\nu}}{p^2-M^2+i\,\Gamma\,M},\label{eq:Spin32Prop1}
\end{eqnarray}
where~\cite{Dicus:2012uh}
\begin{eqnarray}
\Pi^{\mu\nu}=\Pi^{\mu\nu}_{\rm RS}+\left(p^2-M^2\right)\left[\frac{a^2}{6\,M^2}\slashed{p}\gamma^\mu\gamma^\nu-\frac{a\,b}{3\,M}\gamma^\mu\gamma^\nu+\frac{a}{3\,M^2}p^\nu\gamma^\mu+\frac{a\,b}{3\,M^2}p^\mu\gamma^\nu\right],\label{eq:Spin32Prop}
\end{eqnarray}
the propagator for the Rarita-Schwinger spin-3/2 field is
\begin{eqnarray}
\Pi^{\mu\nu}_{\rm RS}=\left(\slashed{p}+M\right)\left[g^{\mu\nu}-\frac{2}{3\,M^2}p^\mu p^\nu-\frac{1}{3}\gamma^\mu\gamma^\nu-\frac{1}{3\,M}\left(p^\nu\gamma^\mu-p^\mu\gamma^\nu\right)\right],
\end{eqnarray}
and the constants are
\begin{eqnarray}
a=\frac{1+A}{1+2\,A},\,b=\frac{A}{1+2\,A}.
\end{eqnarray}
The propagator in Eq.~(\ref{eq:Spin32Prop}) reverts to the propagator for the Rarita-Schwinger field when the particle is on-shell $p^2=M^2$.  

The general propagator can be derived using the usual Green's function method, or using the transformations in Eqs.~(\ref{eq:psiTrans}) and (\ref{eq:ATrans}).  Using the transformation of $\psi_\mu$ we can find the general propagator in terms of the Rarita-Schwinger propagator:
\begin{eqnarray}
\Pi^{\mu\nu}=\left(g^\mu\rho+\kappa\gamma^\mu\gamma^\rho\right)\Pi_{{\rm RS},\rho\sigma} \left(g^{\sigma\nu}+\kappa \gamma^\sigma\gamma^\nu\right)
\end{eqnarray}
Using Eq.~(\ref{eq:ATrans}) with $A=-1$ for the Rarita-Schwinger case, we can solve for the necessary $\kappa$ to make this transformation:
\begin{eqnarray}
\kappa=-\frac{1}{2}\frac{1+A'}{1+2\,A'}.
\end{eqnarray}
From this Eq.~(\ref{eq:Spin32Prop}) can be derived.

The most general form for the lowest order operator that couples a spin-3/2 particle, spin-1/2 particle, and gauge fields has the form~\cite{Hagiwara:2010pi,Christensen:2013aua,Stirling:2011ya,Alhazmi:2018whk,Dicus:2012uh,Burges:1983zg}
\begin{eqnarray}
\bar{\psi}_\mu\left(g^{\mu\nu}+z\,\gamma^\mu\gamma^\nu\right)\gamma^\alpha\frac{1\pm \gamma_5}{2}\,K^a\,\psi\,F^a_{\nu\alpha},\label{eq:Spin32EFT}
\end{eqnarray}
where $K^a$ are Clebsch-Gordan coefficients, $\psi$ is a spin-1/2 field, $F^{a}_{\nu\alpha}$ is a gauge boson field strength tensor, and $z$ is a constant.  For on-shell spin-3/2 resonant amplitudes, the dependence on $z$ will vanish.  If we require that the effective interaction be invariant under the transformation in Eq.~(\ref{eq:psiTrans}), the parameter $z$ must have a transformation:
\begin{eqnarray}
z\rightarrow z'=\frac{z-\kappa}{1+4\,\kappa}.\label{eq:zTrans}
\end{eqnarray}

Another approach is to demand that no additional parameters in the effective interaction have transformations.  In this case, the interaction becomes~\cite{Nath:1971wp,Benmerrouche:1989uc}:
\begin{eqnarray}
\bar{\psi}_\mu\left(g^{\mu\nu}+\left[\frac{1}{2}\left(1+4\,Z\right)A+Z\right]\gamma^\mu\gamma^\nu\right)\gamma^\alpha \frac{1\pm\gamma_5}{2}K^a \psi F_{\nu\alpha}^a,\label{eq:general}
\end{eqnarray}
where $Z$ is an arbitrary constant that does not transform under Eq.~(\ref{eq:psiTrans}) and $A$ is the same constant as appears in Eq.~(\ref{eq:Lammunu}).  When the spin-3/2 particle is on-shell, the parameters $z,Z$ cannot contribute due to the equalities in Eq.~(\ref{eq:Spin32Equalities}).  Also, using the form of the interaction in Eq.~(\ref{eq:general}), matrix elements will be independent of the parameter $A$.  We show this explicitly for the resonance production by calculating the relevant helicity amplitudes in Appendix~\ref{sec:Spin32HelAmp}.

Finally, we note that the combination~\cite{Haberzettl:1998rw} of 
\begin{eqnarray}
\Psi^\mu \equiv \psi^\mu-\frac{1}{d}\gamma^\mu\gamma^\nu\psi_\nu,\label{eq:PsiInv}
\end{eqnarray}
is invariant under the transformation in Eq.~(\ref{eq:psiTrans}):
\begin{eqnarray}
\Psi^\mu \rightarrow \Psi^\mu.
\end{eqnarray}
The number of spacetime dimensions is $d$, which we set to $d=4$ in the following.  
This can be used as the building block to generate the interactions of the spin-3/2 field.  With this language the effective interaction invariant under Eq.~(\ref{eq:psiTrans}) would be~
\begin{eqnarray}
\overline{\Psi}^\nu\gamma^\alpha\frac{1\pm \gamma_5}{2}K^a\psi F^a_{\nu\alpha}=\overline{\psi}_\mu\left(g^{\mu\nu}-\frac{1}{4}\gamma^\mu\gamma^\nu\right)\gamma^\alpha\frac{1\pm \gamma_5}{2}K^a\psi F^a_{\nu\alpha}.\label{eq:invariant}
\end{eqnarray}
This form was proposed by Peccei~\cite{Haberzettl:1998rw}, and is equivalent to Eq.~(\ref{eq:Spin32EFT}) with the choice $z=-1/4$.  Note that unlike Eq.~(\ref{eq:Spin32EFT}), no parameters in this formulation will change with the field transformation of $\psi_\mu$.  Also, unlike Eq.~(\ref{eq:general}), Eq.~(\ref{eq:invariant}) is independent of the unphysical parameter $A$.  However, it has been argued that this choice is too restrictive for off-shell spin-3/2 particles~\cite{Nath:1971wp}.

We can also attempt to reformulate the kinetic term in terms of the invariant combination $\Psi^\mu$ in Eq.~(\ref{eq:PsiInv}).  Note that even for an off-shell field we have $\gamma^\mu\Psi_\mu=0$.  Hence, the most general Lagrangian is~\cite{Haberzettl:1998rw}
\begin{align}
\mathcal{L}_\Psi&=\Psi_\mu(\slashed{p}-M)\Psi^\mu=\bar{\psi}_\mu\left[\left(\slashed{p}-M\right)g^{\mu\nu}-\frac{1}{2}\left(\gamma^\mu p^\nu+p^\mu \gamma^\nu\right)+\frac{3}{8}\gamma^\mu \slashed{p}\gamma^\nu+\frac{1}{4}M\gamma^\mu\gamma^\nu\right]\psi_\nu.
\end{align}
This corresponds to choosing $A=-1/2$.  This has the nice transformation property $A=-1/2\rightarrow A'=-1/2$.  However, $A=-1/2$ is not allowed since the propagator in Eq.~(\ref{eq:Spin32Prop}) would be infinite~\cite{Haberzettl:1998rw,Moldauer:1956zz}.

%%%%%%%%%%%%%%%%%%%%%%%%%%%%%%%
\section{Dijet helicity amplitudes}
\label{HelAmp.app}

Helicity amplitudes are shown explicitly to illustrate the possible angular distributions for $2\rightarrow 2$ processes.  In generality we will label the resonances as $\mathcal{R}$ with mass $M_{\mathcal{R}}$ and width $\Gamma_{\mathcal{R}}$.  For a given initial state and resonance spin, this notation will encompass all possible color and electromagnetic charges of the resonances.  Throughout, we assume the initial state particles are massless but allow for massive final state particles.  We present the helicity amplitudes in terms of Wigner $d$-functions $d^{J}_{j_1,j_2}(\theta)$ and use the conventions of Ref.~\cite{Workman:2022ynf}.  All helicity amplitudes are evaluated in the partonic center of momentum frame.  Finally, we only report non-zero helicity amplitudes, i.e. any missing amplitudes are zero.

\subsection{Initial color states $\rep3\otimes\rep3$}
In this section, we consider $q_iq'_j \to \mathcal{R}  \rightarrow Q_k Q'_l$, where $i,j,k,l$ are the quark color indices.    Let $m_Q$ be the mass of $Q$ and $m_{Q'}$ be the mass of $Q'$.  The energies of $Q$ an $Q'$ in the partonic center of momentum frame are:
\begin{eqnarray}
E_Q=\frac{\hat{s}+m_Q^2-m_{Q'}^2}{2\sqrt{\hat{s}}},\quad{\rm and}\quad E_{Q'}=\frac{\hat{s}+m_{Q'}^2-m_Q^2}{2\sqrt{\hat{s}}},
\end{eqnarray}
respectively.  The $\beta$ factors, the speeds of the final state particles in the partonic c.m.~frame,  are then $\beta_Q=|\mathbf{p}_f|/E_Q$ and $\beta_{Q'}=|\mathbf{p}_f|/E_{Q'}$, where $\mathbf{p}_f$ is the three momentum of one of the final state quarks.

Finally, we only consider the dominant $s$-channel resonant diagrams.  The helicities of the amplitudes are in the order of $\mathcal{M}_s(q_i,q'_j,Q_k,Q'_l)$.

\subsubsection{Spin-0 resonance}
The non-zero helicity amplitudes are
\begin{eqnarray}
\mathcal{M}_s(+,+,+,+)&=&-\frac{1}{2}\left(1+\delta_{qq'}\right)\left(1+\delta_{QQ'}\right)K_{ij}^A\overline{K}_A^{kl}\frac{\sqrt{\hat{s}^2-(m_Q^2-m_{Q'}^2)^2}}{\hat{s}-M_\mathcal{R}^2+i \,\Gamma_\mathcal{R} M_\mathcal{R}}\\
&&\times \lambda^{\mathcal{R},R}_{qq'}\left[\sqrt{(1+\beta_Q)(1+\beta_{Q'})}{\lambda^{\mathcal{R},R}_{QQ'}}^*-\sqrt{(1-\beta_Q)(1-\beta_{Q'})}{\lambda^{\mathcal{R},L}_{QQ'}}^*\right]\nonumber\\
\mathcal{M}_s(-,-,-,-)&=&-\frac{1}{2}\left(1+\delta_{qq'}\right)\left(1+\delta_{QQ'}\right)K_{ij}^A\overline{K}_A^{kl}\frac{\sqrt{\hat{s}^2-(m_Q^2-m_{Q'}^2)^2}}{\hat{s}-M_\mathcal{R}^2+i \Gamma_\mathcal{R} M_\mathcal{R}}\\
&&\times \lambda^{\mathcal{R},L}_{qq'}\left[\sqrt{(1+\beta_Q)(1+\beta_{Q'})}{\lambda^{\mathcal{R},L}_{QQ'}}^*-\sqrt{(1-\beta_Q)(1-\beta_{Q'})}{\lambda^{\mathcal{R},R}_{QQ'}}^*\right]\nonumber\\
\mathcal{M}_s(+,+,-,-)&=&\frac{1}{2}\left(1+\delta_{qq'}\right)\left(1+\delta_{QQ'}\right)K_{ij}^A\overline{K}_A^{kl}\frac{\sqrt{\hat{s}^2-(m_Q^2-m_{Q'}^2)^2}}{\hat{s}-M_\mathcal{R}^2+i \,\Gamma_\mathcal{R} M_\mathcal{R}}\\
&&\times \lambda^{\mathcal{R},R}_{qq'}\left[\sqrt{(1+\beta_Q)(1+\beta_{Q'})}{\lambda^{\mathcal{R},L}_{QQ'}}^*-\sqrt{(1-\beta_Q)(1-\beta_{Q'})}{\lambda^{\mathcal{R},R}_{QQ'}}^*\right]\nonumber\\
\mathcal{M}_s(-,-,+,+)&=&\frac{1}{2}\left(1+\delta_{qq'}\right)\left(1+\delta_{QQ'}\right)K_{ij}^A\overline{K}_A^{kl}\frac{\sqrt{\hat{s}^2-(m_Q^2-m_{Q'}^2)^2}}{\hat{s}-M_\mathcal{R}^2+i\, \Gamma_\mathcal{R} M_\mathcal{R}}\\
&&\times \lambda^{\mathcal{R},L}_{qq'}\left[\sqrt{(1+\beta_Q)(1+\beta_{Q'})}{\lambda^{\mathcal{R},R}_{QQ'}}^*-\sqrt{(1-\beta_Q)(1-\beta_{Q'})}{\lambda^{\mathcal{R},L}_{QQ'}}^*\right]\nonumber
\end{eqnarray}

\subsubsection{Spin-1 resonance}
The non-zero helicity amplitudes for vector diquarks are
\begin{align}
\mathcal{M}_s(+,-,+,-)&=\left(1+\delta_{qq'}\right)\left(1+\delta_{QQ'}\right)K_{ij}^A\overline{K}_A^{kl}\frac{\sqrt{\hat{s}^2-(m_Q^2-m_{Q'}^2)^2}}{\hat{s}-M_{\mathcal{R}}^2+i\,\Gamma_{\mathcal{R}}M_{\mathcal{R}}}\\
&\times \lambda_{qq'}^{\mathcal{R},L}\left[\sqrt{(1+\beta_Q)(1+\beta_{Q'})}{\lambda^{\mathcal{R},R}_{QQ'}}^*+\sqrt{(1-\beta_Q)(1-\beta_{Q'})}{\lambda^{\mathcal{R},L}_{QQ'}}^*\right]d^1_{1,1}(\theta)\nonumber\\
\mathcal{M}_s(-,+,-,+)&=\left(1+\delta_{qq'}\right)\left(1+\delta_{QQ'}\right)K_{ij}^A\overline{K}_A^{kl}\frac{\sqrt{\hat{s}^2-(m_Q^2-m_{Q'}^2)^2}}{\hat{s}-M_{\mathcal{R}}^2+i\,\Gamma_{\mathcal{R}}M_{\mathcal{R}}}\\
&\times \lambda_{qq'}^{\mathcal{R},R}\left[\sqrt{(1+\beta_Q)(1+\beta_{Q'})}{\lambda^{\mathcal{R},L}_{QQ'}}^*+\sqrt{(1-\beta_Q)(1-\beta_{Q'})}{\lambda^{\mathcal{R},R}_{QQ'}}^*\right]d^1_{-1,-1}(\theta)\nonumber\\
\mathcal{M}_s(+,-,-,+)&=-\left(1+\delta_{qq'}\right)\left(1+\delta_{QQ'}\right)K_{ij}^A\overline{K}_A^{kl}\frac{\sqrt{\hat{s}^2-(m_Q^2-m_{Q'}^2)^2}}{\hat{s}-M_{\mathcal{R}}^2+i\,\Gamma_{\mathcal{R}}M_{\mathcal{R}}}\\
&\times \lambda_{qq'}^{\mathcal{R},L}\left[\sqrt{(1+\beta_Q)(1+\beta_{Q'})}{\lambda^{\mathcal{R},L}_{QQ'}}^*+\sqrt{(1-\beta_Q)(1-\beta_{Q'})}{\lambda^{\mathcal{R},R}_{QQ'}}^*\right]d^1_{1,-1}(\theta)\nonumber\\
\mathcal{M}_s(-,+,+,-)&=-\left(1+\delta_{qq'}\right)\left(1+\delta_{QQ'}\right)K_{ij}^A\overline{K}_A^{kl}\frac{\sqrt{\hat{s}^2-(m_Q^2-m_{Q'}^2)^2}}{\hat{s}-M_{\mathcal{R}}^2+i\,\Gamma_{\mathcal{R}}M_{\mathcal{R}}}\\
&\times \lambda_{qq'}^{\mathcal{R},R}\left[\sqrt{(1+\beta_Q)(1+\beta_{Q'})}{\lambda^{\mathcal{R},R}_{QQ'}}^*+\sqrt{(1-\beta_Q)(1-\beta_{Q'})}{\lambda^{\mathcal{R},L}_{QQ'}}^*\right]d^1_{-1,1}(\theta)\nonumber
\end{align}
There are also amplitudes that vanish when both final state quarks are massless:
\begin{align}
\mathcal{M}_s(+,-,-,-)&=\frac{1}{\sqrt{2}}\left(1+\delta_{qq'}\right)\left(1+\delta_{QQ'}\right)K_{ij}^A\overline{K}_A^{kl}\frac{\sqrt{\hat{s}^2-(m_Q^2-m_{Q'}^2)^2}}{\hat{s}-M_{\mathcal{R}}^2+i\,\Gamma_{\mathcal{R}}M_{\mathcal{R}}}\\
&\times \lambda_{qq'}^{\mathcal{R},L}\left[\sqrt{(1+\beta_Q)(1-\beta_{Q'})}{\lambda^{\mathcal{R},L}_{QQ'}}^*+\sqrt{(1-\beta_Q)(1+\beta_{Q'})}{\lambda^{\mathcal{R},R}_{QQ'}}^*\right]d^1_{1,0}(\theta)\nonumber\\
\mathcal{M}_s(-,+,+,+)&=\frac{1}{\sqrt{2}}\left(1+\delta_{qq'}\right)\left(1+\delta_{QQ'}\right)K_{ij}^A\overline{K}_A^{kl}\frac{\sqrt{\hat{s}^2-(m_Q^2-m_{Q'}^2)^2}}{\hat{s}-M_{\mathcal{R}}^2+i\,\Gamma_{\mathcal{R}}M_{\mathcal{R}}}\\
&\times \lambda_{qq'}^{\mathcal{R},R}\left[\sqrt{(1-\beta_Q)(1+\beta_{Q'})}{\lambda^{\mathcal{R},L}_{QQ'}}^*+\sqrt{(1+\beta_Q)(1-\beta_{Q'})}{\lambda^{\mathcal{R},R}_{QQ'}}^*\right]d^1_{-1,0}(\theta)\nonumber\\
\mathcal{M}_s(+,-,+,+)&=-\frac{1}{\sqrt{2}}\left(1+\delta_{qq'}\right)\left(1+\delta_{QQ'}\right)K_{ij}^A\overline{K}_A^{kl}\frac{\sqrt{\hat{s}^2-(m_Q^2-m_{Q'}^2)^2}}{\hat{s}-M_{\mathcal{R}}^2+i\,\Gamma_{\mathcal{R}}M_{\mathcal{R}}}\\
&\times \lambda_{qq'}^{\mathcal{R},L}\left[\sqrt{(1-\beta_Q)(1+\beta_{Q'})}{\lambda^{\mathcal{R},L}_{QQ'}}^*+\sqrt{(1+\beta_Q)(1-\beta_{Q'})}{\lambda^{\mathcal{R},R}_{QQ'}}^*\right]d^1_{1,0}(\theta)\nonumber\\
\mathcal{M}_s(-,+,-,-)&=-\frac{1}{\sqrt{2}}\left(1+\delta_{qq'}\right)\left(1+\delta_{QQ'}\right)K_{ij}^A\overline{K}_A^{kl}\frac{\sqrt{\hat{s}^2-(m_Q^2-m_{Q'}^2)^2}}{\hat{s}-M_{\mathcal{R}}^2+i\,\Gamma_{\mathcal{R}}M_{\mathcal{R}}}\\
&\times \lambda_{qq'}^{\mathcal{R},R}\left[\sqrt{(1+\beta_Q)(1-\beta_{Q'})}{\lambda^{\mathcal{R},L}_{QQ'}}^*+\sqrt{(1-\beta_Q)(1+\beta_{Q'})}{\lambda^{\mathcal{R},R}_{QQ'}}^*\right]d^1_{-1,0}(\theta),\nonumber
\end{align}
where $\theta$ is the angle between $q$ and $Q$ in the resonance rest frame.

%%%%%%%%%%%%%%%%%%%%%%%%%%%%%%%%%%%%%%%%%%%%%%
\subsection{Initial color states $\rep3\otimes\rep8$}
\label{38.SEC}

In this section, we consider $q_ig^A \to  \mathcal{R} \rightarrow Q_jg^B$, where $i,j$ are the quark color indices and $A,B$ are the gluon color indices. For generality, the final state quark is allowed to be massive with mass $m_{Q}$.  Then, the energies of the final state quark and gluon in the partonic center of momentum frame are, respectively,
\begin{eqnarray}
E_Q=\frac{\hat{s}+m_Q^2}{2\sqrt{\hat{s}}},\quad{\rm and}\quad E_g=\frac{\hat{s}-m_Q^2}{2\sqrt{\hat{s}}}.
\end{eqnarray}
The $\beta$ factor is then $\beta=(\hat{s}-m_Q^2)/(\hat{s}+m_Q^2)$.

The helicities of the amplitudes are in the order of $\mathcal{M}_s(q_i,g^A,Q_j,g^B)$.  In the following $\theta$ is the angle between the initial state quark $q$ and final state quark $Q$.  Additionally, initial state couplings are denoted with a subscript $i$ and final state couplings with $f$.

\subsubsection{Spin-1/2 resonance}
We calculate the dominant $s$-channel resonant contributions.  The non-zero helicity amplitudes are
\begin{eqnarray}
\mathcal{M}_s(+,+,+,+)&=&8{(\overline{K}_BK^A)^j}_i \frac{g_S^2}{\Lambda^2}\frac{\hat{s}(\hat{s}-m_Q^2)}{\hat{s}-M_{\mathcal{R}}^2+i\,\Gamma_{\mathcal{R}}M_{\mathcal{R}}}\lambda^{\mathcal{R}}_{i,R}\lambda^{\mathcal{R}*}_{f,R}\,d^{1/2}_{-1/2,-1/2}(\theta)\\
\mathcal{M}_s(-,-,-,-)&=&8{(\overline{K}_BK^A)^j}_i \frac{g_S^2}{\Lambda^2}\frac{\hat{s}(\hat{s}-m_Q^2)}{\hat{s}-M_{\mathcal{R}}^2+i\,\Gamma_{\mathcal{R}}M_{\mathcal{R}}}\lambda^{\mathcal{R}}_{i,L}\lambda^{\mathcal{R}*}_{f,L}\,d^{1/2}_{1/2,1/2}(\theta).\\
\mathcal{M}_s(+,+,-,-)&=&8{(\overline{K}_BK^A)^j}_i \frac{g_S^2}{\Lambda^2}\frac{M_{\mathcal{R}}\sqrt{\hat{s}}(\hat{s}-m_Q^2)}{\hat{s}-M_{\mathcal{R}}^2+i\,\Gamma_{\mathcal{R}}M_{\mathcal{R}}}\lambda^{\mathcal{R}}_{i,R}\lambda^{\mathcal{R}*}_{f,L}\,d^{1/2}_{-1/2,1/2}(\theta)\\
\mathcal{M}_s(-,-,+,+)&=&8{(\overline{K}_BK^A)^j}_i \frac{g_S^2}{\Lambda^2}\frac{M_{\mathcal{R}}\sqrt{\hat{s}}(\hat{s}-m_Q^2)}{\hat{s}-M_{\mathcal{R}}^2+i\,\Gamma_{\mathcal{R}}M_{\mathcal{R}}}\lambda^{\mathcal{R}}_{i,L}\lambda^{\mathcal{R}*}_{f,R}\,d^{1/2}_{1/2,-1/2}(\theta)
\end{eqnarray}

\subsubsection{Spin-3/2 resonance}
\label{sec:Spin32HelAmp}
For the spin-3/2 calculation, we use the general propagator in Eqs.~(\ref{eq:Spin32Prop1}) and (\ref{eq:Spin32Prop}) and the parameterization of the effective interaction in Eqs.~(\ref{eq:qstar}) and (\ref{eq:Spin32EFT}).
The helicity amplitudes for s-channel sextet/triplet spin-$3/2$ fermion that survive when the resonance is on-shell and the final state quark is massless are

\begin{eqnarray}
\mathcal{M}_s(+,-,+,-)&=& \frac{g_S^2}{\Lambda^2}\left(\overline{K}_B K^A\right)^j\,_i\frac{\hat{s}(\hat{s}-m_Q^2)}{\hat{s}-M^2_\mathcal{R}+i\,\Gamma_\mathcal{R}\,M_\mathcal{R}}\lambda^\mathcal{R}_{i,R}\lambda_{f,R}^{\mathcal{R}*}\,d^{3/2}_{3/2,3/2}(\theta)\\
\mathcal{M}_s(-,+,-,+)&=& \frac{g_S^2}{\Lambda^2}\left(\overline{K}_B K^A\right)^j\,_i\frac{\hat{s}(\hat{s}-m_Q^2)}{\hat{s}-M_\mathcal{R}^2+i\,\Gamma_\mathcal{R}\,M_\mathcal{R}}\lambda^\mathcal{R}_{i,L}\lambda_{f,L}^{\mathcal{R}*}\,d^{3/2}_{-3/2,-3/2}(\theta)\\
\mathcal{M}_s(+,-,-,+)&=&\frac{g_S^2}{\Lambda^2}\left(\overline{K}_B K^A\right)^j\,_i\frac{M_\mathcal{R}\sqrt{\hat{s}}(\hat{s}-m_Q^2)}{\hat{s}-M^2_\mathcal{R}+i\,\Gamma_\mathcal{R}\,M_\mathcal{R}}\lambda^\mathcal{R}_{i,R}\lambda_{f,L}^{\mathcal{R}*}\,d^{3/2}_{3/2,-3/2}(\theta)\\
\mathcal{M}_s(-,+,+,-)&=&\frac{g_S^2}{\Lambda^2}\left(\overline{K}_B K^A\right)^j\,_i\frac{M_\mathcal{R}\sqrt{\hat{s}}(\hat{s}-m_Q^2)}{\hat{s}-M_\mathcal{R}^2+i\,\Gamma_\mathcal{R}\,M_\mathcal{R}}\lambda^\mathcal{R}_{i,L}\lambda_{f,R}^{\mathcal{R}*}\,d^{3/2}_{-3/2,3/2}(\theta)
\end{eqnarray}

There are also a set of helicity amplitudes for $s$-channel spin-3/2 particles that vanish when the resonance is on-shell:
\begin{align}
\mathcal{M}_s(+,+,+,+)&=\frac{1}{3}\frac{g_S^2}{\Lambda^2}\left(\overline{K}_B K^A\right)^j\,_i\frac{(\hat{s}-M_\mathcal{R}^2)(\hat{s}-m_Q^2)}{\hat{s}-M_\mathcal{R}^2+i\,\Gamma_\mathcal{R}\,M_\mathcal{R}}\frac{\hat{s}}{M_\mathcal{R}^2}\left(\frac{1+4\,z}{1+2\,A}\right)^2\lambda^\mathcal{R}_{i,R}\lambda^{\mathcal{R}*}_{f,R}\,d^{1/2}_{-1/2,-1/2}(\theta)\\
\mathcal{M}_s(-,-,-,-)&=\frac{1}{3}\frac{g_S^2}{\Lambda^2}\left(\overline{K}_B K^A\right)^j\,_i\frac{(\hat{s}-M_\mathcal{R}^2)(\hat{s}-m_Q^2)}{\hat{s}-M_\mathcal{R}^2+i\,\Gamma_\mathcal{R}\,M_\mathcal{R}}\frac{\hat{s}}{M_\mathcal{R}^2}\left(\frac{1+4\,z}{1+2\,A}\right)^2\lambda^\mathcal{R}_{i,L}\lambda_{f,L}^{\mathcal{R}*}d^{1/2}_{1/2,1/2}(\theta)\\
\mathcal{M}_s(+,+,-,-)&=-\frac{2}{3}\frac{g_S^2}{\Lambda^2}\left(\overline{K}_B K^A\right)^j\,_i\frac{(\hat{s}-M_\mathcal{R}^2)(\hat{s}-m_Q^2)}{\hat{s}-M_\mathcal{R}^2+i\,\Gamma_\mathcal{R}\,M_\mathcal{R}}\frac{\sqrt{\hat{s}}}{M_\mathcal{R}}\left(\frac{1+4\,z}{1+2\,A}\right)^2\nonumber\\
&\quad\quad\times\lambda_{i,R}^\mathcal{R}\left(\lambda_{f,L}^{\mathcal{R}*}-\frac{1}{2}\frac{1+2\,A}{1+4\,z}\frac{m_Q}{M_\mathcal{R}}\lambda^{\mathcal{R}*}_{f,R}\right)d^{1/2}_{-1/2,1/2}(\theta)\\
\mathcal{M}_s(-,-,+,+)&=-\frac{2}{3}\frac{g_S^2}{\Lambda^2}\left(\overline{K}_B K^A\right)^j\,_i\frac{(\hat{s}-M_\mathcal{R}^2)(\hat{s}-m_Q^2)}{\hat{s}-M_\mathcal{R}^2+i\,\Gamma_\mathcal{R}\,M_\mathcal{R}}\frac{\sqrt{\hat{s}}}{M_\mathcal{R}}\left(\frac{1+4\,z}{1+2\,A}\right)^2\nonumber\\
&\times \lambda^\mathcal{R}_{i,L}\left(\lambda_{f,R}^{\mathcal{R}*}-\frac{1}{2}\frac{1+2\,A}{1+4\,z}\frac{m_Q}{M_\mathcal{R}}\lambda^{\mathcal{R}*}_{f,L}\right)d^{1/2}_{1/2,-1/2}(\theta)
\end{align}

Finally, for the $s$-channel there are also amplitudes that vanish when the final state quark is massless:
\begin{align}
\mathcal{M}_s(+,-,+,+)&=\frac{1}{\sqrt{3}}\frac{g_S^2}{\Lambda^2}\left(\overline{K}_B K^A\right)^j\,_i \frac{m_Q M_\mathcal{R} (\hat{s}-m_Q^2)}{\hat{s}-M_\mathcal{R}^2+i\,\Gamma_\mathcal{R} M_\mathcal{R}}\lambda^\mathcal{R}_{i,R}\lambda^{\mathcal{R}*}_{f,L}\,d^{3/2}_{3/2,-1/2}(\theta)\\
\mathcal{M}_s(-,+,-,-)&=\frac{1}{\sqrt{3}}\frac{g_S^2}{\Lambda^2}\left(\overline{K}_B K^A\right)^j\,_i\frac{m_Q M_\mathcal{R} (\hat{s}-m_Q^2)}{\hat{s}-M_\mathcal{R}^2+i\,\Gamma_\mathcal{R} M_\mathcal{R}}\lambda^\mathcal{R}_{i,L}\lambda^{\mathcal{R}*}_{f,R}\,d^{3/2}_{-3/2,1/2}(\theta)\\
\mathcal{M}_s(+,-,-,-)&=\frac{1}{\sqrt{3}}\frac{g_S^2}{\Lambda^2}\left(\overline{K}_B K^A\right)^j\,_i\frac{m_Q\sqrt{\hat{s}}(\hat{s}-m_Q^2)}{\hat{s}-M_\mathcal{R}^2+i\,\Gamma_\mathcal{R} M_\mathcal{R}}\lambda_{i,R}^\mathcal{R}\lambda_{f,R}^{\mathcal{R}*}\,d^{3/2}_{3/2,1/2}(\theta)\\
\mathcal{M}_s(-,+,+,+)&=\frac{1}{\sqrt{3}}\frac{g_S^2}{\Lambda^2}\left(\overline{K}_B K^A\right)^j\,_i \frac{m_Q\sqrt{\hat{s}}(\hat{s}-m_Q^2)}{\hat{s}-M_\mathcal{R}^2+i\,\Gamma_\mathcal{R}M_\mathcal{R}}\lambda_{i,L}^\mathcal{R}\lambda_{f,L}^{\mathcal{R}*}\,d^{3/2}_{-3/2,-1/2}(\theta)
\end{align}

For completeness, and to investigate dependence on the unphysical parameter $A$, for spin-3/2 we also provide the t-channel amplitudes:
\begin{align}
\mathcal{M}_t(+,-,+,-)&=\frac{1}{3}\frac{g_S^2}{\Lambda^2}\left(\overline{K}_AK^B\right)^{j}\,_i\frac{(\hat{s}-m_Q^2)(\hat{t}-M^2_\mathcal{R})}{\hat{t}-M^2_\mathcal{R}+i\,\Gamma_\mathcal{R}\,M_\mathcal{R}}\frac{\hat{s}}{M^2_\mathcal{R}}\left(\frac{1+4\,z}{1+2\,A}\right)^2\lambda^\mathcal{R}_{i,R}\lambda_{f,R}^{\mathcal{R}*}\,d^{3/2}_{3/2,3/2}(\theta)\\
\mathcal{M}_t(-,+,-,+)&=\frac{1}{3}\frac{g_S^2}{\Lambda^2}\left(\overline{K}_AK^B\right)^{j}\,_i\frac{(\hat{s}-m_Q^2)(\hat{t}-M_\mathcal{R}^2)}{\hat{t}-M_\mathcal{R}^2+i\,\Gamma_\mathcal{R}\,M_\mathcal{R}}\frac{\hat{s}}{M_\mathcal{R}^2}\left(\frac{1+4\,z}{1+2\,A}\right)^2\lambda_{i,L}^\mathcal{R}\lambda_{f,L}^{\mathcal{R}*}\,d^{3/2}_{-3/2,-3/2}(\theta)\\
\mathcal{M}_t(+,-,-,+)&=\frac{g_S^2}{\Lambda^2}\left(\overline{K}_AK^B\right)^{j}\,_i\frac{M_{\mathcal{R}}\sqrt{\hat{s}}(\hat{s}-m_Q^2)}{\hat{t}-M_\mathcal{R}^2+i\,\Gamma_\mathcal{R}\,M_\mathcal{R}}\lambda_{i,R}^\mathcal{R}\lambda_{f,L}^{\mathcal{R}*}\,d^{3/2}_{3/2,-3/2}(\theta)\\
\mathcal{M}_t(-,+,+,-)&=\frac{g_S^2}{\Lambda^2}\left(\overline{K}_AK^B\right)^{j}\,_i\frac{M_\mathcal{R}\sqrt{\hat{s}}(\hat{s}-m_Q^2)}{\hat{t}-M_\mathcal{R}^2+i\,\Gamma_\mathcal{R}\,M_\mathcal{R}}\lambda_{i,L}^\mathcal{R}\lambda_{f,R}^{\mathcal{R}*}\,d^{3/2}_{-3/2,3/2}(\theta)\\
\mathcal{M}_t(+,+,+,+)&=\frac{g_S^2}{\Lambda^2}\left(\overline{K}_AK^B\right)^{j}\,_i\frac{\hat{s}(\hat{s}-m_Q^2)}{\hat{t}-M_\mathcal{R}^2+i\,\Gamma_\mathcal{R}\,M_\mathcal{R}}\lambda^\mathcal{R}_{i,R}\lambda_{f,R}^{\mathcal{R}*}\,d^{1/2}_{-1/2,-1/2}(\theta)\\
\mathcal{M}_t(-,-,-,-)&=\frac{g_S^2}{\Lambda^2}\left(\overline{K}_AK^B\right)^{j}\,_i\frac{\hat{s}(\hat{s}-m_Q^2)}{\hat{t}-M_\mathcal{R}^2+i\,\Gamma_\mathcal{R}\,M_\mathcal{R}}\lambda_{i,L}^\mathcal{R}\lambda_{f,L}^{\mathcal{R}*}\,d^{1/2}_{1/2,1/2}(\theta)\\
\mathcal{M}_t(+,+,-,-)&=-\frac{2}{3\sqrt{3}}\frac{g_S^2}{\Lambda^2}\left(\overline{K}_AK^B\right)^{j}\,_i\frac{(\hat{s}-m_Q^2)(\hat{t}-M_\mathcal{R}^2)}{\hat{t}-M_{\mathcal{R}}^2+i\,\Gamma_\mathcal{R}\,M_\mathcal{R}}\frac{\sqrt{\hat{s}}}{M_\mathcal{R}}\left(\frac{1+4\,z}{1+2\,A}\right)^2\nonumber\\
&\quad\quad\times\lambda_{i,R}^\mathcal{R}\left(\lambda_{f,L}^{\mathcal{R}*}-\frac{1}{2}\frac{1+2\,A}{1+4\,z}\frac{m_Q}{M_\mathcal{R}}\lambda^{\mathcal{R}*}_{f,R}\right)\,d^{3/2}_{-1/2,-3/2}(\theta)\\
\mathcal{M}_t(-,-,+,+)&=-\frac{2}{3\sqrt{3}}\frac{g_S^2}{\Lambda^2}\left(\overline{K}_AK^B\right)^{j}\,_i\frac{(\hat{s}-m_Q^2)(\hat{t}-M_\mathcal{R}^2)}{\hat{t}-M_{\mathcal{R}}^2+i\,\Gamma_\mathcal{R}\,M_\mathcal{R}}\frac{\sqrt{\hat{s}}}{M_\mathcal{R}}\left(\frac{1+4\,z}{1+2\,A}\right)^2\nonumber\\
&\quad\quad\times\lambda_{i,L}^\mathcal{R}\left(\lambda_{f,R}^{\mathcal{R}*}-\frac{1}{2}\frac{1+2\,A}{1+4\,z}\frac{m_Q}{M_\mathcal{R}}\lambda^{\mathcal{R}*}_{f,L}\right)\,d^{3/2}_{1/2,3/2}(\theta),
\end{align}
where $\hat{t}=-(\hat{s}-m_Q^2)(1+\cos\theta)/2$.  As with the $s$-chanel diagrams, the $t$-channel diagrams also have a set of amplitudes that go to zero if the final state quark is massless:
\begin{align}
\mathcal{M}_t(+,+,+,-)&=-\frac{2}{3\sqrt{3}}\frac{g_S^2}{\Lambda^2}\left(\overline{K}_AK^B\right)^{j}\,_i\frac{(\hat{s}-M_Q^2)(\hat{t}-M_{\mathcal{R}}^2)}{\hat{t}-M_\mathcal{R}^2+i\,\Gamma_\mathcal{R}M_\mathcal{R}}\frac{m_Q}{M_\mathcal{R}}\left(\frac{1+4\,z}{1+2\,A}\right)^2\nonumber\\
&\quad\quad\times\lambda_{i,R}^\mathcal{R}\left(\lambda_{f,L}^{\mathcal{R}*}-\frac{1}{2}\frac{1+2\,A}{1+4\,z}\frac{m_Q}{M_\mathcal{R}}\lambda_{f,R}^{\mathcal{R}*}\right)\,d^{3/2}_{-1/2,3/2}(\theta)\\
\mathcal{M}_t(-,-,-,+)&=\frac{2}{3\sqrt{3}}\frac{g_S^2}{\Lambda^2}\left(\overline{K}_AK^B\right)^{j}\,_i\frac{(\hat{s}-M_Q^2)(\hat{t}-M_{\mathcal{R}}^2)}{\hat{t}-M_\mathcal{R}^2+i\,\Gamma_\mathcal{R}M_\mathcal{R}}\frac{m_Q}{M_\mathcal{R}}\left(\frac{1+4\,z}{1+2\,A}\right)^2\nonumber\\
&\quad\quad\times\lambda_{i,L}^\mathcal{R}\left(\lambda_{f,R}^{\mathcal{R}*}-\frac{1}{2}\frac{1+2\,A}{1+4\,z}\frac{m_Q}{M_\mathcal{R}}\lambda_{f,L}^{\mathcal{R}*}\right)\,d^{3/2}_{1/2,-3/2}(\theta)\\
\mathcal{M}_t(+,-,+,+)&=\frac{1}{\sqrt{3}}\frac{g_S^2}{\Lambda^2}\left(\overline{K}_AK^B\right)^{j}\,_i \frac{m_QM_\mathcal{R}(\hat{s}-m_Q^2)}{\hat{t}-M_\mathcal{R}^2+i\,\Gamma_\mathcal{R}M_\mathcal{R}}\lambda^\mathcal{R}_{i,R}\lambda^{\mathcal{R}*}_{f,L}\,d^{3/2}_{3/2,-1/2}(\theta)\\
\mathcal{M}_t(-,+,-,-)&=\frac{1}{\sqrt{3}}\frac{g_S^2}{\Lambda^2}\left(\overline{K}_AK^B\right)^{j}\,_i \frac{m_QM_\mathcal{R}(\hat{s}-m_Q^2)}{\hat{t}-M_\mathcal{R}^2+i\,\Gamma_\mathcal{R}M_\mathcal{R}}\lambda^\mathcal{R}_{i,L}\lambda^{\mathcal{R}*}_{f,R}\,d^{3/2}_{-3/2,1/2}(\theta)\\
\mathcal{M}_t(+,-,-,-)&=\frac{1}{3\sqrt{3}}\frac{g_S^2}{\Lambda^2}\left(\overline{K}_AK^B\right)^{j}\,_i\frac{(\hat{s}-m_Q^2)(\hat{t}-M_\mathcal{R}^2)}{\hat{t}-M_\mathcal{R}^2+i\,\Gamma_{\mathcal{R}}M_{\mathcal{R}}}\frac{m_Q\sqrt{\hat{s}}}{M_\mathcal{R}^2}\left(\frac{1+4\,z}{1+2\,A}\right)^2\nonumber\\
&\quad\quad\times\lambda_{i,R}^\mathcal{R}\lambda_{f,R}^{\mathcal{R}*}\,d^{3/2}_{3/2,1/2}(\theta)\\
\mathcal{M}_t(-,+,+,+)&=\frac{1}{3\sqrt{3}}\frac{g_S^2}{\Lambda^2}\left(\overline{K}_AK^B\right)^{j}\,_i\frac{(\hat{s}-m_Q^2)(\hat{t}-M_\mathcal{R}^2)}{\hat{t}-M_\mathcal{R}^2+i\,\Gamma_{\mathcal{R}}M_{\mathcal{R}}}\frac{m_Q\sqrt{\hat{s}}}{M_\mathcal{R}^2}\left(\frac{1+4\,z}{1+2\,A}\right)^2\nonumber\\
&\quad\quad\times\lambda_{i,L}^\mathcal{R}\lambda_{f,L}^{\mathcal{R}*}\,d^{3/2}_{-3/2,-1/2}(\theta)
\end{align}

There are a few things to note about these amplitudes.  First, if we go into the regime where the resonance is on-shell ($\hat{s}=M_\mathcal{R}^2$), all dependence on the parameter $z$ in the effective interaction disappears.  Even if we go into the unphysical regime $\hat{t}=M_\mathcal{R}^2$, the dependence on $z$ disappears.

Now consider the off-shell regime.  There seems to be a problem that the amplitudes appear to depend on the unphysical parameter $A$ from the propagator.  However, what is demanded is that the physical amplitudes are invariant underneath the transformation in Eq.~(\ref{eq:psiTrans}).  As shown above, this results in the transformations of the parameters $z$ and $A$ as given in Eqs.~(\ref{eq:ATrans}) and (\ref{eq:zTrans}). Using these transformations, it can be shown that
\begin{eqnarray}
\frac{1+4\,z}{1+2\,A}=\frac{1+4\,z'}{1+2\,A'}.
\end{eqnarray}
That is, all the amplitudes are invariant under the transformation in Eq.~(\ref{eq:psiTrans}).  However, the particular value of $z$ becomes scheme-dependent.

On the other hand, we can consider the effective interaction formulation in Eq.~(\ref{eq:general}).  This formulation is explicitly invariant under the transformations in Eqs.~(\ref{eq:psiTrans}) and (\ref{eq:ATrans}).  The two formulations can be identified by 
\begin{eqnarray}
z=\frac{1}{2}\left(1+4\,Z\right)+Z.
\end{eqnarray}
Then we find
\begin{eqnarray}
\frac{1+4\,z}{1+2\,A}=1+4\,Z.
\end{eqnarray}
That is, using the interaction in Eq.~(\ref{eq:general}), all dependence on the unphysical parameter $A$ disappears from all of the amplitudes.

Finally, we could use the formulation of the effective interaction in Eq.~(\ref{eq:invariant}).  As stated in the discussion of that equation, this is equivalent to the choice $z=-1/4$.  With this choice, all dependence of the amplitudes on the unphysical parameter $A$ vanishes.  However, man of the off-shell helicity amplitues would vanish as well.

%%%%%%%%%%%%%%%%%%%%%%%%%%%%%%%%%%%%%%%%%%%%%%%%%%%
\subsection{Initial color states $\rep8\otimes\rep8$}
\label{88.SEC}
Now we consider the scattering $g^Ag^B\rightarrow  \mathcal{R} \to  g^Cg^D$, where $A,B,C,D$ label the gluon color indices.
All initial and final state particles are massless in this case.  Hence, in the partonic center of momentum, the individual particle energies are $\sqrt{\hat{s}}/2$.  The helicities of the amplitudes are in the order $\mathcal{M}_s(g^A,g^B,g^C,g^D)$.  We only provide the dominant $s$-channel resonance amplitudes.

\subsubsection{Spin-0 resonance}
The non-zero amplitudes are
\begin{align}
\mathcal{M}_s(+,+,+,+)&=4\frac{g_S^2\kappa_S^2}{\Lambda_S^2}d^{ABE}d^{CDE}\frac{\hat{s}^2}{\hat{s}-M_\mathcal{R}^2+i\,\Gamma_\mathcal{R}M_\mathcal{R}}\\
\mathcal{M}_s(-,-,-,-)&=4\frac{g_S^2\kappa_S^2}{\Lambda_S^2}d^{ABE}d^{CDE}\frac{\hat{s}^2}{\hat{s}-M_\mathcal{R}^2+i\,\Gamma_\mathcal{R}M_\mathcal{R}}\\
\mathcal{M}_s(+,+,-,-)&=4\frac{g_S^2\kappa_S^2}{\Lambda_S^2}d^{ABE}d^{CDE}\frac{\hat{s}^2}{\hat{s}-M_\mathcal{R}^2+i\,\Gamma_\mathcal{R}M_\mathcal{R}}\\
\mathcal{M}_s(-,-,+,+)&=4\frac{g_S^2\kappa_S^2}{\Lambda_S^2}d^{ABE}d^{CDE}\frac{\hat{s}^2}{\hat{s}-M_\mathcal{R}^2+i\,\Gamma_\mathcal{R}M_\mathcal{R}}
\end{align}

\subsubsection{Spin-2 resonance}
The non-zero helicity amplitudes that survive when the resonance is on-shell are 
\begin{align}
\mathcal{M}_s(+,-,+,-)&=\frac{g_S^2\kappa_T^2}{\Lambda_T^2}d^{ABE}d^{CDE}\frac{\hat{s}^2}{\hat{s}-M_\mathcal{R}^2+i\,\Gamma_{\mathcal{R}}M_\mathcal{R}}d^2_{2,2}(\theta)\\
\mathcal{M}_s(-,+,-,+)&=\frac{g_S^2\kappa_T^2}{\Lambda_T^2}d^{ABE}d^{CDE}\frac{\hat{s}^2}{\hat{s}-M_\mathcal{R}^2+i\,\Gamma_{\mathcal{R}}M_\mathcal{R}}d^2_{-2,-2}(\theta)\\
\mathcal{M}_s(+,-,-,+)&=\frac{g_S^2\kappa_T^2}{\Lambda_T^2}d^{ABE}d^{CDE}\frac{\hat{s}^2}{\hat{s}-M_\mathcal{R}^2+i\,\Gamma_{\mathcal{R}}M_\mathcal{R}}d^2_{2,-2}(\theta)\\
\mathcal{M}_s(-,+,+,-)&=\frac{g_S^2\kappa_T^2}{\Lambda_T^2}d^{ABE}d^{CDE}\frac{\hat{s}^2}{\hat{s}-M_\mathcal{R}^2+i\,\Gamma_{\mathcal{R}}M_\mathcal{R}}d^2_{-2,2}(\theta)
\end{align}
There are four amplitudes that vanish when the resonance is on-shell:
\begin{align}
\mathcal{M}_s(+,+,-,-)&=\frac{1}{6}\frac{g_S^2\kappa_T^2}{\Lambda_T^2}d^{ABE}d^{CDE}\left(1+4\,f\right)^2\frac{(\hat{s}+2\,M_\mathcal{R}^2)(\hat{s}-M_\mathcal{R}^2)}{\hat{s}-M_{\mathcal{R}}^2+i\,\Gamma_{\mathcal{R}}M_{\mathcal{R}}}\frac{\hat{s}^2}{M_{\mathcal{R}}^4}\\
\mathcal{M}_s(-,-,+,+)&=\frac{1}{6}\frac{g_S^2\kappa_T^2}{\Lambda_T^2}d^{ABE}d^{CDE}\left(1+4\,f\right)^2\frac{(\hat{s}+2\,M_\mathcal{R}^2)(\hat{s}-M_\mathcal{R}^2)}{\hat{s}-M_{\mathcal{R}}^2+i\,\Gamma_{\mathcal{R}}M_{\mathcal{R}}}\frac{\hat{s}^2}{M_{\mathcal{R}}^4}\\
\mathcal{M}_s(+,+,+,+)&=\frac{1}{6}\frac{g_S^2\kappa_T^2}{\Lambda_T^2}d^{ABE}d^{CDE}\left(1+4\,f\right)^2\frac{(\hat{s}+2\,M_\mathcal{R}^2)(\hat{s}-M_\mathcal{R}^2)}{\hat{s}-M_{\mathcal{R}}^2+i\,\Gamma_{\mathcal{R}}M_{\mathcal{R}}}\frac{\hat{s}^2}{M_{\mathcal{R}}^4}\\
\mathcal{M}_s(-,-,-,-)&=\frac{1}{6}\frac{g_S^2\kappa_T^2}{\Lambda_T^2}d^{ABE}d^{CDE}\left(1+4\,f\right)^2\frac{(\hat{s}+2\,M_\mathcal{R}^2)(\hat{s}-M_\mathcal{R}^2)}{\hat{s}-M_{\mathcal{R}}^2+i\,\Gamma_{\mathcal{R}}M_{\mathcal{R}}}\frac{\hat{s}^2}{M_{\mathcal{R}}^4}
\end{align}
Note that when the resonance is on-shell $\hat{s}=M_\mathcal{R}^2$ all dependence on $f$ vanishes, as expected.

%%%%%%%%%%%%%%%%%%%%%%%%%%%%%%%%%%%%%%%%%%%%%%%%%%%%%%%%%

\subsection{Initial color states $\rep3\otimes\bar{\mathbf{3}}$}
\label{33b.SEC}
We consider the scattering $q_j\bar{q}'_k\rightarrow  \mathcal{R}  \to Q_l \bar{Q}'_n$ through color octet and singlet scalars, where $j,k,l,n$ label the quark color indices.  The helicities of the amplitudes are in the order $\mathcal{M}_s(q_j,\bar{q}'_k,Q_l,\bar{Q}'_n)$.  The amplitudes for the color octets are given.  The color singlet amplitudes can be found with the replacement $T^A_{ij}\rightarrow \delta_{ij}$.

In all amplitudes $\theta$ is the angle between the initial state quark $q$ and final state quark $Q$.  The subscript $i$ on the couplings indicates the initial state and $f$ the final state.

\subsubsection{Neutral spin-1 resonance}
For the neutral vector, we allow the final state particles to be massive but have equal masses $m_Q$.  For the color octet, the non-zero helicity amplitudes that survive for massless final state particles are:
\begin{align}
\mathcal{M}_s(+,-,+,-)&=2\,g_S^2\,T^A_{kj}T^A_{ln}\frac{\hat{s}}{\hat{s}-M_{\mathcal{R}}^2+i\,\Gamma_\mathcal{R}M_\mathcal{R}}\\
&\quad\quad\times g_{i,R}^\mathcal{R}\left[\frac{1}{2}g_{f,R}^\mathcal{R}\left(1+\sqrt{1-\frac{4\,m_Q^2}{\hat{s}}}\right)+\frac{1}{2}g_{f,L}^\mathcal{R}\left(1-\sqrt{1-\frac{4\,m_Q^2}{\hat{s}}}\right)\right]\,d^1_{1,1}(\theta)\nonumber\\
\mathcal{M}_s(-,+,-,+)&=2\,g_S^2\,T^A_{kj}T^A_{ln}\frac{\hat{s}}{\hat{s}-M_{\mathcal{R}}^2+i\,\Gamma_\mathcal{R}M_\mathcal{R}}\\
&\quad\quad\times g_{i,L}^\mathcal{R}\left[\frac{1}{2}g_{f,L}^\mathcal{R}\left(1+\sqrt{1-\frac{4\,m_Q^2}{\hat{s}}}\right)+\frac{1}{2}g_{f,R}^\mathcal{R}\left(1-\sqrt{1-\frac{4\,m_Q^2}{\hat{s}}}\right)\right]\,d^1_{-1,-1}(\theta)\nonumber\\
\mathcal{M}_s(+,-,-,+)&=-2\,g_S^2\,T^A_{kj}T^A_{ln}\frac{\hat{s}}{\hat{s}-M_{\mathcal{R}}^2+i\,\Gamma_\mathcal{R}M_\mathcal{R}}\\
&\quad\quad\times g_{i,R}^\mathcal{R}\left[\frac{1}{2}g_{f,L}^\mathcal{R}\left(1+\sqrt{1-\frac{4\,m_Q^2}{\hat{s}}}\right)+\frac{1}{2}g_{f,R}^\mathcal{R}\left(1-\sqrt{1-\frac{4\,m_Q^2}{\hat{s}}}\right)\right]\,d^1_{1,-1}(\theta)\nonumber\\
\mathcal{M}_s(-,+,+,-)&=-2\,g_S^2\,T^A_{kj}T^A_{ln}\frac{\hat{s}}{\hat{s}-M_{\mathcal{R}}^2+i\,\Gamma_\mathcal{R}M_\mathcal{R}}\\
&\quad\quad\times g_{i,L}^\mathcal{R}\left[\frac{1}{2}g_{f,R}^\mathcal{R}\left(1+\sqrt{1-\frac{4\,m_Q^2}{\hat{s}}}\right)+\frac{1}{2}g_{f,L}^\mathcal{R}\left(1-\sqrt{1-\frac{4\,m_Q^2}{\hat{s}}}\right)\right]\,d^1_{-1,+1}(\theta)\nonumber
\end{align}
The amplitudes that vanish for massless final state quarks are
\begin{align}
\mathcal{M}_s(+,-,+,+)&=-\sqrt{2}\,g_S^2\,T^A_{kj}T^A_{ln}\,\frac{m_Q\sqrt{\hat{s}}}{\hat{s}-M_{\mathcal{R}}^2+i\,\Gamma_\mathcal{R}M_\mathcal{R}}g_{i,R}^\mathcal{R}\left(g_{f,L}^\mathcal{R}+g_{f,R}^\mathcal{R}\right)d^1_{1,0}(\theta)\\
\mathcal{M}_s(-,+,-,-)&=-\sqrt{2}\,g_S^2\,T^A_{kj}T^A_{ln}\,\frac{m_Q\sqrt{\hat{s}}}{\hat{s}-M_{\mathcal{R}}^2+i\,\Gamma_\mathcal{R}M_\mathcal{R}}g_{i,L}^\mathcal{R}\left(g_{f,L}^\mathcal{R}+g_{f,R}^\mathcal{R}\right)d^1_{-1,0}(\theta)\\
\mathcal{M}_s(+,-,-,-)&=\sqrt{2}\,g_S^2\,T^A_{kj}T^A_{ln}\,\frac{m_Q\sqrt{\hat{s}}}{\hat{s}-M_{\mathcal{R}}^2+i\,\Gamma_\mathcal{R}M_\mathcal{R}}g_{i,R}^\mathcal{R}\left(g_{f,L}^\mathcal{R}+g_{f,R}^\mathcal{R}\right)d^1_{1,0}(\theta)\\
\mathcal{M}_s(-,+,+,+)&=\sqrt{2}\,g_S^2\,T^A_{kj}T^A_{ln}\,\frac{m_Q\sqrt{\hat{s}}}{\hat{s}-M_{\mathcal{R}}^2+i\,\Gamma_\mathcal{R}M_\mathcal{R}}g_{i,L}^\mathcal{R}\left(g_{f,L}^\mathcal{R}+g_{f,R}^\mathcal{R}\right)d^1_{-1,0}(\theta)
\end{align}

\subsubsection{Charged spin-1 resonance}

For the charged spin 1-resonance we allow the final state particles to have different masses.  Their energies are then
\begin{eqnarray}
E_Q=\frac{\hat{s}+m_Q^2-m_{Q'}^2}{2\sqrt{\hat{s}}},\quad{\rm and}\quad E_{Q'}=\frac{\hat{s}+m_{Q'}^2-m_Q^2}{2\sqrt{s}}.
\end{eqnarray}
The $\beta$ factors (speed in the partonic center of momenutm frame) are $\beta_Q=|\mathbf{p}_f|/E_Q$ and $\beta_{Q'}=|\mathbf{p}_f|/E_{Q'}$, where $\mathbf{p}_f$ is the three momentum of one of the final state quarks.  We provide amplitudes for when the quarks are up-type and the anti-quarks down-type.

The non-zero helicity amplitudes are then that survive in the zero quark mass limit are
\begin{align}
\mathcal{M}_s(+,-,+,-)&=g_S^2T^A_{kj}T^A_{ln}\frac{\sqrt{\hat{s}^2-(m_Q^2-m_{Q'}^2)^2}}{\hat{s}-M_\mathcal{R}^2+i\,\Gamma_\mathcal{R}M_\mathcal{R}} C_{i,R}^\mathcal{R}V_{R,q'q}^{CKM}\\
&\times\left(\sqrt{(1+\beta_Q)(1+\beta_{Q'})}C_{f,R}^{\mathcal{R}*}V_{R,Q'Q}^{CKM\,*}+\sqrt{(1-\beta_Q)(1-\beta_{Q'})}C_{f,L}^{\mathcal{R}*}V_{L,Q'Q}^{CKM\,*}\right)d^1_{1,1}(\theta)\nonumber\\
\mathcal{M}_s(-,+,-,+)&=g_S^2T^A_{kj}T^A_{ln}\frac{\sqrt{\hat{s}^2-(m_Q^2-m_{Q'}^2)^2}}{\hat{s}-M_\mathcal{R}^2+i\,\Gamma_\mathcal{R}M_\mathcal{R}} C_{i,L}^\mathcal{R}V_{L,q'q}^{CKM}\\
&\times\left(\sqrt{(1+\beta_Q)(1+\beta_{Q'})}C_{f,L}^{\mathcal{R}*}V_{L,Q'Q}^{CKM\,*}+\sqrt{(1-\beta_Q)(1-\beta_{Q'})}C_{f,R}^{\mathcal{R}*}V_{R,Q'Q}^{CKM\,*}\right)d^1_{-1,-1}(\theta)\nonumber\\
\mathcal{M}_s(+,-,-,+)&=-g_S^2T^A_{kj}T^A_{ln}\frac{\sqrt{\hat{s}^2-(m_Q^2-m_{Q'}^2)^2}}{\hat{s}-M_\mathcal{R}^2+i\,\Gamma_\mathcal{R}M_\mathcal{R}} C_{i,R}^\mathcal{R}V_{R,q'q}^{CKM}\\
&\times\left(\sqrt{(1+\beta_Q)(1+\beta_{Q'})}C_{f,L}^{\mathcal{R}*}V_{L,Q'Q}^{CKM\,*}+\sqrt{(1-\beta_Q)(1-\beta_{Q'})}C_{f,R}^{\mathcal{R}*}V_{R,Q'Q}^{CKM\,*}\right)d^1_{1,-1}(\theta)\nonumber\\
\mathcal{M}_s(-,+,+,-)&=-g_S^2T^A_{kj}T^A_{ln}\frac{\sqrt{\hat{s}^2-(m_Q^2-m_{Q'}^2)^2}}{\hat{s}-M_\mathcal{R}^2+i\,\Gamma_\mathcal{R}M_\mathcal{R}} C_{i,L}^\mathcal{R}V_{L,q'q}^{CKM}\\
&\times\left(\sqrt{(1+\beta_Q)(1+\beta_{Q'})}C_{f,R}^{\mathcal{R}*}V_{R,Q'Q}^{CKM\,*}+\sqrt{(1-\beta_Q)(1-\beta_{Q'})}C_{f,L}^{\mathcal{R}*}V_{L,Q'Q}^{CKM\,*}\right)d^1_{-1,1}(\theta)\nonumber
\end{align}
There are also amplitudes that vanish when both final state quarks are massless ($\beta_Q=\beta_{Q'}=1$):
\begin{align}
\mathcal{M}_s(+,-,+,+)&=-\frac{1}{\sqrt{2}}g_S^2T^A_{kj}T^A_{ln}\frac{\sqrt{\hat{s}^2-(m_Q^2-m_{Q'}^2)^2}}{\hat{s}-M_\mathcal{R}^2+i\,\Gamma_\mathcal{R}M_\mathcal{R}} C_{i,R}^\mathcal{R}V_{R,q'q}^{CKM}\\
&\times\left(\sqrt{(1+\beta_Q)(1-\beta_{Q'})}C_{f,R}^{\mathcal{R}*}V_{R,Q'Q}^{CKM\,*}+\sqrt{(1-\beta_Q)(1+\beta_{Q'})}C_{f,L}^{\mathcal{R}*}V_{L,Q'Q}^{CKM\,*}\right)d^1_{1,0}(\theta)\nonumber\\
\mathcal{M}_s(-,+,-,-)&=-\frac{1}{\sqrt{2}}g_S^2T^A_{kj}T^A_{ln}\frac{\sqrt{\hat{s}^2-(m_Q^2-m_{Q'}^2)^2}}{\hat{s}-M_\mathcal{R}^2+i\,\Gamma_\mathcal{R}M_\mathcal{R}} C_{i,L}^\mathcal{R}V_{L,q'q}^{CKM}\\
&\times\left(\sqrt{(1+\beta_Q)(1-\beta_{Q'})}C_{f,L}^{\mathcal{R}*}V_{L,Q'Q}^{CKM\,*}+\sqrt{(1-\beta_Q)(1+\beta_{Q'})}C_{f,R}^{\mathcal{R}*}V_{R,Q'Q}^{CKM\,*}\right)d^1_{-1,0}(\theta)\nonumber\\
\mathcal{M}_s(+,-,-,-)&=\frac{1}{\sqrt{2}}g_S^2T^A_{kj}T^A_{ln}\frac{\sqrt{\hat{s}^2-(m_Q^2-m_{Q'}^2)^2}}{\hat{s}-M_\mathcal{R}^2+i\,\Gamma_\mathcal{R}M_\mathcal{R}} C_{i,R}^\mathcal{R}V_{R,q'q}^{CKM}\\
&\times\left(\sqrt{(1+\beta_Q)(1-\beta_{Q'})}C_{f,L}^{\mathcal{R}*}V_{L,Q'Q}^{CKM\,*}+\sqrt{(1-\beta_Q)(1+\beta_{Q'})}C_{f,R}^{\mathcal{R}*}V_{R,Q'Q}^{CKM\,*}\right)d^1_{1,0}(\theta)\nonumber\\
\mathcal{M}_s(-,+,+,+)&=\frac{1}{\sqrt{2}}g_S^2T^A_{kj}T^A_{ln}\frac{\sqrt{\hat{s}^2-(m_Q^2-m_{Q'}^2)^2}}{\hat{s}-M_\mathcal{R}^2+i\,\Gamma_\mathcal{R}M_\mathcal{R}} C_{i,L}^\mathcal{R}V_{L,q'q}^{CKM}\\
&\times\left(\sqrt{(1+\beta_Q)(1-\beta_{Q'})}C_{f,R}^{\mathcal{R}*}V_{R,Q'Q}^{CKM\,*}+\sqrt{(1-\beta_Q)(1+\beta_{Q'})}C_{f,L}^{\mathcal{R}*}V_{L,Q'Q}^{CKM\,*}\right)d^1_{-1,0}(\theta)\nonumber
\end{align}

%%%%%%%%%%%%%%%%%%%%%%%%%%%%%%%%%%%%%%%%%%%%%%%
%\section{Color flow diagrams}
%
%\clearpage
%
%%%%%%%%%%%%%%%%%%%%%%%%%%%%%%%%%%%%%%%%%%%%%%%
%%
\bibliographystyle{JHEP}
\bibliography{ref}
\end{document}